\RequirePackage{fix-cm}
\documentclass{svjour3}                     
\smartqed  
\usepackage{graphicx}
\usepackage{url}
\usepackage{comment}
\usepackage{multirow}
\usepackage{float}
\usepackage{longtable}
\usepackage{cite}
\usepackage{csquotes}
\usepackage{mathptmx}
\usepackage[table,xcdraw]{xcolor}
\usepackage[title]{appendix}

\usepackage[colorinlistoftodos]{todonotes}

\usepackage{authorbiography}%
\usepackage{hyperref}%
\usepackage{blindtext}%

\newcommand{\RQone}{What are the existing criteria to identify whether a software project uses CI?}
\newcommand{\RQtwo}{What are the reported claims regarding the effects of CI on software development?}
\newcommand{\RQthree}{Which empirical methods, projects and artifacts are used in the studies that investigate the effects of CI on software development?}

\newenvironment{boxed}
    {\begin{center}
    \begin{tabular}{|p{0.9\textwidth}|}
    \hline\\
    }
    { 
    \\\hline
    \end{tabular} 
    \end{center}
    }

\begin{document}

\title{The Effects of Continuous Integration on Software Development: a Systematic Literature Review
}


\author{Eliezio Soares         \and
        Gustavo Sizilio        \and
        Jadson Santos        \and
        Daniel Alencar        \and
        Uirá Kulesza
}


\institute{Federal Institute of Rio Grande do Norte - IFRN \\
            Federal University of Rio Grande do Norte - UFRN \at
            Natal-RN, Brasil\\
              \email{eliezio.soares@ifrn.edu.br}             
           \and
           Federal Institute of Rio Grande do Norte - IFRN \\
            Federal University of Rio Grande do Norte - UFRN \at
            Natal-RN, Brasil\\
              \email{gustavo.sizilio@ifrn.edu.br}
          \and
          Federal University of Rio Grande do Norte - UFRN \at
          Natal-RN, Brasil\\
          \email{jadson.santos@ufrn.br}      
          \and
          University of Otago \at
          Dunedin, New Zealand   \\
          \email{danielcalencar@otago.ac.nz}      
          \and
          Federal University of Rio Grande do Norte - UFRN \at
          Natal-RN, Brasil\\
          \email{uira@dimap.ufrn.br}    
}

\date{Received: date / Accepted: date}

\maketitle

\begin{abstract}

\textbf{Context:} Continuous integration (CI) is a software engineering technique that proclaims frequent activities to assure the software product health. Researchers and practitioners mention several benefits related to CI. However, no systematic study surveys state of the art regarding such benefits or cons. 
\textbf{Objective:} This study aims to identify and interpret empirical evidence regarding how CI impacts software development.  
\textbf{Method:} Through a Systematic Literature Review, we search for studies in six digital libraries. Starting from 479 studies, we select 101 empirical studies that evaluate CI in the context of software development. We thoroughly read and extract information regarding (i) the CI environment, (ii) findings related to the effects of CI, and (iii) the employed research methods. We apply a {\em thematic synthesis} to group and summarize the findings.
\textbf{Results:} Existing research has explored the positive effects of CI, such as better cooperation, or negative effects, such as adding technical and process challenges. From our thematic synthesis, we identify six themes: {\em development activities}, {\em software process}, {\em quality assurance}, {\em integration patterns}, {\em issues \& defects}, and {\em build patterns}. 
\textbf{Conclusions:} Empirical research in CI has been increasing over recent years. We found that much of the existing research reveals that CI brings positive effects to software development. However, CI may also bring technical challenges to software development teams. Despite the overall positive outlook regarding CI, we still find room for improvements in the existing empirical research that evaluates the effects of CI.

\end{abstract}

\section{Introduction \label{intro}}	

Continuous integration (CI) is a software engineering practice that became increasingly popular with the eXtreme Programming methodology, which was proposed by Beck K. \cite{beck2005} in the late 90s. CI, as a practice, proposes the usage of a set of sub-practices that are synergic, e.g., frequent code commits,  automated tests, frequent builds, immediately fixing a broken build, among others \cite{fowler2006, duvall2013,stahl2014}. Thereby, CI aims to reduce the cost and risk of work integration among distributed teams \cite{fowler2006}. The potential of CI stands out even more in a global world with increasing distributed software development. This scenario demands strong coordination and control from software development teams facing temporal, geographical, and socio-cultural challenges \cite{homstrom2006}.

As Continuous Integration (CI) gained popularity, several benefits related to CI were proclaimed, such as risks reduction, improvement of project visibility and predictability, greater confidence in the software product, easiness to locate and fix bugs, improvement in team communication, among others \cite{fowler2006, duvall2013, stahl2013}.


Given the increasing popularity of CI along with CI's claimed benefits, there have been substantial research efforts related to CI. Researchers have investigated CI practices~\cite{vassalo2018, Yu2016, pinto2018}, environments \& tools \cite{stahl2014b, zampetti2017, Johanssen2018}, potential benefits, \cite{Bernardo2018, Embury2019, stahl2013}, potential problems \cite{Rausch2017, Ghaleb2019, vassalo2019, Debbiche2014}, and even new practices \cite{rogers2004, Volf2017, Meedeniya2019} in different project settings. Considering these research contributions and the increasing need for distributed software development, our community needs a clear map of the empirical advantages or disadvantages of using CI. 

Existing research has summarized the findings in the literature concerning agile methods \cite{dikert2016}, continuous integration, delivery, and deployment \cite{laukkanen2017, shahin2017, stahl2013}. However, what is missing is a systematic study that covers and summarizes all the potential benefits and cons pertaining to the usage of CI (i.e., the effects of adopting CI on the development process). Such a study can better inform practitioners and researchers about the potential of using CI as well as future avenues for research.

In this work, we investigate: (i) how CI has been evaluated by existing research along with their results; (ii) the criteria used to identify whether a given project uses CI or not—which is essential for designing empirical studies related to CI—and (iii) which research methodologies have been used in existing studies to evaluate the potential effects of adopting CI. 

Our work is a {\em Systematic Literature Review} (SLR) \cite{kitchenham2007} of the existing empirical evidence regarding the effects of CI in diverse software development activities. Accordingly, we encompass diverse empirical methods and various associations with CI, i.e., the effects of CI in different variables such as test coverage, bugs reported, team communication, time to deliver pull-requests, among others. Considering this variability in the empirical methods and diversity of variables (i.e., bugs, tests, pull-requests, and others), we do not intend to perform a meta-analysis. Instead, we present an interpretive SLR. As such, we intend to draw a picture of the reported benefits and cons of adopting CI and collate the claims made about Continuous Integration in the existing literature, assessing the strength of these claims in a systematic manner.  In this way, this work offers meaningful and relevant evidence-based support for practitioners, organizations, and researchers.

Given the goal of our SLR, we investigate the following research questions:

\begin{itemize}
    \item \textbf{RQ1.} \RQone 
    \item \textbf{RQ2.} \RQtwo
    \item \textbf{RQ3.} \RQthree
\end{itemize}


The findings regarding the effects of CI and their evidence are discussed across six themes: {\em development activities}, {\em software process}, {\em quality assurance}, {\em integration patterns}, {\em issues \& defects}, and {\em build patterns}. Our paper provides researchers and practitioners with state-of-the-art empirical claims related to the effects of CI while collating their existing evidence and with insights regarding the interrelation between research methodologies, quality assessment, and themes, which delineate potential future studies in these themes (see Section~\ref{conclusion}).

The remainder of this paper is organized as follows: Section~\ref{back} provides the background regarding continuous integration and investigates related SLRs. In Section~\ref{meth}, we describe our study. We present the quantitative and qualitative results of the research questions in Section~\ref{results}. In Section~\ref{discussion}, we discuss our main findings and provide research insights. We discuss the threats to validity in Section~\ref{threats}, and we draw conclusions and future directions in Section~\ref{conclusion}.

\section{Background and related work}
\label{back}

In this section, we provide the background material regarding CI (section \ref{continuous_integration}) and discuss the existing systematic literature reviews that are related to our work (section \ref{related}).

\subsection{Continuous Integration \label{continuous_integration}}

Continuous Integration (CI) is one of the practices of eXtreme Programming (XP) methodology proposed by Beck K \cite{beck2005}. The overarching goal of CI is to reduce the cost of integrating the code developed by different developers in a team (or different teams) by making integration a daily practice. For example, there must be no more than a couple of hours between code integration. While CI compels the code to be collective and the knowledge to be shared more, CI's main benefit is the reduced risk of a big and cumbersome integration (e.g., after days, weeks, or months of work developed) \cite{beck2005}.  

To properly employ CI, at least four mechanisms are required: (i) a version control system, (ii) a build script, (iii) a feedback mechanism, and (iv) a process for integrating the source code changes \cite{duvall2013}. Modern distributed version control systems (VCS), especially those based on \textsc{Git}, have grown in popularity because of social coding platforms, such as \textsc{GitHub} \cite{vasilescu2015}, which have fostered collaborative software development. Within these popular social coding platforms, several services have been proposed to support CI (e.g., \textsc{TravisCI}, \textsc{CircleCI} and \textsc{Jenkins}), easing the automation of build pipelines, which are triggered by source code changes on the VCS \cite{hilton2016}.

Studies have reported an increasing number of projects adopting the continuous integration practice~\cite{hilton2016}, and some of such studies bring up evidence showing changes in the practice of these projects, such as higher commit frequency and an increase in test automation \cite{zhao2017}.

Duvall et al.~\cite{duvall2013} advocate that CI is the centerpiece of software development, ensuring the health and determining the quality of software. To get the benefits of CI, the authors argue that developers should implement a set of sub practices on a daily basis, whereas implementing only a fraction of the practices is not enough to employ CI. Fowler, in his definition of CI, also mentions a series of critical practices to make CI effective \cite{fowler2006} (see Table \ref{tab:ci_practices}). 


Nevertheless, some authors have studied differences in implemented CI processes and demonstrated a lack of consensus regarding these CI processes, which results in many CI variants \cite{stahl2014,viggiato2019}. Ståhl \& Bosch \cite{stahl2014} identified variation points from 16 out of 22 clusters of CI practices and argue that it is necessary to investigate which kind of continuous integration a project applies when analyzing or assessing projects. Viggiato et al. \cite{viggiato2019} suggested that continuous integration may not always be homogeneous, i.e., CI may have different usages across different domains. Studies still suggest the inclusion of other practices to potentialize benefits, Vassalo C et al. \cite{vassalo2018}, for example, suggests adding ``\textit{continuous refactoring}'' as a CI best practice as it is useful to control the increasing complexity of the changes.

On top of that, there is a discussion regarding an existing confusion around the definition of Continuous Integration (CI), Continuous Delivery (CDE), and Continuous Deployment (CD), or still, the recent emphasis on DevOps shedding light on the integration between software development and its operational deployment \cite{shahin2017,fitzgerald2017}. A conservative perspective presents these continuous practices as sequential and well-defined techniques, i.e., CI as a foundation for CDE in such a manner that an organization should implement a reliable CI practice to adopt CDE, in the same way, to implement CD an organization should implement CDE \cite{shahin2017}. Fitzgerald B \& Stol K \cite{fitzgerald2017}, in turn, defends a holistic view ---``Continuous  \(\displaystyle *\)''--- including Business Strategy \& Planning, Development, and Operations, in which CI incorporates CDE, and CD.

Therefore the variability around the continuous practices and the dynamic nature of the employed practices in continuous integration leads to a potentially endless variation of CI implementations and a lack of consensus on an exact definition of CI. Considering the lack of consensus regarding an exact definition of CI, in our research, we focus on the practices discussed by Duvall et al.~\cite{duvall2013} and Fowler~\cite{fowler2006}for one main reason.  While other authors reveal the variability around continuous integration, they often do not provide concrete guidelines as to what should be considered CI or not. Conversely, Duvall et al.~\cite{duvall2013} and Fowler~\cite{fowler2006} present a concrete minimum number of practices that projects should adopt in order to use CI.

\subsubsection{Continuous Integration Practices\label{ci_practices}}

Table \ref{tab:ci_practices} shows an overview of the practices proposed by Duvall et al. ~\cite{duvall2013} and those reported by Fowler \cite{fowler2006}. The practices proposed by Duvall are shown in the second column, while the third column shows the practices reported by Fowler. In the first column, we organize the CI practices into four groups: (i) integration, (ii) test, (iii) build, and (iv) feedback. 
\begin{table}[H]
    \caption{Continuous integration practices enumarated by Duvall et al. \cite{duvall2013} and Fowler \cite{fowler2006}.}
    \label{tab:ci_practices}
    \begin{tabular}{ p{.13\textwidth} p{.38\textwidth} p{.49\textwidth}}
    \hline
                                                                &   \multicolumn{1}{c}{\textbf{Duvall et al. practices \cite{duvall2013}}} & \multicolumn{1}{c}{\textbf{Fowler practices \cite{fowler2006}}}                                                               \\ \hline 
    \multirow{2}{*}{\begin{tabular}[c]{@{}l@{}}Integration\\ Practices\end{tabular}}                       &      Commit code frequently                                & Everyone commits to the mainline every day                                                                  \\ \cline{2-3} 
                                                                &        -                                                     & Maintain a single source repository                                                                        \\ \hline
    \multirow{2}{*}{\begin{tabular}[c]{@{}l@{}}\\ \\Test\\ Practices\end{tabular}}                            &        Write automated developer tests                       & Make your build self-testing                                                                                \\ \cline{2-3} 
                                                                &        All tests and inspections must pass                   & Test in a clone of the production environment                                                                                                           \\ \cline{2-3} 
                                                                &        -                                                     & \begin{tabular}[c]{@{}l@{}}Make it easy for anyone to get the \\latest executable\end{tabular}             \\ \cline{2-3} 
                                                                &        -                                                     & Automate deployment                                                                                       \\ \hline
    \multirow{2}{*}{\begin{tabular}[c]{@{}l@{}}\\ \\Build\\ Practices\end{tabular}}                            &        Don’t commit broken code                              & Automate the build                                                                                          \\ \cline{2-3} 
                                                                &        Run private builds                                                     & \begin{tabular}[c]{@{}l@{}}Every commit should build the mainline \\on an integration machine\end{tabular} \\ \cline{2-3} 
                                                                &        Fix broken builds immediately                         & Fix broken builds immediately                                                                               \\ \cline{2-3} 
                                                                &        -                                                     & Keep the build fast                                                                                         \\ \hline
    Feedback Practices                                      &        Avoid getting broken code                             & Everyone can see what's happening                                                                           \\ \hline
\end{tabular}
\end{table}

``\textit{Commit code frequently}'' is the practice of integrating code changes as ``{\em early and often}'' as possible to a ``\textit{single source code repository}'' (e.g., \textsc{GitHub}, \textsc{GitLab}, or \textsc{Bitbucket}). This practice is central to CI because it prevents a complex integration---an integration that requires more time and effort---while treating potential integrations problems~\cite{duvall2013, fowler2006}.

When it comes to testing, CI bears the principle that ``\textit{all tests and inspections must pass}''. This practice advocates that not only tests must pass but also the inspections related to coding and design standards (e.g., test coverage, cyclomatic complexity, or others). Ideally, the tests and inspections should be triggered in an automated fashion. Therefore, CI requires developers to ``\textit{write automated development tests}'', ``\textit{making the builds to become self-testing}'', which enables a fully automated build process that provides meaningful feedback \cite{duvall2013, fowler2006}. 

Still regarding tests, Fowler recommends to ``\textit{test the software in a clone of the production environment}'' to mitigate the risk of not identifying problems that would occur only within the production environment. For this reason, Fowler also proposes the ``\textit{automated deployment}''---to prepare test-environments automatically---and the practice of ``\textit{making it easy for anyone to get the latest executable}''---so that anyone has easy access to the current state of development \cite{fowler2006}.

With respect to build practices, the team must follow the ``\textit{don't commit broken code}'' practice. To do so, it is vital to employ the ``\textit{automate the build}'' practice. The build automation consists of empowering the team with scripts that fully manage the build process, from dependency managers (e.g., \textsc{Maven}, \textsc{Gradle}, \textsc{NuGet}, or \textsc{Bundler}) and tests to database schema, or other required tool. Once a consistent build script is set, developers should ``\textit{run private builds}'' that emulate an integration build in their workstation, ensuring a well-succeeded build process before integrating their changes into the central repository (i.e., the mainline)~\cite{duvall2013, fowler2006}.

Additionally, Fowler recommends that ``\textit{every commit should build the mainline on an integration machine}'', i.e., a change sent to the mainline repository must trigger a build process in a dedicated server. It is also important to ``\textit{keep the build fast}'', so the dedicated server can be effective to give rapid feedback, helping developers to ``\textit{fix broken builds immediately}''. Regarding build duration, the eXtreme Programming (XP) recommends the limit of 10 minutes. Builds that take more than 10 minutes may lead the development team to give up on using CI \cite{fowler2006, beck2005}.

``\textit{Fix broken builds immediately}'' is cited both by Fowler and Duvall et al. A build may break due to a compilation error, a failed test, or several other reasons. When a build is broken, the development team must focus on fixing the build before any other implementation activity---the build should be always {\em on green}.  

There is also CI practices related to feedback. One example is the practice ``\textit{everyone can see what's happening}'', which makes the communication clear and transparent within or across development teams. The immediate feedback from CI allows the development team to ``\textit{avoid getting broken code}''. In other words, a developer can check the current build status before performing a checkout (or pull)~\cite{fowler2006, beck2005}.


\subsection{Related Work \label{related}}

In this section, we discuss {\em Systematic Literature Reviews} (SLR) that are related to our work. We highlight the main differences in contributions and findings of five others SLRs, as shown in Table~\ref{tab:related_works}. In particular, Dikert K et al. studied agile methods~\cite{dikert2016}, Laukkanen E et al. studied continuous delivery~\cite{laukkanen2017}, Shahin et al. studied continuous integration, delivery and deployment \cite{shahin2017}. Two other studies by Ståhl \& Bosch investigated the existing literature regarding CI \cite{stahl2013,stahl2014}.

Ståhl \& Bosch \cite{stahl2013} investigated which known benefits of CI are experienced in the industry. They conducted a systematic literature review, including 33 articles with 7 explicit claims regarding the benefits of CI. They interviewed 22 individuals (developers, testers, project managers and line managers) from 4 projects to complement the study. Their results reveal high standard deviations in answers, indicating disparate reported experiences. 

Another study by Ståhl \& Bosch \cite{stahl2014}, motivated by their previous work, performed a literature review on CI to better understand the different benefits of CI adoption. The new study included 46 articles to find differing practices, supporting identifying potential CI variation points. They synthesized the extracted statements in 22 clusters, of which only six do not have disagreements. In addition, the study proposes a descriptive model for documenting these variations and highlights the need for better documenting such CI variants to understand any benefit or disadvantage of them better. 

Dikert et al. \cite{dikert2016}, conducted an SLR on large-scale agile transformations (i.e., changes in practices or organizational culture in companies with 50 or more people, or at least six teams) to identify success and challenge factors. They searched for papers describing industrial cases in agile development adoption, including 52 publications in a thematic synthesis. The authors documented 35 challenges in 9 categories and 29 success factors distributed into 11 categories. 

Shahin et al. \cite{shahin2017} studied 69 papers in a SLR to classify approaches/tools and to identify challenges and practices in Continuous Integration, Continuous Delivery, and Continuous Deployment. The contributions of the study include the classification of approaches/tools, a list of critical factors to implement continuous practices, a guide to select approaches/tools, and a list of research directions.

Laukkanen et al. \cite{laukkanen2017}, also performed a SLR to explore the reported problems when adopting Continuous Delivery. Their study also identified causes and solutions to these problems. The study selected 30 articles in which they found 40 problems and 29 solutions. The problems and solutions were classified into seven themes, e.g., integration, testing, and build design. 

Some of these studies focus on a more general perspective, such as Dikert et al. \cite{dikert2016} which focused on agile methods adoption, and Shahin et al. \cite{shahin2017} which studied continuous integration, delivery, and deployment. Laukkanen et al. \cite{laukkanen2017} studied continuous delivery, which differs from the implications of using continuous integration. Our work focuses strictly on continuous integration, and we explore studies in-depth, analyzing and comparing their findings.

Ståhl \& Bosch \cite{stahl2013} also perform analyses strictly related to CI investigating the experienced benefits in the industry. However, while our study exhaustively explores the literature to analyze the claims related to CI (benefits and cons),  Ståhl \& Bosch \cite{stahl2013} do not provide an exhaustive list of CI benefits nor explore the potential adverse effects (or challenges) of adopting CI. In their subsequent work, Ståhl \& Bosch \cite{stahl2014} cataloged CI variation points. Our systematic literature review complements their work because we discuss the empirical claims related to CI across six different themes (each representing an area of software development). Our work helps practitioners and researchers to obtain a holistic view of the implications of using continuous integration in different areas of software development and different granularities of CI practices.

Differently from the studies mentioned above, our SLR analyzes a substantially larger sample of articles, i.e., 101 studies ranging from 2003 to 2019. Considering the years of the related research presented is noticeable that the newest (i.e.  Laukkanen et al. \cite{laukkanen2017} and Shahin et al. \cite{shahin2017}) is dated to 2017, including mostly primary from 2016 \cite{shahin2017}. Our work contributes to the community by advancing at least three years in the related literature. Our study also innovates not only by studying the effects of CI but also by considering the specific CI practices within different CI settings. Moreover, our work also analyzes the methodologies of our 101 selected studies.


\begin{table}[H]
	\caption{Systematic Literature Reviews (SLRs) that related to our work. We show the authors, focus, findings, number of included articles, and the year of publication.}
    \label{tab:related_works} 
    \begin{tabular}{lllcc}
    \hline
    \multicolumn{1}{c}{\textbf{Study}}                                                               & \multicolumn{1}{c}{\textbf{Focus}}                                           & \multicolumn{1}{c}{\textbf{Findings}}                                                  & \textbf{\# Papers} & \textbf{Year} \\ \hline
    \begin{tabular}[c]{@{}l@{}}Ståhl and Bosh \cite{stahl2013}\end{tabular}      & \begin{tabular}[c]{@{}l@{}}Continuous\\ integration\end{tabular}                                                                            & Benefits of CI                                                                         & 33                 & 2013          \\ \hline
    \begin{tabular}[c]{@{}l@{}}Ståhl and Bosh \cite{stahl2014}\end{tabular}      & \begin{tabular}[c]{@{}l@{}}Continuous\\ integration\end{tabular}                                                                          & \begin{tabular}[c]{@{}l@{}}Differences in \\ CI practices\end{tabular}                 & 46                 & 2014          \\ \hline
    \begin{tabular}[c]{@{}l@{}}Dikert et al. \cite{dikert2016}\end{tabular}      & \begin{tabular}[c]{@{}l@{}}Large-scale \\ agile transformations\end{tabular} & \begin{tabular}[c]{@{}l@{}}Challenges and \\ success of agile \\ adoption\end{tabular} & 52                 & 2016          \\ \hline
    \begin{tabular}[c]{@{}l@{}}Shahin et al. \cite{shahin2017}\end{tabular}      & \begin{tabular}[c]{@{}l@{}}Continuous \\integration, \\delivery, \\and deployment\end{tabular}                            & \begin{tabular}[c]{@{}l@{}}Approaches, \\tools, challenges \\and practices \end{tabular}                                        & 69                   & 2017          \\ \hline
    \begin{tabular}[c]{@{}l@{}}Laukkanen et al. \cite{laukkanen2017}\end{tabular} & \begin{tabular}[c]{@{}l@{}}Continuous\\ delivery\end{tabular}                               & \begin{tabular}[c]{@{}l@{}}Problems, causes\\ and solutions.\end{tabular}                                         & 30                   & 2017          \\ \hline
    \end{tabular}
\end{table}

\section{Research Method \label{meth}}

The main goal of our study is to provide a holistic view for researchers and practitioners regarding how Continuous Integration (CI) can influence the software development phenomena (both in terms of potential benefits and cons). Therefore, we conduct a {\em Systematic Literature Review} (SLR) of studies that investigated the potential effects of CI on software development. To evaluate the scientific rigour of our target studies, we investigate the methodologies that were employed in these studies. The purpose of this investigation is to better understand the strength of the existing scientific claims and inform the reader accordingly. We also consider how our target studies determined whether their subject projects used CI or not. Identifying whether a project uses CI is a crucial step in any study evaluating the effects of adopting CI as this is how empirical comparisons regarding CI vs. non-CI can be performed. To conduct our SLR, we follow the guidelines provided by Kitchenham \& Charters \cite{kitchenham2007}. 

The next subsections describe our review protocol \cite{kitchenham2007}. Section \ref{rqs} describes the rationale behind our research questions. Section \ref{search_strategy} details the search mechanisms that we perform. Section \ref{selection} describes the inclusion and exclusion criteria and the screening process. Section \ref{extraction} describes the data extraction details, while Section \ref{synthesis} reveals the procedures that we use to synthesize the collected data. Finally, Section \ref{quality_assessment} explains how we assess the quality of the studies.


\subsection{Research questions \label{rqs}}

To fulfill the goal of our study, we address the following research questions (RQs):

\textbf{RQ1: \RQone}

\textbf{Rationale.} Several authors have listed a set of practices or principles related to CI~\cite{duvall2013,fowler2017,stahl2014,zhao2017,viggiato2019}. Some of these practices include: ``commit code frequently'', ``test automation'', ``run private builds'', ``all tests and inspections must pass'', and ``fix broken builds immediately''. However, there exists evidence that many CI projects do not adopt many of these practices.

For example, Felidré et al.~\cite{felidre2019} analyzed 1,270 open-source projects using \textsc{TravisCI} (the most used CI server). They observed that about 60\% of the projects do not follow proper CI practices. For example, some projects have infrequent commits, low test coverage, and 85\% of projects take more than four days to fix certain builds. Therefore, in RQ1, we investigate which criteria have been applied in the studies to identify whether the subject projects employ CI or not. This investigation is important because it has a direct impact on the quality of the data. For example, if a project is deemed to be using CI, but performs infrequent commits and takes a long time to fix builds, the empirical results observed to such a project would not reflect proper CI usage.

\textbf{RQ2: \RQtwo}

\textbf{Rationale.} Most of practitioners adopt CI practices with the expectation of increasing the quality of software development \cite{leppanen2015}. Researchers have reported the benefits of applying CI~\cite{fowler2006, duvall2013}, such as risk reduction, decrease in repetitive manual processes, readily deployable software, improved project visibility, greater confidence in the software product, and easiness to locate and remove defects.

To help practitioners and researchers, from an evidence-based software engineering effort \cite{kitchenham2004}, this RQ aims to collect, organize, and compare the empirical investigations related to CI that were performed by existing studies, while highlighting the assumptions and claims associated with these empirical investigations.

\textbf{RQ3: \RQthree}

\textbf{Rationale.} As observed by Easterbrook S et al.~\cite{easterbrook2008}, there is a lack of guidance regarding which methods to apply in {\em Empirical Software Engineering} (ESE) studies---which leads many researchers to select an inappropriate methodology. Rodríguez-Pérez et al.~\cite{perez2018} investigated the reproducibility aspects of ESE through a case study. According to their investigations, 39\% of the papers that were analyzed did not provide sufficient data or documentation to support the reproduction of the studies. To better understand the methodologies that are applied in the ESE field with respect to CI, in this RQ, we shed light on the methods, evaluations, domains, and kind of projects that are investigated in our target studies.

\begin{figure}[H]
  \begin{center}
\includegraphics[scale=0.2]{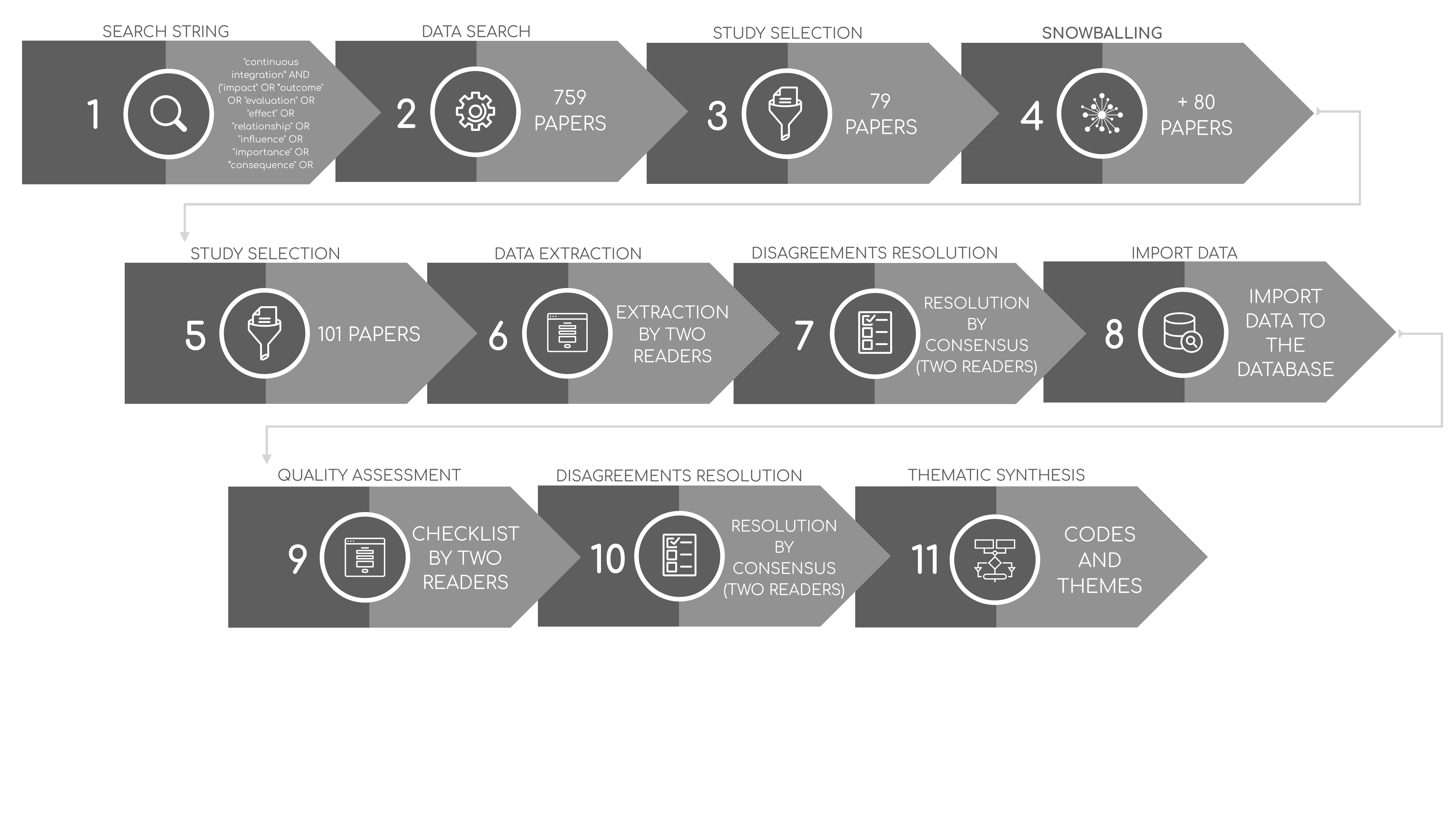}
\caption{Research methodology. Step 1: Search string definition; Step 2: Data search; Step 3: Study selection; Step 4: snowballing; Step 5: snowballing study selection; Step 6: Data extraction; Step 7: Disagreements resolution; Step 8: Database import; Step 9: Quality Assessment; Step 10: Quality assessment disagreements resolution; Step 11: Thematic synthesis.}
\label{fig:methodology}
  \end{center}
\end{figure}


\subsection{Search strategy \label{search_strategy}}

The search process of our SLR consists of the first six steps shown in Figure~\ref{fig:methodology}. Step 1--Definition of the search string (section \ref{string}); Step 2--Delimitation of the search mechanisms 
(section \ref{mechanisms}); Steps 3 to 5--Papers screening (section \ref{screening}).

\label{string}
\textit{\textbf{Search String.}} Our goal is to find studies that evaluate continuous integration and find pros or cons of adopting CI in any software development activity. Therefore, we use generic words that express the act of evaluating CI. We craft a string to fetch papers containing the term ``continuous integration'' and another word that expresses ``impact'' or ``effect'' in the title, abstract, or keywords. The terms we used were:

\begin{enumerate}
  \item ``continuous integration''
  \item (``impact'' \textbf{OR} ``outcome'' \textbf{OR} ``evaluation'' \textbf{OR} ``effect'' \textbf{OR} ``relationship'' \textbf{OR} ``influence'' \textbf{OR} ``importance'' \textbf{OR} ``consequence'' \textbf{OR} ``study'')
\end{enumerate}

Items 1 and 2 were combined with a boolean operator ``AND'' to match studies with both item 1 and at least one term from item 2. In this way, our search is denoted by the logical expression:
\begin{itemize}
  \item 1 AND 2
\end{itemize}

To operationalize our search string in different search engines, we first perform our search using only item 1. Once the results are obtained, i.e., papers containing ``continuous integration'' on the title, abstract, or keywords, we use scripts to filter out papers not satisfying item 2. The scripts used in this process are available in our digital appendix~\cite{soares_eliezio_2020}.

\label{mechanisms}
\textit{\textbf{Data Search.}} Regarding the selection of digital libraries, we considered Chen et al. \cite{chen2010} recommendations and included the main publishers’ sites and one index engine. Table \ref{tab:databases} shows the number of papers that we retrieved from each digital library. We apply the search string in each digital library separately and store the results in spreadsheets. As a result, our first search (i.e., step 2 from Fig. \ref{fig:methodology}) resulted in 759 papers.

\begin{table}[H]
	\caption{The digital libraries included in our search along with the number of matches (before and after removing duplicates).}
\label{tab:databases}    
\begin{tabular}{lcccc}
\hline
\multicolumn{1}{c}{\textbf{Database}} & \textbf{\# of matches} & \textbf{\%} & \textbf{\# without duplicates} & \textbf{\%}          \\ \hline
IEEE Xplore                           & 169                    & 22.27       & 130                            & 27.14                \\
ACM Digital Library                   & 121                    & 15.94       & 117                            & 27.14                \\
SpringerLink                          & 53                     & 6.98        & 53                             & 24.43                \\
Wiley Online Library                  & 4                      & 0.53        & 4                              & 0.84                 \\
ScienceDirect                         & 12                     & 1.58        & 12                             & 2.51                 \\
SCOPUS                                & 400                    & 52.70       & 163                            & 34.03                \\ \hline
                                    & 759                       &           & 479                           &
\multicolumn{1}{l}{}
\end{tabular}
\end{table}

\subsection{Study Selection \label{selection}}

After performing the first search, we proceed with the study selection step. In this section, we present our selection criteria (Section \ref{selection_criteria}) and the process of paper screening (Section~\ref{screening}).

\subsubsection{Selection Criteria \label{selection_criteria}}

In this step, we apply the inclusion and exclusion criteria based on our RQs. This step is necessary to aim for relevant papers retrieved from the studied digital libraries. We apply our inclusion and exclusion criteria in steps 3, 4, and 5 (see Figure \ref{fig:methodology}).

As our work aims to collect evidence reported in the literature regarding the effects of continuous integration (CI) on software development, we are strictly interested in finding empirical studies reporting an evaluation of CI projects or CI project settings (e.g., employees and organization characteristics). To maintain a rigour in our analyses, our selected studies must meet a minimum set of quality criteria to provide our review with reliable evidence (we present our quality criteria with more details in Section~\ref{quality_assessment}). Given that most international and high-quality research venues in software engineering use English as their official language (e.g., ICSE, FSE and TSE), we excluded papers not written in English. Our aim for high-quality and international venues in software engineering is also a mechanism to maintain the rigour of our analyses.

\begin{figure}[H]
  \begin{center}
\includegraphics[scale=0.4]{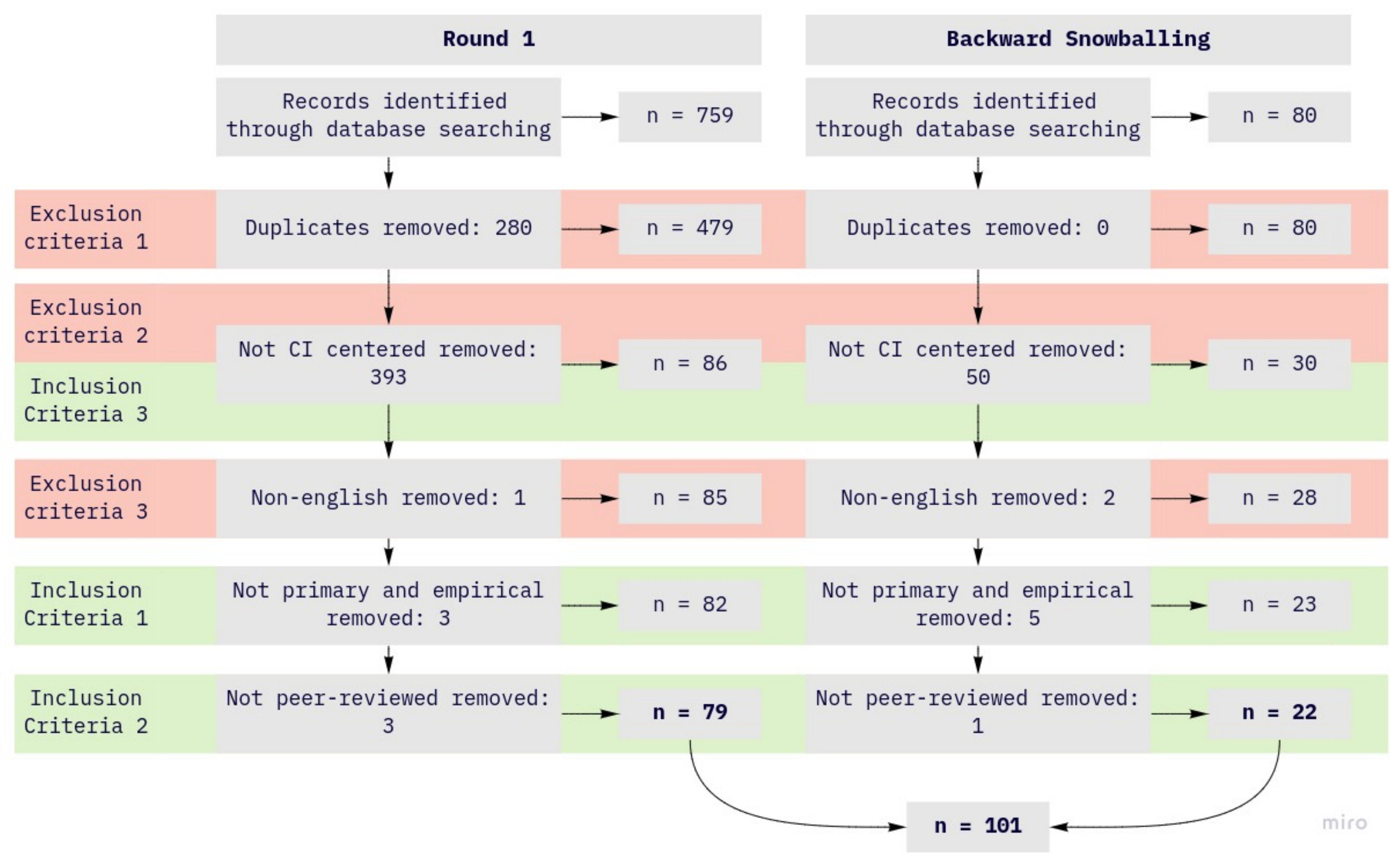}
\caption{Diagram illustrating the inclusion and exclusion criteria employment, presenting the number of remaining papers after each stage.}
\label{fig:prisma}
  \end{center}
\end{figure}

Our inclusion criteria are the following: (i) the studies must be empirical primary studies; (ii) be peer-reviewed papers; and (iii) show that CI adoption may (or may not) have an effect on any aspect of software development. Our exclusion criteria are the following: (i) studies must not be duplicates; (ii) studies must investigate the effects of CI instead of proposing a new tool or a new practice for CI; (iii) papers must be written in English. Figure~\ref{fig:prisma} shows the number of papers removed after the application of each criterion.

\subsubsection{Screening of papers \label{screening}}
Figure~\ref{fig:methodology} shows an overview of our screening steps. In Steps 1 and 2, we apply our search string (Section \ref{search_strategy}) onto the referred digital libraries, obtaining 759 papers. By applying the exclusion criteria 1 in Step 3, we obtain 479 distinct papers (see Table~\ref{tab:databases} and Figure \ref{fig:prisma}).

In Step 3, we perform a reading of the 479 papers. Two authors read the title and abstract of each study and judge them based on the inclusion and exclusion criteria. By using the Cohen Kappa statistic~\cite{cohen}, we obtain a score of 0.72, which represents a substantial agreement. Afterward, a third author checks the disagreements (there are 34 disagreements) and resolves each one. As a result, a total of 79 papers were obtained at the end of Step 3.

Since the screening of papers is a step based on explicit inclusion and exclusion criteria, we decide to apply an arbitration disagreement resolution strategy involving a third researcher to check and break a tie in each disagreement. At this step, most of the disagreements (there are 34) are about whether CI is the primary investigation topic of the study. For example, we have an occurrence regarding the study entitled ``Moving from Closed to Open Source: Observations from Six Transitioned Projects to GitHub'' (P69) on the inclusion criteria 3, since apparently, it does not investigate CI directly. However, the arbiter voted for inclusion, and the paper does present findings on CI.

In the next step (Step 4 in Fig. \ref{fig:methodology}), we perform a backward snowballing, collecting 80 references of the selected studies, which contain the term ``continuous integration''---both in the title or abstract. Next, in step 5, two authors read the title and abstract and apply the inclusion and exclusion criteria. At this stage, we include 22 additional papers. We achieve an agreement rate of 0.76 (Cohen Kappa), signaling a substantial agreement between authors. Afterward, we repeated the dispute resolution process with the arbitration of the third researcher, which resulted in 101 studies at the end of Step 5. Figure~\ref{fig:prisma} presents this process in detail with a column for the first cycle --- column ``round 1'', and another to the snowballing process.

Appendix~\ref{appendix_studies} lists the selected papers. The files containing the lists of papers on each step are available in our digital repository. A backup of the relational database that we use in our SLR is also available~\cite{soares_eliezio_2020}.

\begin{table}[H]
  \caption{Fields of the extraction form.}
  \label{tab:extraction} 
  \begin{tabular}{lll}
  \hline
  \multicolumn{2}{|c|}{{ \textbf{Extraction Form Fields}}}                                                                        & \multicolumn{1}{c|}{\textbf{Sub items}}                \\ \hline
  F1                   & \multicolumn{2}{l}{\textbf{\begin{tabular}[c]{@{}l@{}}What are the claims presented \\in the paper?\end{tabular}}}                                                                                                              \\ \hline
  F2                   & \multicolumn{2}{l}{\textbf{\begin{tabular}[c]{@{}l@{}}What are the variables related \\to each claim?\\ (And, if it is not clear, what \\is the meaning of each variable?)\end{tabular}}} \\ \hline
                       &                                                                                                                              & Controlled Experiments                                 \\
                       &                                                                                                                              & Case Studies                                           \\
                       &                                                                                                                              & Survey Research                                        \\
                       &                                                                                                                              & Action Research                                        \\
  \multirow{-5}{*}{F3} & \multirow{-5}{*}{\textbf{\begin{tabular}[c]{@{}l@{}}What kind of study was performed \\to evaluate the claim?\end{tabular}}}                                           & MSR                                                    \\ \hline
  F4                   & \multicolumn{2}{l}{\textbf{\begin{tabular}[c]{@{}l@{}}How the claim was evaluated?\end{tabular}}}                                                                                               \\ \hline
  F5                   & \multicolumn{2}{l}{\textbf{\begin{tabular}[c]{@{}l@{}}How many projects were involved \\in the study?\end{tabular}}}                                                                                                            \\ \hline
  F6                   & \multicolumn{2}{l}{\textbf{\begin{tabular}[c]{@{}l@{}}Are there open-source, industry, \\or both classes \\ of projects involved in the study?\end{tabular}}}                           \\ \hline
  F7                   & \multicolumn{2}{l}{\textbf{\begin{tabular}[c]{@{}l@{}}Is the study focused on a specific \\domain area? which one?\end{tabular}}}                                                                                               \\ \hline
  F8                   & \multicolumn{2}{l}{\textbf{\begin{tabular}[c]{@{}l@{}}Does the study have the artefacts \\available?\end{tabular}}}                                                                                                 \\ \hline
  F9.1                 & \multicolumn{1}{c}{}                                                                                                         & Integration Frequency                                  \\
  F9.2                 & \multicolumn{1}{c}{}                                                                                                         & Automatic Build                                        \\
  F9.3                 & \multicolumn{1}{c}{}                                                                                                         & Build Duration                                         \\
  F9.4                 & \multicolumn{1}{c}{}                                                                                                         & Automated Tests                                        \\
  F9.5                 & \multicolumn{1}{c}{}                                                                                                         & Test Coverage                                          \\
  F9.6                 & \multicolumn{1}{c}{}                                                                                                         & Integration on Master                                  \\
  F9.7                 & \multicolumn{1}{c}{\multirow{-7}{*}{\textbf{\begin{tabular}[c]{@{}l@{}}What kind of criteria was considered \\to determine CI adoption?\end{tabular}}}}                & CI SERVICE                                             \\ \hline
  \end{tabular}
  \end{table}

\subsection{Data Extraction \label{extraction}}

The extraction process consists of three steps: meta-data retrieval, data extraction, and disagreement resolution. An automated process retrieves the meta-data, which includes the title, authors, year, and publication venue of the studies. We use a reference management tool named Mendeley\footnote{Available at https://www.mendeley.com/} to support the meta-data extraction. Mendeley exports the meta-data in an XML format. We then use a script to read Mendeley's XML files and store the meta-data into our database.

Two authors perform the data extraction by reading all 101 studies while collecting relevant data (Step 6 in Figure~\ref{fig:methodology}). When a paper is completely read by each author, they both submit a form containing the data extracted from that paper. For this purpose, we use a web form containing the fields that are shown in Table \ref{tab:extraction} \cite{kitchenham2007}. Next, we export the data from the forms into a .csv file. Then, we run a script to import the extracted data into our database.

Our script automatically checks for the consistency of data provided by the authors. If our script identifies that the two authors extracted different data for a given paper, the script generates a \texttt{diff} containing the different content beside each other. The diff files support the resolution of disagreements (Step 7 on Figure~\ref{fig:methodology}), in which both authors would check the diff files and reach consensus regarding which data should be extracted and imported into the database (step 8 on Figure~\ref{fig:methodology}). 

Examples of inconsistencies include typing errors, misunderstandings of extracting the data, or regarding the study interpretation. Given this interpretative nature, we adopt a consensus disagreement resolution strategy involving both researchers in this step--- data extraction. They assess the paper in a virtual meeting to discuss item by item the paper details and then confirm the extracted data in a new form to import.

\subsection{Quality Assessment \label{quality_assessment}}

Following the recommendations from Kitchenham \& Charters \cite{kitchenham2007}, we developed a quality checklist to assess the quality of each of the individual selected primary studies.
Our quality assessment aims to understand the quality differences in the collected evidence, supporting the weighting of their claims. Thus, considering the heterogeneity of the selected studies, in terms of study types and the outcomes investigated, we adopt the framework proposed by Dybå et al. \cite{dyba2007}. This framework was proposed for the quality assessment of both qualitative and quantitative empirical research.

Therefore, we adapt the Dybå et al. \cite{dyba2007} checklist (see Table \ref{tab:quality_checklist}) composed of 11 questions among 4 quality criteria: (i) quality of reporting (3 questions --- Q2 to Q4); (ii) rigour (4 questions --- Q5 to Q8); (iii) credibility (3 questions --- Q1,  Q9, and Q10); (iv) relevance (1 question --- Q11). Questions Q2, Q3, Q5, Q7, Q8, and Q11 are verbatim from Dybå et al. \cite{dyba2007}. The remaining five questions were inspired by examples from Kitchenham \& Charters \cite{kitchenham2007} and Dybå et al. \cite{dyba2007}, maintaining the adequacy to the framework structure.

In this checklist, quality of reporting means the clarity with which it communicates its context, motivation, and goals. The transparency and unambiguity of a study enable readers to extract information and accurate conclusions from it. In this sense, we apply three questions (Q2 to Q4 on Table \ref{tab:quality_checklist}) assessing these issues.

We designated four questions for the rigor criterion (Q5 to Q8 on Table \ref{tab:quality_checklist}), constituting the heaviest factor of this checklist. The questions about rigor highlight the methodological decisions of the studies and their rationale. We analyze whether participants/projects selection is suitable or not (e.g., Has the study justified the selection procedures?). We also look for the metrics and measures and if they are provided/explained. We observe if the research design is appropriate to the research goals (e.g., Has the researcher justified the research design? Has the researcher presented and explained the statistical tests applied?). Furthermore, in Q8, we look for comparison or control groups as indicative of analytical rigor. 

The credibility factor comprises three items  (see Q1,  Q9, and Q10 on Table \ref{tab:quality_checklist}) assessing acceptability and the coherence between the presented findings and applied methods. The first question filters peer-reviewed approved studies. We ask if empirical data and experiment results support the findings and conclusions (e.g., Are the findings explicit? Are limitations of the study discussed explicitly? Are the findings discussed concerning the original research questions?). Lastly, we check whether data is available (or scripts or detailed descriptions to obtain it) for reproduction or replication.

Finally, we assess the relevance of contributions (see Q11 on Table \ref{tab:quality_checklist}) for industry or academy as an indicator of the study quality. This criterion has the lightest weight. In question 11, we check if the researchers discuss the impact of their study to the state-of-the-art and state-of-the-practice (e.g., do they consider the findings concerning current practice or relevant research-based literature?).

These 11 questions behave as binary variables (1 - yes; 0 - no), and together they provide a metric of quality and reliability of the findings. To cover a broad set of empirical evidence and draw a big picture of continuous integration reported effects, just the Q1 was used as an inclusion criterion (section \ref{selection_criteria}). The remaining questions compound a checklist to assess the strength of the body of evidence in Sections \ref{rq_themes} and \ref{rq_method}. For that reason, we only evaluated studies with collected findings in our approach (i.e., papers from which we find claims).

The sum of 11 questions allows us to compute a quality score per study (see Section \ref{sub: results_methods}). We consider this score a measure of the reliability of the extracted claims, i.e., claims originated from studies with high scores are more reliable than those with lower scores. We built our checklist with the goal of rewarding a greater variety of methods to support a claim. 
For example, the value of method variability is clearly seen in the higher scores obtained by mixed-methods studies (MSR and survey). Mixed-methods studies score better than other types with a median of 10 points (see section \ref{sub: results_methods}). The overall median score is 9.

Furthermore, we consider certain codes (see section \ref{synthesis}) more reliable if they are supported by a higher number of studies and a higher variety of study types. In sections \ref{rq_themes} and \ref{discussion}, we assess and discuss CI claims by examining: the set of studies supporting these claims, the variety of methods to support these claims, the complementarity between findings, and the respective quality scores of the studies. For example, in section \ref{sec:integration_patterns}, we present a code describing an association between CI and a {\em ``change in commit patterns.''} This code represents five claims over three studies with various methods and quality scores (one Case Study, one Mining Software Repository --- MSR, and one MSR/Survey). Although the case study scores 6 points of quality, the MSR scores 9 points, and the MSR/Survey scores 10 points, i.e., different methodologies combined with an overall higher quality score support the claim that CI promotes a {\em ``change in commit patterns.''} Therefore, this code is more reliable than if it was supported by only a case study or other studies of the same type and similar quality scores.

Similar to what was exposed in section \ref{extraction} for data extraction, after reading, two authors independently assessed the quality of the study using a web form, achieving a Kappa score of 0.55, which indicates a moderate agreement. To subsidize this quality assessment, we added guiding questions for each item in the quality checklist. Later, each divergence was discussed between the pair and settled by consensus after revisiting the study (step 10 in Figure~\ref{fig:methodology}).

\begin{table}[H]
  \caption{Quality Assessment checklist}
  \label{tab:quality_checklist} 
  \begin{tabular}{cl}
  \hline
  \multicolumn{2}{c}{\textbf{Question}}                                                  \\ \hline
  \textbf{Q1}           & Was the paper peer-reviewed?                                                           \\
  \textbf{Q2}           & Is there a clear statement of the aims of the research?                                \\
  \textbf{Q3}           & Is there an adequate description of the context in which the research was carried out? \\
  \textbf{Q4}           & Is the size of the data set stated?                                                    \\
  \textbf{Q5}           & Was the recruitment strategy appropriate to the aims of the research?                  \\
  \textbf{Q6}           & Are the definitions for the measures or metrics provided?                              \\
  \textbf{Q7}           & Was the research design appropriate to address the aims of the research?               \\
  \textbf{Q8}           & Is there a comparison or control group?                                                \\
  \textbf{Q9}           & Does the empirical data and results support the findings?                              \\
  \textbf{Q10}          & Is the data available?                                                                 \\
  \textbf{Q11}          & Is the study of value for research or practice?                                        \\ \hline
  \end{tabular}
\end{table}

\subsection{Synthesis \label{synthesis}}
In step 11 of Figure~\ref{fig:methodology}, we use the data extracted from our extraction form (Table \ref{tab:extraction}) to address RQ1, RQ2, and RQ3 (Section \ref{rqs}). We first analyze the demographic data (see Appendix~\ref{appendix_demographic}). Next, we perform the analyses to answer the Research Questions.  

To answer {\em RQ1--\RQone}, we use the F9 field. To answer {\em RQ2--\RQtwo}, we run a {\em thematic synthesis} \cite{cruzes2011} using the fields F1 and F2. To answer {\em RQ3--\RQthree}, we use fields from F3-to-F8 (see Table \ref{tab:extraction}).

In the {\em thematic synthesis} to answer RQ2, we follow the steps recommended by Cruzes \& Dyba \cite{cruzes2011}.
The thematic synthesis consists of identifying patterns (themes) within the data, which provides a systematic manner to report the findings of a study. The thematic synthesis consists of five steps:

\begin{enumerate}
    \item Extract data,
    \item Code data,
    \item Translate codes into themes,
    \item Create a model of higher-order themes, and
    \item Assess the trustworthiness of the synthesis.
\end{enumerate}

The {\em Extract Data} is the  first step of the thematic synthesis (Section \ref{extraction}). To answer RQ2, we analyze the data from fields F1 and F2 (see Table~\ref{tab:extraction}), which are {\em claims regarding the effects of CI}, i.e., any consideration in a study indicating a positive or negative effect of CI on the software development phenomena. Therefore, we do not consider to be a {\em claim} statements that are indirect or unrelated to the effects of CI on software development---even if CI is used by the software projects under investigation. Table~\ref{tab:claims_example} shows two examples of claims.

\begin{table}[H]
  \caption{Extracted claims from studies P25 and P74, from fields F1 and F2 of the extraction form.}
  \label{tab:claims_example} 
  \begin{tabular}{ |p{.5\textwidth}| p{.4\textwidth} | p{.06\textwidth}|}
    \hline
    \multicolumn{1}{|c|}{\textbf{Claim}}                                                                       & \multicolumn{1}{c|}{\textbf{Variables}}                     & \textbf{Paper id} \\ \hline
    CI increases normalized collaboration amount between programmers (OSS and proprietary projects)            & Normalized median in-degree (NMID)                          & 25                \\ \hline
    Core developers in teams using CI are able to discover significantly more bugs than in teams not using CI. & Number of bug reports (i.e. issues clearly labeled as bugs) & 74                \\ \hline
    \end{tabular}
\end{table}

We group the information from fields F1 and F2 in a spreadsheet. Next, we code the information through an inductive approach~\cite{cruzes2011}, i.e., two authors analyze all the {\em claims} together and collaboratively assign one or two codes to each of the claims. The assigned codes are based on the central message within a claim. Therefore, the two authors create an established {\em list of representative codes}. 

Once the list of codes is created, two other authors are debriefed regarding the codes to understand their meanings. These two other authors revisit every claim independently and select one or more codes from the list of codes to assign to the claims. As an example, consider the following finding in study P25: ``After adoption of CI, normalized collaboration amount between programmers significantly increases for our set of OSS and proprietary projects. [...]''. Both authors assign the code {\em ``CI IS ASSOCIATED WITH AN INCREASE IN COOPERATION''} to such a claim. At this stage, we obtain a Cohen Kappa statistic \cite{cohen} of 0.73, which indicates a substantial agreement. 

At this step, since it is a task of synthesizing ideas, we adopt an arbitration disagreement resolution strategy to explore contributions from a more experienced author. All disagreements were solved by a third author. As an example of disagreement, consider the following claim in study P74:  ``Core developers in teams using CI are able to discover significantly more bugs than in teams not using CI. [...]''. One author assigned the code {\em ``CI IS ASSOCIATED WITH DEFECT REDUCTION''}, while the other author assigned {\em ``CI IS ASSOCIATED WITH A DECREASE IN TIME TO ADDRESS DEFECTS''}. In this case, the third author analyzed the claim and decided to maintain the code {\em ``CI IS ASSOCIATED WITH DEFECT REDUCTION''}. 

In the third step of the thematic synthesis (i.e., {\em Translate codes into themes}) we compute the frequency of each code and propose overarching themes. Finally, we develop a thematic network to express the relationship between codes and themes (Step 4 of the thematic synthesis). Once the thematic network was developed we performed two meetings with all authors to discuss the meaningfulness of the network and codes (Step 5 of the thematic synthesis). After 4 hours of discussion (each meeting having 2 hours), we refined the thematic network and the codes and themes within it (see Section~\ref{rq_themes} and Figure~\ref{fig:thematic_analysis}).

\section{Results \label{results}}

The appendix \ref{appendix_demographic} present some demographic information about the studies. In this Sectionwe present the results of our systematic literature review (SLR). The following subsections explores the results to our research questions.

\subsection{\bfseries RQ1: \RQone \label{sub: rq_criteria}}

To answer this research question, we analyze in the primary studies which criteria (e.g., CI practices or attributes) were considered when describing or selecting the analyzed projects---For example, how are the projects using CI deemed as such? More specifically, we do not investigate the analyzed projects themselves. Instead, we investigate whether our primary studies select (or classify) their analyzed projects based on the following criteria: integration frequency, automated build, build duration, automated tests, test coverage threshold, integration on the mainline, and CI service~\cite{duvall2013,fowler2017}. 

As discussed in section~\ref{back} there is no consensus around the definition of CI or a homogeneous set of CI practices~\cite{duvall2013,fowler2017,stahl2014,zhao2017,shahin2017,fitzgerald2017,viggiato2019}. Therefore, considering the literature diversity, we adopt a set of criteria based on Duvall's seven cornerstones \cite{duvall2013}, and CI practices highlighted by Fowler \cite{fowler2006} for this analysis as they present a prescriptive minimal number of practices, instead of discussing the variability among diverse CI implementations.
\begin{figure}[h!]
  \includegraphics[scale=0.68]{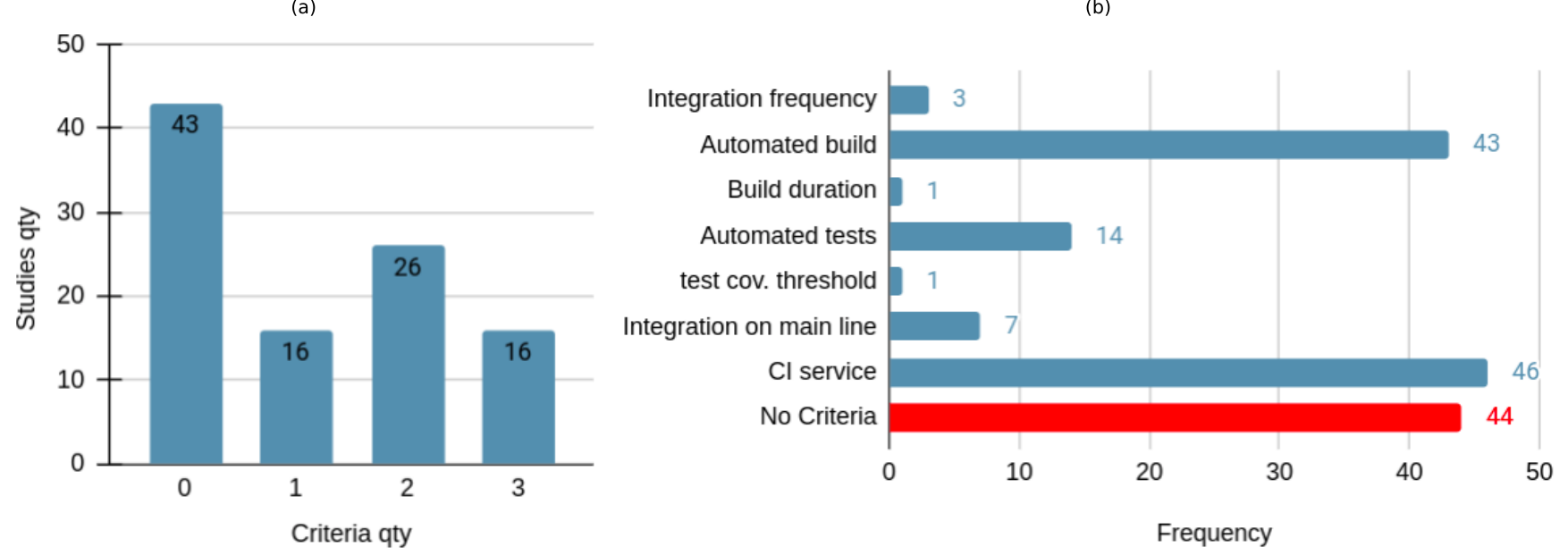}
  \caption{(a) Histogram representing the proportion of primary studies using a number of CI criteria; (b) Frequency of usage of each criteria;}
  \label{fig:criteria_qty}
\end{figure}

Figure~\ref{fig:criteria_qty} (a) shows the number of criteria considered in the primary studies to identify whether a project uses CI. From the seven considered criteria that we expect to see, 43 (42.5\%) of the primary studies, surprisingly, did not apply or determine any of them. On the other hand, 26 (25.7\%) of the primary studies used two criteria, while 16 (15.8\%) of the projects and another 16 (15.8\%) of them used one and three criteria, respectively. 

By inspecting the 43 (42.5\%) studies without clear criteria for determining whether a project uses CI, we observe that: (i) 28 of the studies do not analyze data related directly to the project's development. Instead, they are studies based on interviews, surveys, companies, or other data sources, e.g., build logs; (ii) some of the studies present experience reports without further details regarding how projects adopt the CI practices. In addition, (iii) a few studies (P49, P52, P60, P69) analyze both projects and self-described declarations, like interviews or surveys. 

Although it is understandable that the criteria we are looking for (e.g., integration frequency) may not be applied in such studies, it would still be valuable to perform certain checks during the interviews or surveys. For example, questions such as {\em ``on a scale of 1 to 7, how would you classify that your project adheres to CI?''} along with a definition of CI could help such studies to gauge the quality the CI practices that are implemented by the subjects. Regarding the studies that investigate build logs only (e.g., build logs from \textsc{TravisCI}), it would also be desirable to be more restrictive regarding these logs, since not every build log from \textsc{TravisCI} may come from a project that properly employs CI. Therefore, solely relying on the fact that build logs are generated from a CI server does not necessarily imply that the derived observations can be associated with the adoption of CI practices.   

Concerning studies applying only one criterion to identify whether a project uses CI, the CI server configuration is the most common criterion (9/16 studies - 56,25\%). We observe in Figure \ref{fig:criteria_qty} (b) that the usage of a CI service is the most common criterion applied. This criterion consists of checking whether subject projects have used a CI service (e.g., \textsc{TravisCI}). The second most frequent criterion is checking whether subject projects perform automatic builds. Table \ref{tab:ci_services} shows the CI services cited in the included studies, revealing that \textsc{TravisCI} \footnote{https://travis-ci.com/} is the most used CI service, confirming the finding by Hilton et al. \cite{hilton2016}. 
    
\begin{table}[]
\caption{CI Services cited in the included studies.}
\label{tab:ci_services} 
\begin{tabular}{lc}
\hline
\multicolumn{1}{c}{\textbf{CI Services}}                                  & \textbf{Studies} \\ \hline
TRAVIS CI                                                                 & 37               \\ \hline
JENKINS                                                                   & 8                \\ \hline
CUSTOMIZED                                                                & 4                \\ \hline
CIRCLE CI                                                                 & 3                \\ \hline
APPVEYOR, TEAM CITY, WERKCER                                              & 2                \\ \hline
BUILDBOT, CONCOURSE, CRUISECONTROL, \\
GERRIT, CLOUDBEES, XCODE BOTS, GITLAB & 1                \\ \hline
\end{tabular}
\end{table}

%
%
%

\subsection{\bfseries RQ2: \RQtwo \label{rq_themes}}

\begin{figure}[h!]
  \centering
  \includegraphics[scale=0.48]{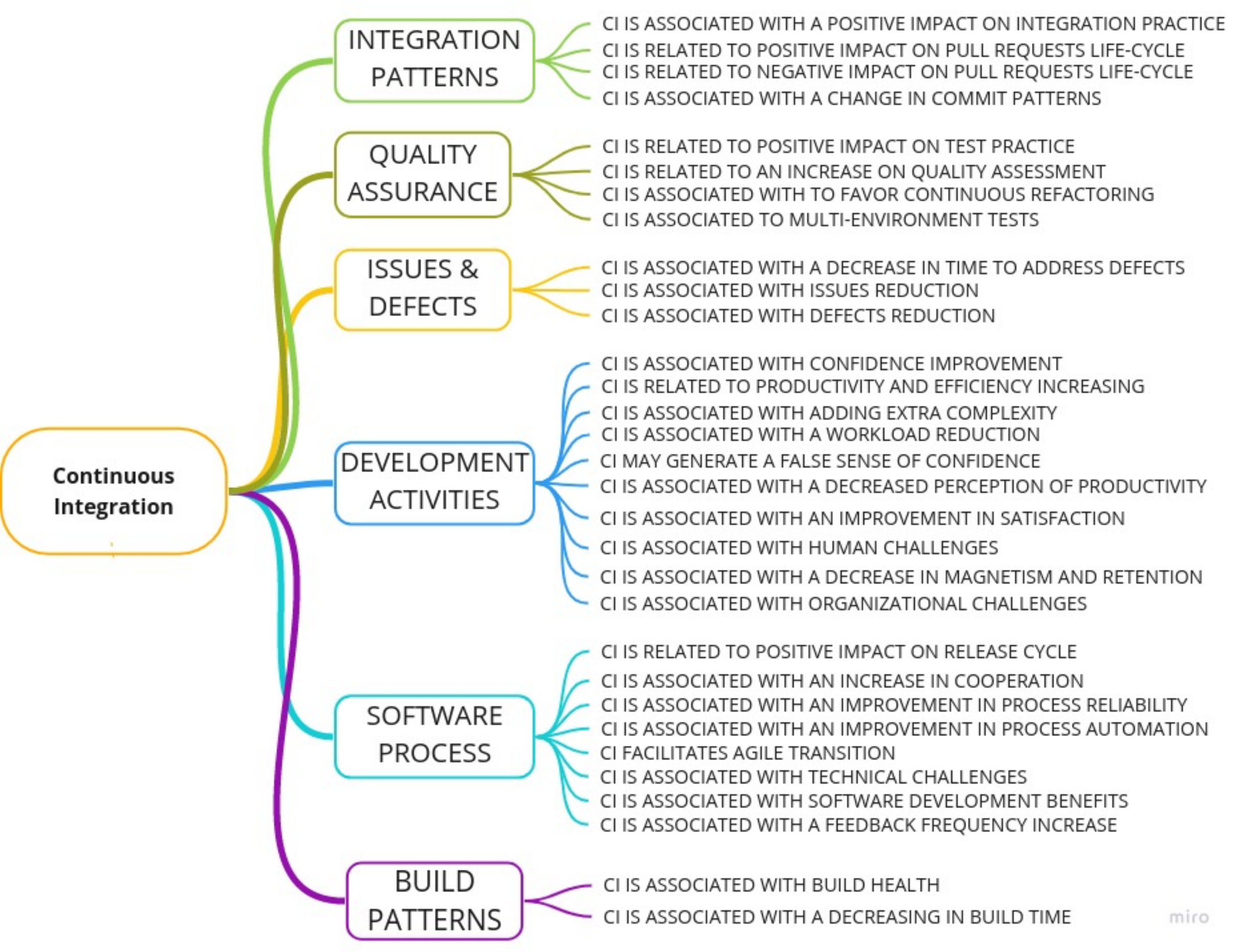}
  \caption{Themes and codes representing the studies claims.}
  \label{fig:thematic_analysis}
\end{figure}

To answer RQ2, we collect the claims from our primary studies and proceed with the thematic synthesis to produce the codes and themes regarding the claims. As explained in Section~\ref{synthesis}, a claim is a statement regarding any positive or negative effect of CI on the software development phenomena. We found 125 claims regarding the effects of CI in 38 out of 101 papers (37.6\% of studies). From the thematic synthesis, we produce 31 codes from the 125 extracted claims. Figure~\ref{fig:thematic_analysis} shows the produced codes organized into 6 overarching themes. 

Table~\ref{tab:themes_frequency} shows the following information: a) the themes, b) the number of claims pertaining to a theme, and (c) the primary studies that make the claims. The most common themes in the primary studies are: \textit{``development activities''}--- having 35 claims across 18 papers--- and \textit{``software processes''}--- with 35 claims across 18 papers. 
Although we observe in RQ1 that automated builds is a common criterion to check whether CI is used by subject projects, the theme \textit{``build patterns''} has only 7 claims from 4 primary studies.

\begin{table}[H]
	\caption{Number of claims and studies that pertain to a theme. }
\label{tab:themes_frequency} 
\begin{tabular}{lcl}
\hline
\multicolumn{1}{c}{\textbf{Theme}} & \textbf{\begin{tabular}[c]{@{}c@{}}Number of \\ Claims\end{tabular}} & \multicolumn{1}{c}{\textbf{Primary Studies}}                                                                                         \\ \hline
Development activities                     & 35                                                                   & \begin{tabular}[c]{@{}l@{}}P7, P9, P31, P39, P52, P58, P59, P73, P79,\\ P74, P90, P91, P93, P97, P99, P100, P102, \\ P106\end{tabular} \\ \hline
Sofware process                & 35                                                                   & \begin{tabular}[c]{@{}l@{}}P4, P24, P25, P38, P40, P46, P52, P58, P64, \\ P79, P81, P92, P93, P97, P100, P102, P105, P106\end{tabular}      \\ \hline
Quality assurance                 & 23                                                                   & \begin{tabular}[c]{@{}l@{}}P14, P29, P44, P52, P58, P64, P79, \\ P89, P93, P97, P100, P102, P106\end{tabular}                     \\ \hline
Integration patterns               & 22                                                                   & \begin{tabular}[c]{@{}l@{}}P25, P47, P59, P69, P74, P79, P81, P89, \\ P97, P100, P102, P104\end{tabular}                               \\ \hline
Issues \& defects                 & 14                                                                   & \begin{tabular}[c]{@{}l@{}}P25, P31, P50, P59, P74, P79, P97, P100, \\ P89, P106\end{tabular}                                          \\ \hline
Build Patterns                     & 7                                                                    & P29,P49, P97, P102                                                                                                                       \\ \hline
\end{tabular}
\end{table}

\begin{figure}[h!]
  \centering
  \includegraphics[scale=0.7]{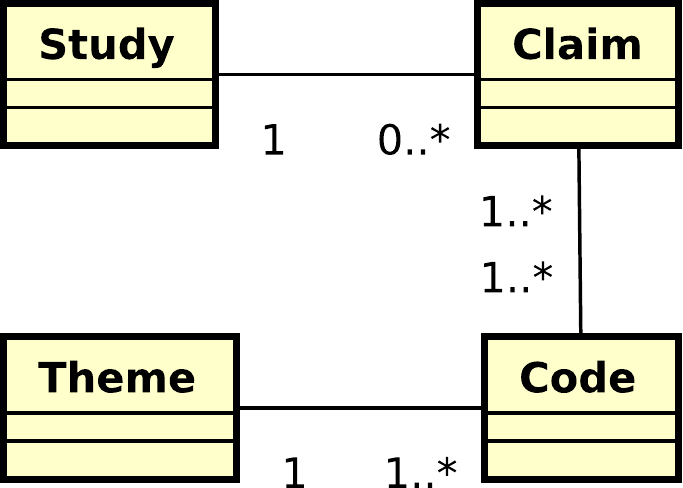}
  \caption{Conceptual class diagram of relationships between studies, claims, codes, and themes.}
  \label{fig:synthesis_concept}
\end{figure}

The link between a theme and a paper does not necessarily mean that the theme is the paper's main topic. A paper may have one or more claims related to a theme, but the same paper may have other claims related to other themes. Figure \ref{fig:synthesis_concept} shows a conceptual class diagram expressing how a study can have none or several claims, while each claim can be related to one or more codes. Each code is related to a theme. Section~\ref{synthesis} describes the entire process of our synthesis. Therefore: (i) each code sentence shown at the right side of Figure~\ref{fig:thematic_analysis} is representative of a set of claims extracted from primary studies and mapped to such code; (ii) a study is not necessarily linked directly to a theme but may be related to several themes by transitivity.

To further evaluate the reliability of the findings within the themes, we add an earlier discussion about quality scores in the subsequent Section \ref{rq_method}. For the purpose of our analysis, we consider the mean (\(\displaystyle \bar{Qscore} \)) and the median (\(\displaystyle \tilde{Qscore} \)) as the quality measurement for our body of evidence.

\subsubsection{Development Activities \label{sec:developer_work}}

\begin{table}[H]
  \caption{Codes from the ``development activities'' theme. We show the number of claims related to the code, the primary studies supporting it, the mean and median of quality scores of such studies.}
  \label{tab:codes_developer_work} 
  \begin{tabular}{lclcc}
  \hline
  \multicolumn{1}{c}{\textbf{Code}} & \textbf{\begin{tabular}[c]{@{}c@{}}Number of\\ Claims\end{tabular}} & \multicolumn{1}{c}{\textbf{Primary Studies}} & \(\displaystyle \bar{Qscore} \) & \(\displaystyle \tilde{Qscore} \) \\ \hline
  \begin{tabular}[c]{@{}l@{}}CI is related to productivity and \\ efficiency increase\end{tabular}      & 12                 & \begin{tabular}[c]{@{}l@{}}P31, P39, P52, P59, P73, P79, \\ P74, P97, P100, P102, P106\end{tabular} & 8.8 & 8   \\ \hline
  \begin{tabular}[c]{@{}l@{}}CI is associated with Confidence \\ improvement\end{tabular}                 & 6                  & P39, P58, P93, P97, P100, P106                                                                    & 7.0 & 7  \\ \hline
  \begin{tabular}[c]{@{}l@{}}CI is associated with adding extra \\ complexity\end{tabular}                & 5                  & P7, P58, P106                                                                                     & 6.2 & 7  \\ \hline
  \begin{tabular}[c]{@{}l@{}}CI may generate a false sense of \\ confidence\end{tabular}                  & 3                  & P58, P106                                                                                         & 6.0 & 7  \\ \hline
  \begin{tabular}[c]{@{}l@{}}CI is associated with a workload \\ reduction\end{tabular}                   & 3                  & P6, P79, P93                                                                                      & 5.6 & 5       \\ \hline
  \begin{tabular}[c]{@{}l@{}}CI is associated with human \\ challenges\end{tabular}                       & 2                  & P58, P59                                                                                          & 5.5 & 4  \\ \hline
  \begin{tabular}[c]{@{}l@{}}CI is associated with a decreased \\ perception of productivity\end{tabular} & 1                  & P91                                                                                               & 8.0 & 8  \\ \hline
  \begin{tabular}[c]{@{}l@{}}CI is associated with an \\ improvement in satisfaction\end{tabular}         & 1                  & P9                                                                                                & 8.0 & 8  \\ \hline
  \begin{tabular}[c]{@{}l@{}}CI is associated with \\ organizational challenges\end{tabular}              & 1                  & P99                                                                                               & 9.0 & 9  \\ \hline
  \begin{tabular}[c]{@{}l@{}}CI is associated with a decrease \\ in magnetism and retention\end{tabular}  & 1                  & P90                                                                                               & 9.0 & 9  \\ \hline
  \multicolumn{3}{r}{\textbf{Overall}}                                                                                                                                                                              & \textbf{7.1}      & \textbf{7}\\ 
\end{tabular}
  \end{table}

Several primary studies have claims regarding the effects of CI on development activities. Table \ref{tab:codes_developer_work} shows the claims, the development activity related to the claim, the ID of the primary studies, the mean \(\displaystyle \bar{Qscore} \) and the median \(\displaystyle \tilde{Qscore} \). We observe 4 positive effects of CI on productivity, efficiency, confidence, satisfaction, and reduction in the workload. 

There are several mentions in the primary studies claiming an increase in productivity and efficiency when using CI (12 occurrences in 11 studies). As reported in study P97: 
\begin{displayquote}
\textit{``According to our interview participants, CI allows developers to focus more on being productive, and to let the CI take care of boring, repetitive steps, which can be handled by automation.''} (p. 203) 
\end{displayquote} 
And 
\begin{displayquote}
\textit{``Another reason [...] was that CI allows for faster iterations, which helps developers be more productive.''} (p. 204) 
\end{displayquote} 

Several studies also mention an improvement in confidence after using CI (6 occurrences in 6 papers). Paper P58 states the following: 
\begin{displayquote}
\textit{``Depends on the coverage, but some sort of confidence that introduced changes don't break the current behavior. [...] I assume the most critical parts of the system have been covered by test cases.''} (p. 76) 
\end{displayquote} 

These positive associations with CI have the support of numerous and diverse studies reported in Table~\ref{tab:codes_developer_work} with a substantial quality profile. However, there are still low scores on the rigor criterion in our quality assessment (Q5 to Q8 on Table~\ref{tab:quality_checklist}). For example, regarding CI increasing productivity and efficiency, P79 does not substantiate its assumptions with statistical tests, and P106 does not apply any control or comparison group (Q7 and Q8 of the quality checklist). In addition, 5 out of 11 studies scored 2 or fewer points on the rigor criterion (out of 4 points). Regarding ``confidence improvement'', P58 and P106 do not satisfy Q7, and 4 out of 6 studies scored 2 or fewer out of 4 points on the rigor criterion.

The association of CI with a workload reduction has low-quality scores of \(\displaystyle \bar{Qscore}=5.6\) and \(\displaystyle \tilde{Qscore}=5\), and the association of CI with an improvement in satisfaction has only one supporting study, although with a high-quality profile (\(\displaystyle Qscore=8.0\)).

Nevertheless, not everything seems to be positive in terms of development activities. We found six negative aspects stated in the primary studies: extra complexity added, the existence of a false sense of confidence, human and organizational challenges, a decreased perception of productivity, and a decrease in magnetism and retention of collaborators in projects. The most endorsed negative effects of CI are the addition of extra complexity (5 occurrences in 3 papers), and the generation of a false sense of confidence (3 occurrences in 2 studies). P7 mentions the extra complexity related to using CI: 
\begin{displayquote}
\textit{``Results of our study [...] highlights the complexity of dealing with CI in certain situations, e.g., when dealing with emulated environments, non-deterministic (flaky) tests, or different environments exhibiting an inconsistent behavior''.} (p. 47) 
\end{displayquote} 

P106 explains the false sense of confidence: 
\begin{displayquote}
\textit{``As opposed to the confidence benefit, respondents described the false sense of confidence as a situation of which developers blindly trust in tests''} (p. 2232) 
\end{displayquote}

However, some of these negative effects obtain low-quality scores, especially on the rigor criterion (e.g., P58 scored 0 out of 4 points), and need to be further investigated by our research community. Regarding the increase in human challenges, for example, studies obtain scores of  \(\displaystyle \bar{Qscore}=5.5\) and \(\displaystyle \tilde{Qscore}=4\) (see Table \ref{tab:codes_developer_work}). The claims suggesting a “false sense of confidence” as an effect of CI come from only two studies conducted by the same authors.

On the other hand, other negative associations, such as ``CI is associated with a decreased perception of productivity'', ``CI is associated with organizational challenges'', and ``CI is associated with a decrease in magnetism and retention'' obtain the highest quality scores. However,  all these negative associations are supported by only one study each. The association ``CI is associated with adding extra complexity'' obtained scores of \(\displaystyle \bar{Qscore}=6.2\) and \(\displaystyle \tilde{Qscore}=7\), particularly one study --- P7 --- conducted a mixed-method study (MSR + Survey) and has a high quality score (Qscore = 9).

These adverse effects seem to point in the same direction: CI introduces complexity, challenges the organizational environment, and influences developers’ perception of productivity. Such statements contradict the assumption that CI promotes a workload reduction. We argue that more studies are required by our community to better understand the context and the extent of such adverse effects, given the small variety and generalizability of studies supporting them. Furthermore, the studies' overall rigor mean is near 50\% of the max score. This result suggests that we need more efforts in performing reliable studies on the \textit{Development Activities} theme.

\begin{boxed}
  According to the literature, there is reliable evidence of the association between CI and improved productivity, efficiency, and developer confidence. CI may create a positive impact on the stakeholders’ satisfaction. On the other hand, findings suggest that CI introduces complexity to the development environment, demanding more developer effort and discipline, negatively impacting developers’ perception of their productivity. Given the low number of studies related to some of these evaluation aspects, there is room for further studies on these aspects.
\end{boxed}

\subsubsection{Software Process \label{sec:software_process}}

\begin{table}[H]
  \caption{Codes from theme ``Sofware Processes''. We show the number of claims related to the code, the primary studies supporting it, the mean and median of quality scores of such studies.}
  \label{tab:codes_development_process} 
  \begin{tabular}{lclcc}
    \hline
    \multicolumn{1}{c}{\textbf{Code}} & \textbf{\begin{tabular}[c]{@{}c@{}}Number of\\ Claims\end{tabular}} & \multicolumn{1}{c}{\textbf{Primary Studies}} & \(\displaystyle \bar{Qscore} \) & \(\displaystyle \tilde{Qscore} \) \\ \hline
  \begin{tabular}[c]{@{}l@{}}CI is related to positive impact \\ on release cycle\end{tabular}         & 7                  & P58,P64, P97, P100, P102                                                              & 7.8 & 10         \\ \hline
  \begin{tabular}[c]{@{}l@{}}CI is associated with an increase\\ in cooperation\end{tabular}            & 8                  & P25, P52, P79, P81, P93, P106                                                        & 8.0 & 8         \\ \hline
  \begin{tabular}[c]{@{}l@{}}CI is associated with an improvement\\ in process reliability\end{tabular} & 7                  & \begin{tabular}[c]{@{}l@{}}P46, P52, P79, P92, P105, \\P106\end{tabular}             & 7.8 & 8                    \\ \hline
  \begin{tabular}[c]{@{}l@{}}CI is associated with technical \\ challenges\end{tabular}                 & 6                  & P4, P24, P38, P58, P97                                                               & 6.8 & 6.5                            \\ \hline
  \begin{tabular}[c]{@{}l@{}}CI is associated with an improvement\\ in process automation\end{tabular}  & 3                  & P58, P97, P106                                                                       & 7.0 & 7                     \\ \hline
  \begin{tabular}[c]{@{}l@{}}CI is associated with software \\ development benefits\end{tabular}        & 3                  & P40, P92, P105                                                                       & 9.0 & 9                           \\ \hline
  \begin{tabular}[c]{@{}l@{}}CI is associated with an increase \\ in the feedback frequency\end{tabular}       & 1                  & P46                                                                                   & 5.0 & 5                    \\ \hline
  CI facilitates the transition to agile                                                               & 1                  & P46                                                                                   & 5.0 & 5                    \\ \hline
  \multicolumn{3}{r}{\textbf{Overall}}                                                                                                                                                                              & \textbf{7.5}      & \textbf{8}\\      
  \end{tabular}
  \end{table}

In this theme, we group codes related to the effects of CI on the software processes. Table 10 presents the codes, the number of claims supporting the code, the primary studies in which they appear, the mean (\(\displaystyle \bar{Qscore}\)) and median (\(\displaystyle \tilde{Qscore}\)) scores. We map seven codes representing findings of positive effects of CI, and one negative effect. 

Three studies (see the quality scores in Table~\ref{tab:codes_development_process}) claim that ``CI is associated with software development benefits'', being considered a factor of success and contributing to a decrease in the rate of project failure. Moreover, ``CI is associated with an improvement in process reliability'' in six different studies pointing out progress in transparency, stability, predictability, and support for a quantitative view of progress. These studies are diverse in their claims, and there is room to further investigate what benefits and how CI contributes to process reliability. Moreover, some of these studies did not perform well in our quality assessment. P46 obtains a \(\displaystyle Qscore = 5\), P79 a \(\displaystyle Qscore = 8\) and P106 a \(\displaystyle Qscore = 7\), all of them have issues with the rigor criterion. For example, P106 scored 1 out of 4 on the rigor criterion, and P79 only shows descriptive statistics as a means to support its claims.

There is evidence that ``CI is associated with an increase in cooperation'', e.g., improving inter-team and intra-team communication (P52). Regarding this association, there are three studies with a low \(\displaystyle Qscore \), due to the rigor criterion. For example, P79 does not ground its claims on statistical tests. Also, P93 and P106 obtain 0 out of 4 points on the rigor criterion. On the other hand, P25, P52, and P81 convey methodological confidence. These studies perform well in all quality criteria and provide more reliability to the association between CI and increased cooperation. Concerning cooperation, the primary study P25 states:
\begin{displayquote}
  \textit{``After adoption of CI, normalized collaboration amount between programmers significantly increases for our set of OSS and proprietary projects.''}. (p. 12) 
\end{displayquote}

Some studies still suggest ``an improvement in process automation'', regarding this, the P97 discusses:

\begin{displayquote}
  \textit{``CI allows developers to focus more on being productive, and to let the CI take care of boring, repetitive steps, which can be handled by automation.''}. (p. 203) 
\end{displayquote}
The increase in automation mentioned by P97 is believed to lower the developers’ workload. However, the increase in automation also introduces technical challenges (shown below) to the development process that are associated with a perceived decrease in productivity (see Section~\ref{sec:developer_work}).

Five studies shed light on the relationship between CI with and a ``positive impact on release cycle''. Such studies  show that continuous integration promotes fast iterations supporting fast and regular releases. For example, P100 states:

\begin{displayquote}
  \textit{``We found that projects that use CI do indeed release more often than either (1) the same projects before they used CI or (2) the projects that do not use CI.''}. (p. 432)
\end{displayquote}

We can note in Table~\ref{tab:codes_development_process} a significant variety and high-quality scores (\(\displaystyle Qscore \)) of the studies claiming an association between CI and positive impacts on the release cycle, cooperation, process reliability, and software development benefits. On the other hand, few studies suggest that CI encourages process automation, and the codes ``CI is associated with an increase in feedback frequency'' and ``CI facilitates the transition to agile'' are presented by just one study, which is P46, with \(\displaystyle Qscore = 5\) (below the average).

A potential negative effect of CI in the {\em Software Process} are the technical challenges associated with adopting CI (6 occurrences in 5 studies), confirming the addition of extra complexity (see \ref{sec:developer_work}). For example, study P58 states: 

\begin{displayquote}
\textit{``As regarding the hidden problems associated with continuous integration usage, we found that 31 respondents are having a hard time configuring the build environment''.} (p. 76)
\end{displayquote}

Conversely, it is essential to highlight the low scores obtained by some studies supporting the association with technical challenges. For example, P38 and P58 obtained low scores mainly because of the rigor criterion (scores of 0 out of 4). This result suggests poor methodological reliability from these studies.  On the other hand, P4 and P97 obtained a high \(\displaystyle Qscore\) in all aspects and support the association between CI and technical challenges. Overall, studies highlight the difficulty in implementing CI as well as setting up the environment, especially to newcomers (by confirming problems of magnetism and retention of developers, see Section~\ref{sec:developer_work}). Moreover, some studies state that the lack of maturity of technology may contribute to the abandonment of CI, as stated in study P24: 

\begin{displayquote}
\textit{``Results show that all of the 13 interviewees mentioned challenges related to tools and infrastructure such as code review, regression feedback time when adopting to CI. The maturity of the tools and infrastructure was found to be a major issue.''}(p. 29) 
\end{displayquote}

These studies are diverse, being two case studies, two surveys, and one MSR/Survey, with quality scores of \(\displaystyle \bar{Qscore}=6.8\) and \(\displaystyle \tilde{Qscore}=6.5 \). As in the previous theme, the studies of this theme also highlight the human challenges and the extra complexity added, i.e., CI adds some level of complexity to practitioners (see Section \ref{sec:developer_work}). Additionally, the technical challenges seem to be related to CI configurations and seem to impact newcomers and specific domains (such as embedded systems). P58, for example, argues that newcomers may face barriers to create a build due to a lack of experience with the project. In turn, P38 shows that some embedded systems contain complex user scenarios that require manual testing.

\begin{boxed}
CI is mentioned as a success factor in software projects, positively affecting software processes, promoting faster iterations, more stability, predictability, and transparency. However, CI may also bring technical challenges to the team related to the build environment and tools. Practitioners and researchers may consider such challenges and elaborate strategies to mitigate them. Moreover, there is still space for studies to answer questions about automation and productivity.
\end{boxed}

\subsubsection{Quality Assurance \label{sec:quality_assurance}}

\begin{table}[H]
  \caption{Codes from the ``Quality Assurance'' theme. We show the number of claims related to the code, the primary studies supporting it, the mean and median of quality scores of such studies.}
  \label{tab:codes_quality} 
  \begin{tabular}{lclcc}
    \hline
    \multicolumn{1}{c}{\textbf{Code}} & \textbf{\begin{tabular}[c]{@{}c@{}}Number of\\ Claims\end{tabular}} & \multicolumn{1}{c}{\textbf{Primary Studies}} & \(\displaystyle \bar{Qscore} \) & \(\displaystyle \tilde{Qscore} \)                                             \\ \hline
  \begin{tabular}[c]{@{}l@{}}CI is related to positive impact \\ on test practice\end{tabular}    & 10                                     & \begin{tabular}[c]{@{}l@{}}P14, P29, P52, P89, P93, P97, \\ P100, P102, P106\end{tabular}    & 8.6   & 9                                         \\ \hline
  \begin{tabular}[c]{@{}l@{}}CI is related to an increase on \\ quality assessment\end{tabular}       & 8                                      & P44, P58, P64, P79, P97, P106                                                             & 6.7   & 7                                             \\ \hline
  \begin{tabular}[c]{@{}l@{}}CI is associated with to favor \\ continuous refactoring\end{tabular} & 1                                      & P44                                                                                         & 9.0   & 9                                          \\ \hline
  \begin{tabular}[c]{@{}l@{}}CI is associated to \\ multi-environment tests\end{tabular}           & 4                                      & P58, P97, P100, P106                                                                        & 7.7   & 7                                          \\ \hline
  \multicolumn{3}{r}{\textbf{Overall}}                                                                                                                                                                                                     & \textbf{7.8}      & \textbf{9}\\
  \end{tabular}
  \end{table}
  
Another significant theme that emerged from our primary studies is ``Quality Assurance''. As shown in Table~\ref{tab:codes_quality}, in general, CI is associated with continuous practice of quality assessment (P44), refactoring (P44), finding problems earlier (P58), and improving the code quality (P58, P64, P97, P106).  These associations emerge from the perception that CI can be used as a quality assessment, providing transparency and supporting multi-environment tests. One of these studies (P44) also suggests that CI provides an adequate context for employing continuous refactoring. 
  
Under the code ``CI is related to an increase on quality assessment'', there is little diversity of study types. In addition, P58 (\(\displaystyle Qscore = 4\)) and P64 (\(\displaystyle Qscore = 5\)) have low \(\displaystyle Qscores\). On the other hand, there is a convergence between the studies regarding greater awareness of code and product quality (P44, P58, P64, P79, P97, P106). 

Other studies with a wider variety of methods and high-quality scores provide reliable support for the association between CI and good test practices (at least in terms of the number of tests and test coverage). The exceptions are studies P58, P93, and P106 with a low \(\displaystyle Qscore\), especially regarding the fragility in validating their claims in their respective data. However, although P58, P93, and P106 obtain low scores, the median \(\displaystyle Qscore\) for the association between CI and good test practices is still strong.

According to the primary studies, the adoption of CI tends to enforce automated software testing, increasing the volume and coverage of tests. CI also encourages best practices of automated tests ranging from tests within private builds to functional tests on the cloud. The support to multi-environment tests is also mentioned as a support to a ``real-world environment''. For example, study P89 states: 

\begin{displayquote}
\textit{``After some (expected) initial adjustments, the amount (and potentially the quality) of automated tests seems to increase.'' }(p. 69) 
\end{displayquote}

In the same way, study P97 states: 
\begin{displayquote}
\textit{``Developers believe that using CI leads to higher code quality. By writing a good automated test suite, and running it after every change, developers can quickly identify when they make a change that does not behave as anticipated, or breaks some other part of the code.'' }(p. 203)
\end{displayquote}

\begin{boxed}
  CI is perceived as a provider of transparency and continuous quality assessment through enforcing the test practices and supporting multi-environmental tests. There is reliable evidence on the association between CI and test increasing and coverage. There is room to further investigation on the test quality and test effort in CI projects.
\end{boxed}

\subsubsection{Integration Patterns \label{sec:integration_patterns}}

\begin{table}[H]
  \caption{Codes from the ``Integration Patterns'' theme. We show the number of claims related to the code, the primary studies supporting it, the mean and median of quality scores of such studies.}
  \label{tab:codes_integration} 
  \begin{tabular}{lclcc}
    \hline
    \multicolumn{1}{c}{\textbf{Code}} & \textbf{\begin{tabular}[c]{@{}c@{}}Number of\\ Claims\end{tabular}} & \multicolumn{1}{c}{\textbf{Primary Studies}} & \(\displaystyle \bar{Qscore} \) & \(\displaystyle \tilde{Qscore} \) \\ \hline
    \begin{tabular}[c]{@{}l@{}}CI is related to positive impact \\ on pull request life-cycle\end{tabular}   & 10                  & P47, P69, P74, P81, P89, P100, P104                                                    & 10.0 & 10                                          \\ \hline
  \begin{tabular}[c]{@{}l@{}}CI is associated with a commit \\ pattern change\end{tabular}                      & 5                  & P25, P89, P102                                                              & 8.8 & 9                                                \\  \hline
  \begin{tabular}[c]{@{}l@{}}CI is associated with a positive \\ impact on integration practice\end{tabular} & 5                  & P59, P79, P97, P100, P102                                                      & 8.2 & 8                                             \\ \hline
  \begin{tabular}[c]{@{}l@{}}CI is related to negative impact \\ on pull request life-cycle\end{tabular}   & 3                  & P81, P89                                                                        & 10.0 & 10                                           \\ \hline
  \multicolumn{3}{r}{\textbf{Overall}}                                                                                                                                                                              & \textbf{9.3}      & \textbf{10}\\
  \end{tabular}
  \end{table}

  The ``Integration Patterns'' theme consists of claims related to commits and pull requests, as shown in Table~\ref{tab:codes_integration}. The association between CI and a ``positive impact on integration practice'' is observed by five studies through three case studies (P59, P79, P102) and two surveys (P97, P100). There are mentions of CI as a facilitator to the integration practice, making the code integration easier (P97, P100), reducing the stress (P59), and supporting a faster integration (P79, P102). Such benefits could also be a motivation for CI adoption, as the authors of P100 explains:
\begin{displayquote}
  \textit{``One reason developers gave for using CI is that it makes integration easier. One respondent added `To be more confident when merging PRs.' '' }(p. 433)
\end{displayquote}

The CI association with ``a change in commit patterns'', i.e., the way developers commit, can confirm the perceived benefits through three studies: one MSR (P25 - \(\displaystyle Qscore=9 \)), one MSR/survey (P89 - \(\displaystyle Qscore=10 \)), and one case study (P102 - \(\displaystyle Qscore=6 \)). When analyzing the evolution of projects, studies identify an increasing frequency and a decreasing size of commits. For example, study P89 applied an (\textit{Regression Discontinuity Design (RDD)}) associated with a survey to understand the longitudinal effect of CI adoption (TRAVISCI adoption) and found: an increasing number of merge commits as an indicator of a workflow change (e.g., migration to a pull-based model); and a decrease in size of merge commits as an indicator of more frequent integration.

Some primary studies with high \(\displaystyle Qscore\) and high variety of methods report a ``positive impact on pull request life-cycle'', such as (i) an increase in the number of pull requests (PRs) submissions (P81); (ii) an increase in the number of PRs closed (P89); (iii) acceleration in the integration process (P100); (iv) higher support to identify and reject problematic code submissions more quickly (P47, P69, P74, P100); (v) a higher contribution from external collaboration (P74); (vi) a higher delivery of PRs (P81). However, these studies also reveal a ``negative impact on pull request life-cycle'', i.e., merging pull requests might become slower after adopting CI (P81, P89). P81 states that: 


\begin{displayquote}
  \textit{``Open source projects that plan to adopt CI should be aware that the adoption of CI will not necessarily deliver merged PRs more quickly. On the other hand, as the pull-based development can attract the interest of external contributors, and hence, increase the project's workload, CI may help in other aspects, e.g., delivering more functionalities to end-users.'' } (p. 140)
\end{displayquote}
The study P81 also indicates that CI can be associated with an increase in the lifetime of pull request:
\begin{displayquote}
  \textit{``We  observe  that  in  54\%  (47/87)  of  our projects, PRs have a larger lifetime after adopting CI.''}. (p. 134)
\end{displayquote}
We discuss this apparent contradiction in more details in Section~\ref{disc: clarify_integration}.

Although the three case studies P59 (\(\displaystyle Qscore=7\)), P79 (\(\displaystyle Qscore=8\)), and P102 (\(\displaystyle Qscore=6\)) obtain lower \(\displaystyle Qscores\) than the rest of studies in  the {\em Integration Patterns} theme, the variety of methods and the overall mean quality of the studies in this theme is high --- \(\displaystyle \bar{Qscore}  = 9.3\). Additionally, the studies seem reliable as their overall rigor mean is of 2.83 (out of 4 points). 

\begin{boxed}
The studies report a perception of CI as a facilitator to the integration practice, influencing positively the way developers perform commit even the workflow. CI can benefit the pull-based development by improving and accelerating the integration process. However, there is evidence that CI may increase the lifetime of pull requests.
\end{boxed}

\subsubsection{Issues \& defects \label{sec:issues_defects}}

\begin{table}[H]
  \caption{Codes from the ``Issues \& Defects'' themes. We show the number of claims related to the code, the primary studies supporting it, the mean and median of quality scores of such studies.}
  \label{tab:codes_issues} 
  \begin{tabular}{lclcc}
    \hline
    \multicolumn{1}{c}{\textbf{Code}} & \textbf{\begin{tabular}[c]{@{}c@{}}Number of\\ Claims\end{tabular}} & \multicolumn{1}{c}{\textbf{Primary Studies}} & \(\displaystyle \bar{Qscore} \) & \(\displaystyle \tilde{Qscore} \) \\ \hline
  CI is associated with defect reduction                                                             & 6                  & P25, P31, P50, P74, P79                                                                                       & 8.5 & 9                                                    \\ \hline
  CI is associated with issues reduction                                                              & 2                  & P25, P89                                                                                                     & 9.5 & 9                                                     \\ \hline
  \begin{tabular}[c]{@{}l@{}}CI is associated with a decrease \\ in time to address defects\end{tabular} & 6                  & \begin{tabular}[c]{@{}l@{}}P50, P59, P79, P97, P100, \\ P106\end{tabular}                                    & 8.6 & 8                                                     \\ \hline
  \multicolumn{3}{r}{\textbf{Overall}}                                                                                                                                                                                                    & \textbf{8.7}      & \textbf{9}\\
  \end{tabular}
  \end{table}

The manner by which CI projects address defects, bugs, and issues is grouped under the ``issues and defects'' theme. Table \ref{tab:codes_issues} shows three codes representing these aspects and summarizes their occurrences. These studies consistently indicate that CI enables teams to detect and address issues earlier, which is related to an overall decrease in the number of issues and defects reported, i.e., an external quality improvement. For instance, study P50 states that:

\begin{displayquote}
\textit{``The descriptive statistics point to an overall improvement in not only finding more defects (defect reduction), but also in shortening the time required to fix the defects (defect lead and throughput).'' } (p. 8)
\end{displayquote}

We observe other statements regarding CI helping development teams to find and fix bugs and broken builds, shortening the time to fix these bugs and builds. In particular P59 states:
\begin{displayquote}
\textit{``An indirect, but important, advantage of CI is related to the following human factor: the earlier the developer is notified of an issue with the patch that was just committed, the easier it is for him or her to associate this regression with specific changes in code that could have caused the problem and fix it.'' }(p. 9)
\end{displayquote}

However, regarding issues or bugs resolution rate, while study P25 identified an increase in the resolution rate after CI adoption for OSS projects, P89 identifies that issue resolution tends to be slower after CI adoption. 

The overall mean quality score of studies in this theme is high --- \(\displaystyle \bar{Qscore}  = 8.7\) and \( \displaystyle \tilde{Qscore}  = 9\), having also an overall mean rigor of 2.6 (out of 4 points). The quality weaknesses are compensated by other studies with higher quality scores and methodological rigor. For example, for the code ``CI is associated with a decrease in time to address defects'', P79 and P106 claim that CI supports the team to catch issues earlier. Similarly, P50, P59, P97, and P100 corroborate the claim that CI helps to find and fix problems earlier. Regarding the code ``CI is associated with defect reduction'', P31 and P79 reveal an association with reduced defects, which P50 confirms. Additionally, P74 claims that CI yields a higher bug discovery, and P25 claims that CI yields a higher bug resolved rate.

\begin{boxed}
  Studies suggest that CI can improve the time to find and address issues. They also observed a decrease in defects reported.
\end{boxed}

\subsubsection{Build Patterns \label{sec:build_patterns}}

\begin{table}[H]
  \caption{Codes from the ``Build Patterns'' theme. We show the number of claims related to the code, the primary studies supporting it, the mean and median of quality scores of such studies.}
  \label{tab:codes_build} 
  \begin{tabular}{lclcc}
    \hline
    \multicolumn{1}{c}{\textbf{Code}} & \textbf{\begin{tabular}[c]{@{}c@{}}Number of\\ Claims\end{tabular}} & \multicolumn{1}{c}{\textbf{Primary Studies}} & \(\displaystyle \bar{Qscore} \) & \(\displaystyle \tilde{Qscore} \) \\ \hline
  CI is associated with build health                                                          & 6                  & P29,P49, P97, P102            & 8.0  & 7                          \\ \hline
  \begin{tabular}[c]{@{}l@{}}CI is associated with a decreasing \\ in build time\end{tabular} & 1                  & P102                          & 6.0  & 6                           \\ \hline
  \multicolumn{3}{r}{\textbf{Overall}}                                                                                                             & \textbf{7.7}      & \textbf{7}\\
  \end{tabular}
\end{table}

The ``Build Patterns" theme encompasses the reported associations between CI and build metrics. Table \ref{tab:codes_build} shows two codes representing the ``Build Patterns'' theme.  Under the code ``CI is associated with build health'', 6 mentions in 4 studies report developers' good practices encouraged by CI, as well as an improvement in build success rate (P49, P102). CI encourages good practices that contribute to build health, such as testing in private builds (P29), prioritization to fix broken builds (P29), and supporting a shared build environment (P97). P97 states the following:

\begin{displayquote}
\textit{``Several developers told us that in their team if the code does not build on the CI server, then the build is considered broken, regardless of how it behaves on an individual developer’s machine. For example, S5 said:`...If it doesn’t work here (on the CI), it doesn’t matter if it works on your machine.'''} (p. 202)
\end{displayquote}

One study, P102, also reports a decrease in the build time. On the other hand, P97 registers long build time as a common barrier faced by CI developers. Therefore, we argue that there is scope for further investigation of the factors that influence build time, build health, and relationships with other variables in CI projects. For example, only one of these studies analyzes build data (P49) in a controlled context. In addition, the overall mean quality score of studies in this theme is not high --- \(\displaystyle \bar{Qscore}  = 7.7\) and \( \displaystyle \tilde{Qscore}  = 7\), with an overall mean rigor of 2.5 (out of 4 points).

\begin{boxed}
  CI promotes good practices related to build health and contributes to an increase in successful builds.
\end{boxed}

\subsection{\bfseries RQ3: \RQthree \label{rq_method}}
In this RQ we investigate the methodologies applied in the primary studies. In particular, we analyze: the kind of projects that our primary studies analyze (Section \ref{sub: projects}), the study methodologies, i.e., the kind of the studies and their quality scores (Section \ref{sub: results_methods}), and the availability of the artifacts produced as part of these studies  (Section \ref{sub: reproducibility}).

\subsubsection{Projects analyzed \label{sub: projects}}
\begin{figure}[h!]
  \centering
  \includegraphics[scale=.5]{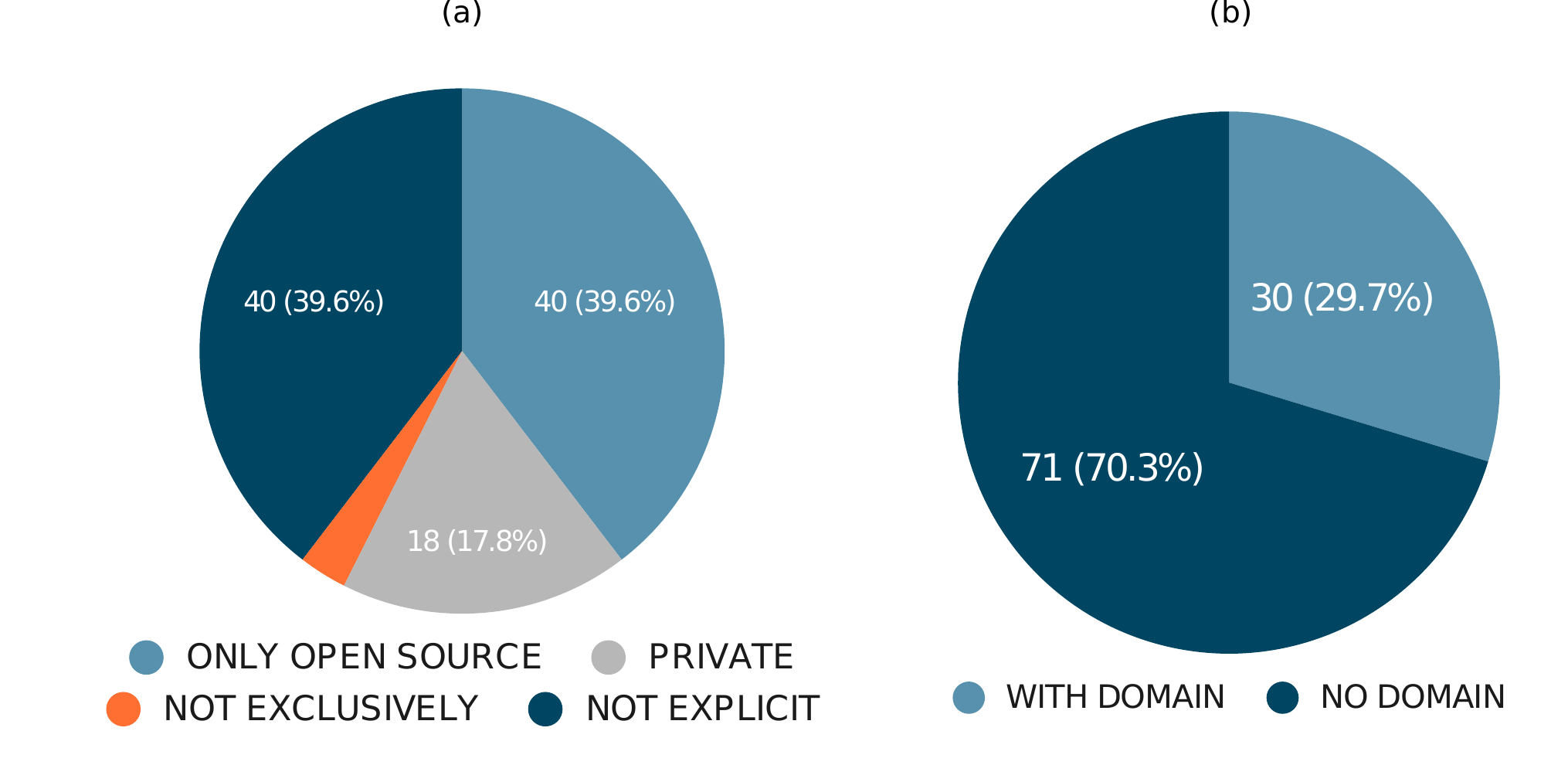}
	\caption{(a) Proportion of studies based on the type of projects they analyze; (b) Proportion of studies that analyzed projects from specific domains (and vice versa);}
  \label{fig:pub_projects}
\end{figure}

Figure~\ref{fig:pub_projects} (a) shows information about the type of projects that were investigated in CI studies.
We can observe that 40 of 101 studies (39.6\%), analyze only open source projects. In contrast, 18 studies (17.8\%) investigate private projects. On the other hand, 40 of the studies (39.6\%) are not explicit about licenses of the projects, and 3 (3\%) studies analyze mixed project settings, i.e., both open source and private projects.

Figure~\ref{fig:pub_projects} (b) shows that 71 out of 101 studies (70.3\%) do not investigate projects from a specific domain. On the other hand, 30 studies (29.7\%) investigate domain-specific projects. For studies investigating specific project domains, we catalog 17 different domains (see Table~\ref{tab:application_domains}). The most frequent domains are transports (4 occurrences), embedded systems (4 occurrences), telecommunications (3 occurrences), and software development (3 occurrences).

\begin{table}[H]
\caption{Applications domains investigated in primary studies.}
\label{tab:application_domains} 
\begin{tabular}{lc}
\hline
\multicolumn{1}{c}{\textbf{Domain}} & \textbf{Ocurrences} \\ \hline
Transports                          & 4                \\
Embedded systems                    & 4                \\
Telecommuntications                 & 3      \\
Software Development                & 3      \\
Web Application                     & 2      \\
Finance                             & 2      \\
Cloud Computing                     & 2      \\
Military Systems                    & 2      \\
Home and office solutions           & 1      \\
Bookmaking company                  & 1      \\
Mobile software and social networks & 1      \\
Health care                         & 1      \\
HPC environment                     & 1      \\
Serverless applications             & 1      \\
Neuroinformatics                    & 1      \\
Databases migration                 & 1      \\ \hline
\end{tabular}
\end{table}

\begin{figure}[h!]
  \centering
  \includegraphics[scale=.5]{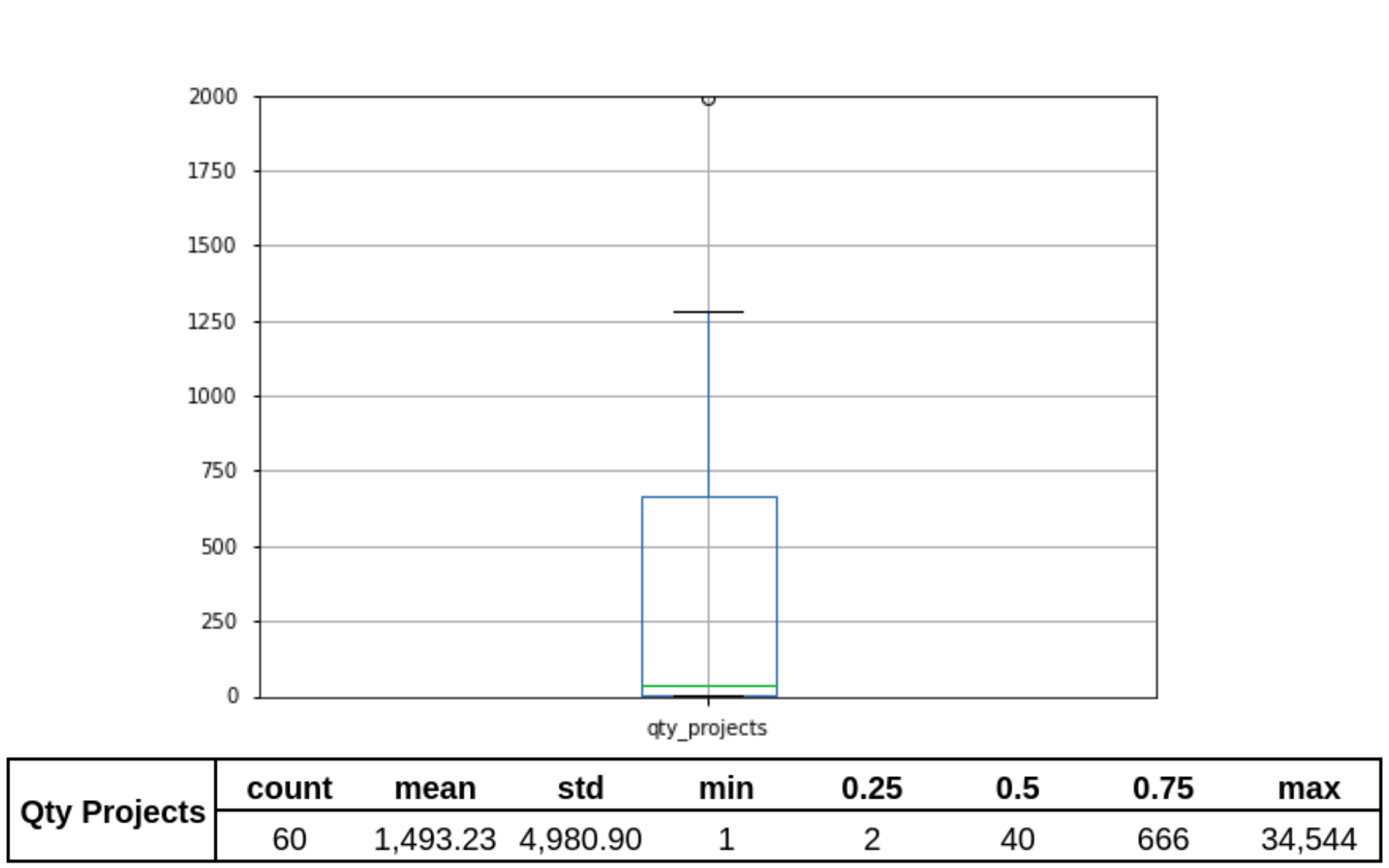}
	\caption{Boxplot and descriptive statistics of the projects that were analyzed by our primary studies.}
  \label{fig:qty_projects}
\end{figure}

60 out of 101 studies focus on analyzing the historical data of software projects (e.g., production code or tests). The other remaining studies conducted interviews, surveys, or analyzed other units of information different from projects' source code, e.g., builds or companies. 
Figure \ref{fig:qty_projects} shows the descriptive statistics of the projects that were analyzed per study. We hide outliers for readability purposes---the highest outlier has 34,544 projects. While the mean of analyzed projects is 1,493 projects, the median is just 40, with a high frequency of studies analyzing just 1 or 2 projects. Some studies seem to be outliers, such as P72, which investigated 13,590 projects, and P100, which investigated 34,544 projects. 

The P100 study uses a large corpus of projects (i.e., 34,544 projects) for specific investigations. For example, a large corpus of projects is used to identify which CI services are mostly used. However, to perform more specific investigations, P100 uses only a subset of the total corpus of projects (i.e., 1,000 projects). P72 highlights the \textsc{TravisTorrent}~\cite{travistorrent} dataset, which is a widely known dataset of projects from \textsc{GitHub} that collates build logs from \textsc{TravisCI}.

With the presented data, we can observe that, in general, the studies are distributed in various contexts, with the majority without a specific domain. Despite that, considering the 29.7\% of the studies investigate domain-specific projects and the variance of CI among different domains and implementations \cite{stahl2014,viggiato2019}, we suggest that the community employ studies that explore the differences in the CI implementations among such domains and how it influences the CI outcomes. We still draw attention to the high frequency of studies analyzing low-size samples.

\subsubsection{Availability of Artifacts \label{sub: reproducibility}}

Robles~\cite{robles2010} investigated the MSR conference papers from 2004 to 2009. He found that the majority of published papers are hard to replicate. For example, although 64 out of 171 papers (37.4\%) are based on publicly available datasets, these datasets are in the {\em ``raw''} form and the papers do not provide the processed version of the datasets nor the tools that were used to process these datasets. Another 18.12\% of papers (i.e., 31 papers) do not even provide the {\em ``raw''} data to begin with.

\begin{figure}[H]
  \includegraphics[scale=0.5]{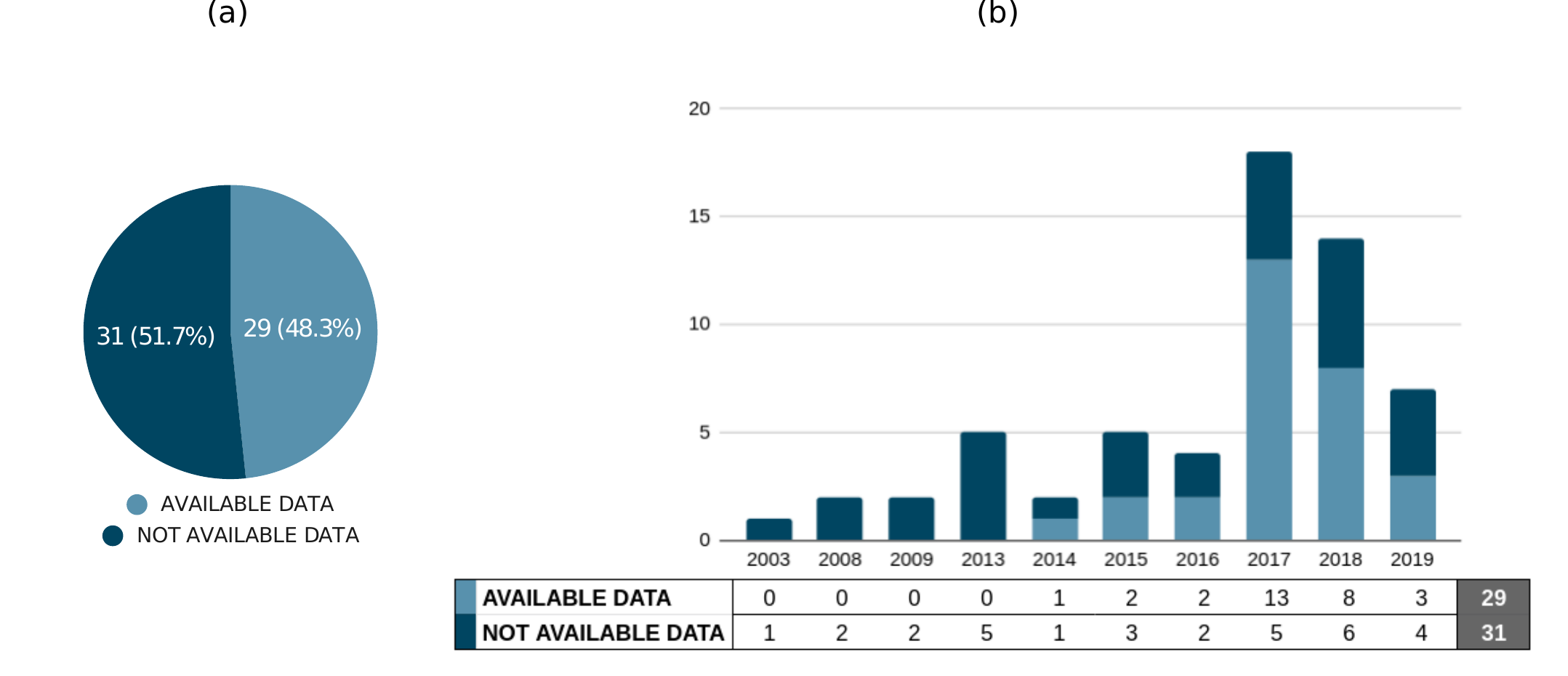}
  \caption{(a) Proportion of studies according to data availability; (b) Proportion of transparency over the years.}
  \label{fig:pub_reproducibility}
\end{figure}

Rodríguez-Pérez et al. \cite{perez2018} capitalize on the same issue of data availability and raise the concern about the reliability of the results from studies that are not reproducible studies. On the other hand, they \cite{perez2018} also report the increasing attention that the community has given to the issue of data availability over the last years.
Therefore, in our study, we collect information about the availability of the artifacts used or produced in the primary studies, considering those that analyze projects. Figure~\ref{fig:pub_reproducibility} (a) reveals that 29 studies out of 60 (48.33\%) provide publicly available datasets, while 31 studies (51.66\%) do not provide publicly available datasets. From the studies providing publicly available datasets, all of them are studies using open source projects. On the other hand, those studies investigating private projects do not present dataset nor anonymized nor in a raw manner, while some of the studies with mixed---both private and open source---projects provide only partial data referring to OSS projects. Other studies are not explicit regarding whether the dataset comes from private or OSS projects and does not present it.

Figure~\ref{fig:pub_reproducibility} (b) shows an increasing trend of pushing for data transparency in CI studies. Over recent years, we observe that the proportion of studies providing a publicly available dataset is higher than 50\% (we consider only studies that analyze projects data). This increase in transparency might be due to initiatives from prominent conferences, such as the {\em artifact tracks}, in which authors are provided with special {\em badges} as a credit for their effort invested in sharing their artifacts. Nowadays, there are even awards to encourage the sharing of reproducible artifacts~\footnote{\url{https://icsme2020.github.io/cfp/ArtifactROSETrackCFP.html}}.

\subsubsection{Study Quality and Methodologies \label{sub: results_methods}}

Using the quality checklist and procedures presented in Section\ref{quality_assessment}, we assessed the methodologies of primary studies from which we extracted the findings discussed in Section\ref{rq_themes} (38 papers). Similar to the approach applied by Dybå \& Dingsøyr \cite{dyba2008}, such questions allow us to measure the reliability of the findings using the \textit{quality score} as a proxy. By answering each question in the checklist (see Table \ref{tab:quality_checklist}) with a 1 (yes) or 0 (no), the sum of these values produces a quality score, as shown in the last column of Table \ref{tab:quality_assessment}. The remaining columns show the answers to each question of the checklist, while the rows represent each paper with claims discussed in this work.

As peer-review is an exclusion criterion, all of selected papers obtain 1 to Q1. The values for Q2 indicate whether studies have clearly defined aims. While the entire set of responses to Q1 through Q4 shows that the papers have, in general, a good report quality, in terms of rigour (Q5 to Q8), the obtained scores are lower, especially regarding Q8 - ``Is there a comparison or control group?''. Q8 has the lowest rate among primary studies with only five papers (P4, P14, P49, P50, and P79) having applied control groups to compare their results and findings. 
                                                                                                                                                                                                                                
The few studies with data available (12 out of the 38) impacts the credibility assessed by questions Q9 and Q10, i.e., ``Does the empirical data and results support the findings?'', respectively. The relevance (as expressed by Q11) of 30 out of 38 studies is clear and well discussed in terms of contributions to research and practice.

\begin{figure}[H]
  \includegraphics[scale=.65]{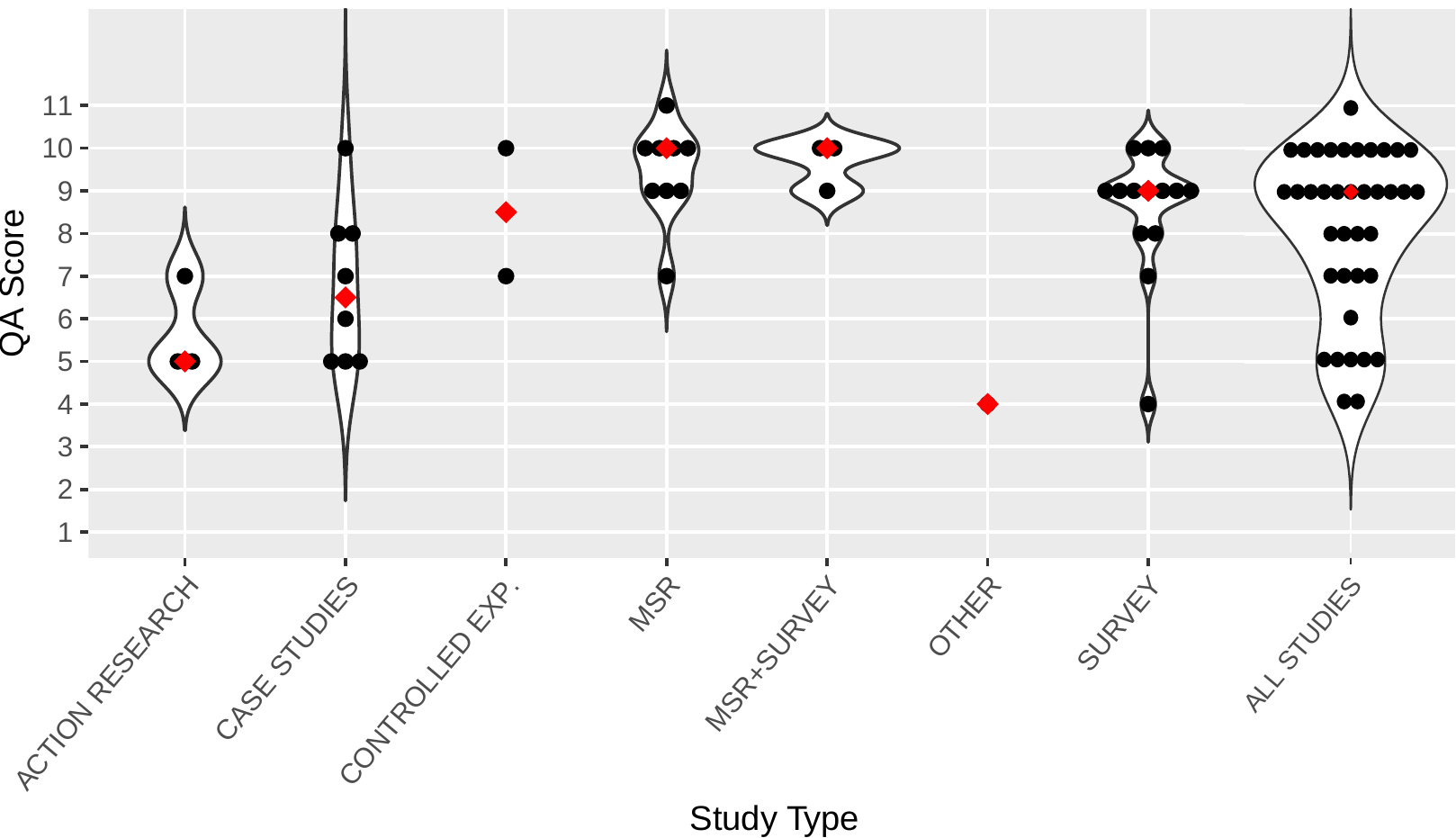}
  \caption{Quality assessment scores per study type.}
  \label{fig:qa_kind}
\end{figure}

Figure \ref{fig:qa_kind} shows the scores grouped by type of study. The categories are (i) {\em mining software repository} (MSR) studies, and the other four classical empirical methods for software engineering: (ii) {\em controlled experiment}, (iii) {\em case study}, (iv) {\em survey}, and (v) {\em action research}~\cite{easterbrook2008}. We consider claims extracted from studies with higher scores to be more reliable than claims from studies with lower scores. We observe that MSR studies have higher quality scores, with a median of 10 points. The mixed-methods studies (MSR and survey) have the same median also with high scores (and even less variation). The last plot in Figure \ref{fig:qa_kind} shows the overall quality of the studies, with 4 as minimal score and a maximum of 11. The median score is 9.

Table~\ref{tab:QA_kindStudy_QAspect} shows the average performance of the studies grouped by study type and the four quality criteria within our quality assessment checklist (see section~\ref{quality_assessment}). All types of studies performed well in quality of reporting, bordering the maximum score. In rigor, on the other hand, studies from action research and case studies have, on average, less than 2 points. This lower performance is significantly affected by question 6 regarding the metrics or concepts considered in the studies. Question 8, which concerns control or comparison groups, also contributes to the lower performance in quality scores. 

We suggest the software engineering community to strengthen the metrics and concepts and to employ comparison strategies. For example, when studying productivity, the studies should clarify the definitions of productivity (e.g., developer perception, story points, worked hours), how productivity is measured (e.g., questionnaires, issue trackers, management systems, work time), and the rationale behind this concept definition and measurement.

In the credibility criterion, case study and action research studies performed poorly, mainly in questions Q9 and Q10, which are regarding the evidence supporting findings and data availability. Lastly, regarding the relevance criterion (Q11), the selected controlled experiments scored low. Considering the results of the relevance criterion, we also recommend special attention to the development of case study and action research studies to be more transparent regarding data and to provide the rationale behind their conclusions. For example, although P50 performed a high-quality case study, P50 did not share its data for verification or reproducibility. 

\begin{table}[H]
  \caption{Quality Assessment per Kind of Study}
  \label{tab:QA_kindStudy_QAspect} 
  \begin{tabular}{lcccc}
                          & \textbf{\begin{tabular}[c]{@{}c@{}}Quality of Reporting\\ (0..3)\end{tabular}} & \textbf{\begin{tabular}[c]{@{}c@{}}Rigor\\ (0..4)\end{tabular}} & \textbf{\begin{tabular}[c]{@{}c@{}}Credibility\\ (0..3)\end{tabular}} & \textbf{\begin{tabular}[c]{@{}c@{}}Relevance\\ (0..1)\end{tabular}} \\ \hline
  \textbf{ACTION RESEARCH} & 2.66                                                                           & 1.33                                                            & 1.00                                                                  & 0.66                                                                \\ \hline
  \textbf{SURVEY}          & 3.00                                                                           & 2.50                                                            & 2.14                                                                  & 0.92                                                                \\ \hline
  \textbf{CONTROLLED EXP.} & 3.00                                                                           & 2.50                                                            & 2.50                                                                  & 0.50                                                                \\ \hline
  \textbf{CASE STUDIES}    & 2.75                                                                           & 1.87                                                            & 1.50                                                                  & 0.62                                                                \\ \hline
  \textbf{MSR}             & 2.88                                                                           & 3.00                                                            & 2.66                                                                  & 0.88                                                                \\ \hline
  \textbf{MSR+SURVEY}      & 3.00                                                                           & 3.33                                                            & 2.66                                                                  & 0.66                                                                \\ \hline
  \end{tabular}
\end{table}

\begin{table}[H]
  \caption{Quality Assessment.}
  \label{tab:quality_assessment}
  \begin{tabular}{|c|c|c|c|c|c|c|c|c|c|c|c|c|}
  \hline
  \textbf{Study} & \textbf{Q1} & \textbf{Q2} & \textbf{Q3} & \textbf{Q4} & \textbf{Q5} & \textbf{Q6} & \textbf{Q7} & \textbf{Q8} & \textbf{Q9} & \textbf{Q10} & \textbf{Q11} & \textbf{Total} \\
  \hline
  P4           & 1           & 1           & 1           & 1           & 1           & 1           & 1           & 1           & 1           & 0            & 1            & \textbf{10}             \\ \hline
  P6           & 1           & 1           & 1           & 1           & 0           & 0           & 0           & 0           & 0           & 0            & 1            & \textbf{5}              \\ \hline
  P7           & 1           & 1           & 1           & 1           & 1           & 1           & 1           & 0           & 1           & 1            & 0            & \textbf{9}              \\ \hline
  P9           & 1           & 1           & 1           & 1           & 0           & 1           & 1           & 0           & 1           & 0            & 1            & \textbf{8}              \\ \hline
  P14          & 1           & 1           & 1           & 1           & 1           & 1           & 1           & 1           & 1           & 1            & 1            & \textbf{11}             \\ \hline
  P24          & 1           & 1           & 1           & 1           & 0           & 1           & 1           & 0           & 1           & 0            & 1            & \textbf{8}              \\ \hline
  P25          & 1           & 1           & 1           & 1           & 1           & 1           & 1           & 0           & 1           & 0            & 1            & \textbf{9}              \\ \hline
  P29          & 1           & 1           & 1           & 1           & 1           & 1           & 1           & 0           & 1           & 0            & 1             & \textbf{9}              \\ \hline
  P31          & 1           & 1           & 1           & 0           & 0           & 0           & 1           & 0           & 0           & 0            & 1             & \textbf{5}              \\ \hline
  P38          & 1           & 1           & 1           & 1           & 0           & 0           & 0           & 0           & 0           & 0            & 1            & \textbf{5}              \\ \hline
  P39          & 1           & 1           & 1           & 1           & 1           & 0           & 1           & 0           & 0           & 0            & 1            & \textbf{7}              \\ \hline
  P40          & 1           & 1           & 1           & 1           & 1           & 1           & 1           & 0           & 1           & 0            & 1            & \textbf{9}              \\ \hline
  P44          & 1           & 1           & 1           & 1           & 1           & 1           & 1           & 0           & 1           & 0            & 1            & \textbf{9}              \\ \hline
  P46          & 1           & 1           & 1           & 0           & 1           & 0           & 1           & 0           & 0           & 0            & 0            & \textbf{5}              \\ \hline
  P47          & 1           & 1           & 1           & 1           & 1           & 1           & 1           & 0           & 1           & 1            & 1            & \textbf{10}             \\ \hline
  P49          & 1           & 1           & 1           & 1           & 0           & 1           & 0           & 1           & 1           & 0            & 0            & \textbf{7}              \\ \hline
  P50          & 1           & 1           & 1           & 1           & 1           & 1           & 1           & 1           & 1           & 0            & 1            & \textbf{10}             \\ \hline
  P52          & 1           & 1           & 1           & 1           & 1           & 1           & 1           & 0           & 1           & 0            & 1            & \textbf{9}              \\ \hline
  P58          & 1           & 1           & 1           & 1           & 0           & 0           & 0           & 0           & 0           & 0            & 0            & \textbf{4}              \\ \hline
  P59          & 1           & 1           & 1           & 0           & 0           & 1           & 1           & 0           & 1           & 0            & 1            & \textbf{7}              \\ \hline
  P64          & 1           & 1           & 1           & 1           & 0           & 0           & 1           & 0           & 0           & 0            & 0            & \textbf{5}              \\ \hline
  P69          & 1           & 1           & 1           & 1           & 1           & 1           & 1           & 0           & 1           & 1            & 1            & \textbf{10}             \\ \hline
  P73          & 1           & 1           & 1           & 1           & 1           & 1           & 1           & 0           & 1           & 0            & 1            & \textbf{9}              \\ \hline
  P74          & 1           & 1           & 1           & 1           & 1           & 1           & 1           & 0           & 1           & 1            & 1            & \textbf{10}             \\ \hline
  P79          & 1           & 1           & 1           & 1           & 1           & 1           & 0           & 1           & 1           & 0            & 0            & \textbf{8}              \\ \hline
  P81          & 1           & 1           & 1           & 1           & 1           & 1           & 1           & 0           & 1           & 1            & 1            & \textbf{10}            \\ \hline
  P89          & 1           & 1           & 1           & 1           & 1           & 1           & 1           & 0           & 1           & 1            & 1            & \textbf{10}             \\ \hline
  P90          & 1           & 1           & 1           & 1           & 1           & 1           & 1           & 0           & 1           & 1            & 0            & \textbf{9}              \\ \hline
  P91          & 1           & 1           & 1           & 1           & 0           & 1           & 1           & 0           & 1           & 0            & 1            & \textbf{8}              \\ \hline
  P92          & 1           & 1           & 1           & 1           & 1           & 1           & 1           & 0           & 1           & 0            & 1            & \textbf{9}              \\ \hline
  P93          & 1           & 1           & 0           & 0           & 0           & 0           & 1           & 0           & 0           & 0            & 1            & \textbf{4}              \\ \hline
  P97          & 1           & 1           & 1           & 1           & 1           & 1           & 1           & 0           & 1           & 1            & 1            & \textbf{10}             \\ \hline
  P99          & 1           & 1           & 1           & 1           & 1           & 1           & 1           & 0           & 1           & 0            & 1            & \textbf{9}              \\ \hline
  P100         & 1           & 1           & 1           & 1           & 1           & 1           & 1           & 0           & 1           & 1            & 1            & \textbf{10}             \\ \hline
  P102         & 1           & 1           & 1           & 1           & 1           & 0           & 1           & 0           & 0           & 0            & 0            & \textbf{6}              \\ \hline
  P104         & 1           & 1           & 1           & 1           & 1           & 1           & 1           & 0           & 1           & 1            & 1            & \textbf{10}             \\ \hline
  P105         & 1           & 1           & 1           & 1           & 1           & 1           & 1           & 0           & 1           & 0            & 1            & \textbf{9}              \\ \hline
  P106         & 1           & 1           & 1           & 1           & 1           & 0           & 0           & 0           & 0           & 1            & 1            & \textbf{7}              \\ \hline
  \textbf{Total} & \textbf{38} & \textbf{38} & \textbf{37} & \textbf{34} & \textbf{27} & \textbf{28} & \textbf{32} & \textbf{5} & \textbf{28} & \textbf{12} & \textbf{30} &  \\ \hline
  \end{tabular}
  \end{table}

\begin{figure}[H]
  \includegraphics[scale=.8]{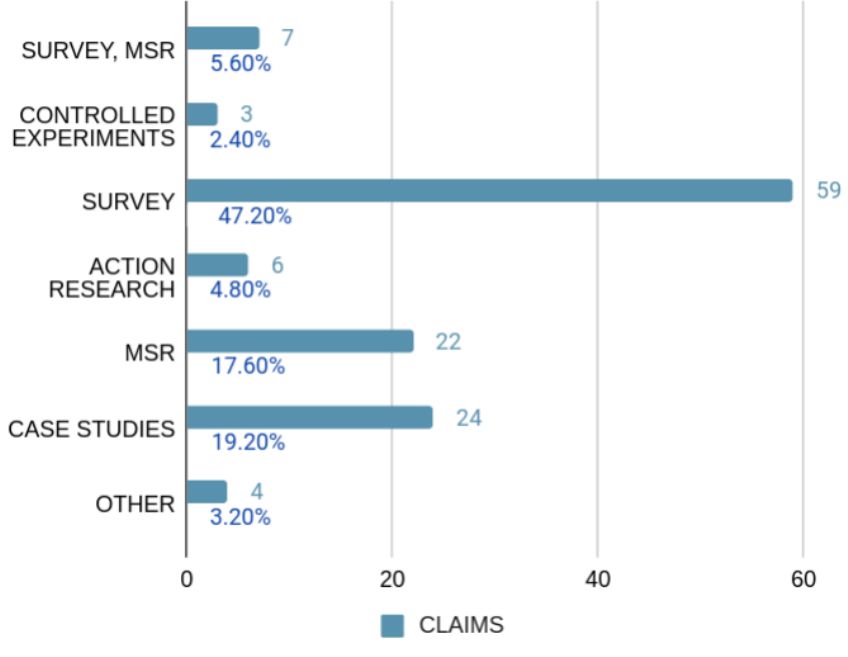}
  \caption{Proportion and quantity of claims per study type;}
  \label{fig:claims_kind}
\end{figure}

Figure~\ref{fig:claims_kind} shows the proportion of claims per study type, revealing that survey research was the method applied the most concerning the extracted claims (47.2\% - 59 claims)---followed by case studies (19.2\% - 24 claims) and MSR (17.6\% - 22 claims). Only 7 (5.6\%) of the claims are associated with a mixed-methods approach, emerging from 3 studies (P4, P7, and P89). The largest proportion of the claims --- 70.4\% emerges from MSR, Surveys, and mixed-methods studies, which is a promising given that such categories of studies obtain the highest quality scores.

\begin{figure}[H]
  \includegraphics[scale=.6]{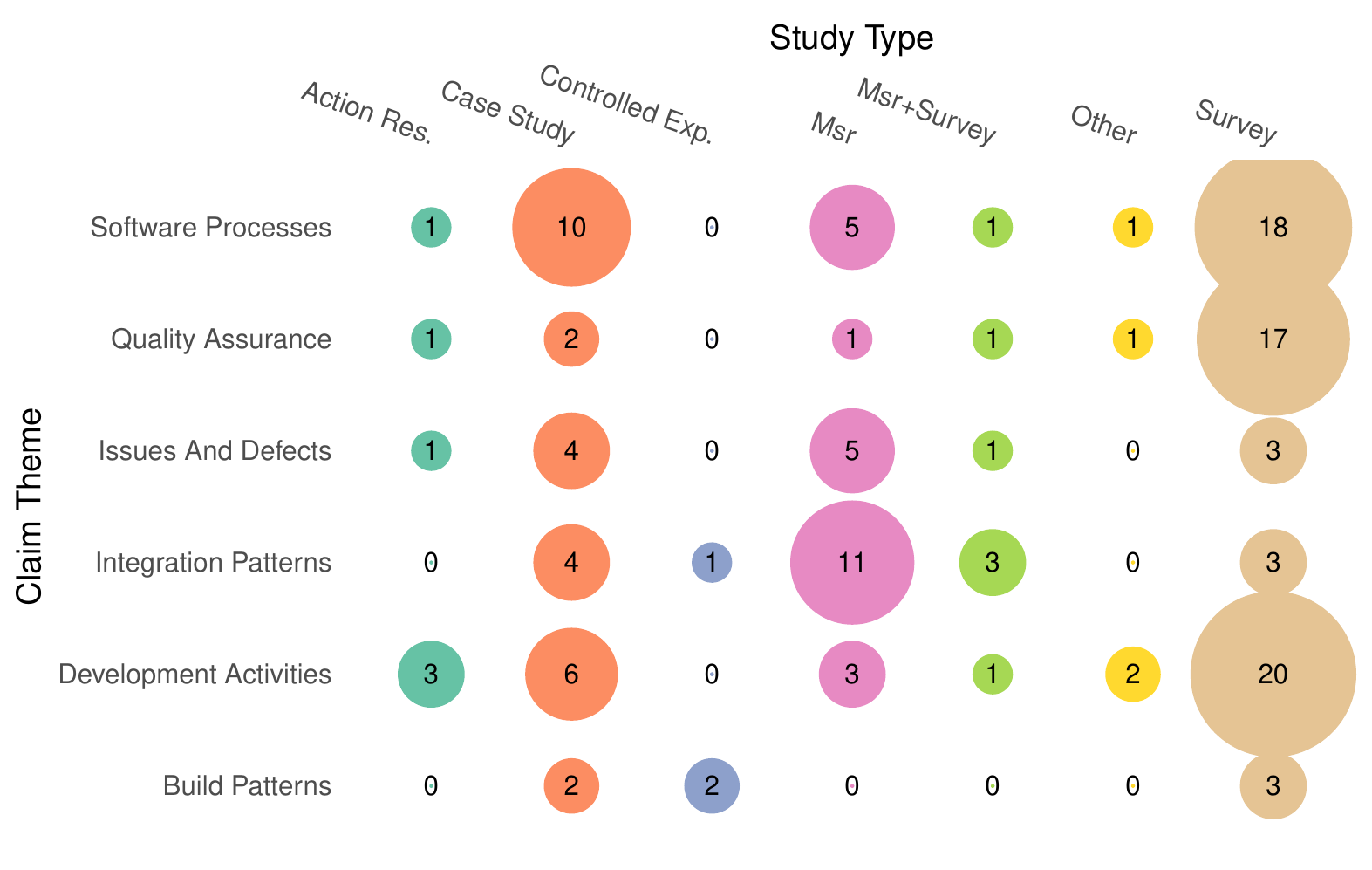}
  \caption{Claims quantity for each theme and study type.}
  \label{fig:themes_kind}
\end{figure}

In Figure \ref{fig:themes_kind}, we analyze the occurrence of each study type within the studied themes. We observe that the {\em Integration Patterns} theme has 14 out of 22 claims (63.64\%) made by mining software repositories (MSR) or mixed-methods (which are categories with the highest quality scores).

The {\em Issues \& Defects} theme also has a significant frequency (6 out of 14 - 42.86\%) of claims from MSR and mixed-methods. With respect to Quality Assurance, the theme has 23 claims, but there is a huge concentration (73.91\%) of claims made from surveys, which can suggest the need for more complementary study types.

Lastly, during the extraction phase (Section \ref{extraction}), we recorded methodological instruments used by the primary studies to confirm their findings. Table \ref{tab:evaluation_methods} reports a summary of statistical tests, models, and qualitative methodological instruments identified in the selected studies. We extract methods ranging from statistical tests, such as Cliff’s delta, and Mann-Whitney-Wilcoxon, to qualitative methods such as thematic analysis and interviews.

\begin{table}[H]
  \caption{Methodological instruments applied in the studies to confirm findings.}
  \label{tab:evaluation_methods} 
  \begin{tabular}{ll}
  \hline
  \multicolumn{1}{c}{\textbf{Instruments}}               & \multicolumn{1}{c}{\textbf{Studies}} \\ \hline
  ANOVA                                                        & P91, P104                                  \\
  Cliff's delta                                                & P14,P25, P81, P90, P105                                  \\
  Cohorts comparison                                           & P49                                  \\
  Cronbach alphas and Factor analysis                          & P9                                   \\
  Fisher’s exact test                                          & P100                                 \\
  Interview                                                    & P97, P99                                  \\
  Linear Regression and ANOVA                                  & P91                                  \\
  Logistic Regression                                          & P4, P74                                  \\
  K-Means                                                      & P14                                  \\
  Mann-Whitney-Wilcoxon test (MWW)                             & P14, P25, P81, P90, P100, P105                                  \\
  Mixed-effects RDD Model                                      & P89                                  \\
  Multiple Linear Regression                                   & P47, P104                                  \\
  Survey                                                       & P7, P52, P89, P97, P99                                  \\
  Survey Average Score and Standard Deviation                  & P52                                  \\
  Thematic Analysis                                            & P24                                  \\
  \end{tabular}
\end{table}

\section{Discussion \label{discussion} }

In the previous sections, we presented the findings from this SLR related to the research questions. In this section, we discuss the results, beginning with some methodological aspects. Then, we identify and discuss limitations on literature concerning the considered setup of continuous integration. Finally, we highlight some research opportunities.

\subsection{CI Environment and Study Results \label{discussion_ci_criteria}}
\label{disc_ci_environment}

As discussed in Section \ref{continuous_integration}, we adopt practices based on Duvall et al.~\cite{duvall2013} and Fowler~\cite{fowler2006} to identify implementation of CI. From these criteria, we find (see Section \ref{sub: rq_criteria}) that 42.5\% (43) of the CI studies do not discuss or present any of these specific criteria. On the other hand, 15.8\% (16) of studies apply one criterion. The results is an alarming proportion of 58.3\% of primary studies having none or only one criterion to identify whether CI has been implemented in a project. This is alarming because this suggests that most of the existing claims regarding CI might be biased towards projects that do not consistently implement CI. The most frequent criterion specified for 45.5\% of the studies is the usage of an online CI service (see Figure \ref{fig:criteria_qty} (b)), which allows implementing a CI pipeline for existing projects. However, this finding represents a challenge to be overcome by the research community since other studies revealed that CI usage may be inconsistent, sporadical, or discontinued \cite{vasilescu2015}.

Vasilescu et al. \cite{vasilescu2015} investigated CI quality and productivity outcomes. From a dataset of 918 \textsc{GitHub}  projects that used \textsc{TravisCI}, they found that only 246 projects have a good level of activity using \textsc{TravisCI}, while the other 672 projects have used \textsc{TravisCI} only for a few months. Thus, it suggests that solely relying on CI service configuration is not enough to determine a proper CI adoption.

Vassalo et al. \cite{vassalo2019} performed a survey with 124 professional developers confirming that deviations from CI best practices occur in practice and can be the cause of CI degradation. They mined 36 projects and verified relevant instances of four anti-patterns \cite{duvall2018}: late merging, slow build, broken release branch, and skip failed test.

Felidre et al. \cite{felidre2019} also investigated CI bad practices, beyond slow build and broken release branch, they shed light on infrequent commits and poor test coverage. In addition, their analysis of 1,270 open source projects confirmed the existence of a phenomenon known by practitioners as {\em CI Theater}, which refers to self-proclaimed CI projects that do not really implement CI~\cite{citheatre2017}.

Considering the findings observed in Section \ref{sub: rq_criteria}, and the studies mentioned above, we suspect that there are few studies considering a more robust number of criteria in order to perform a more rigorous evaluation of CI adoption. In line with Ståhl \& Bosch \cite{stahl2014}, we observe that simply stating that projects use continuous integration is not sufficient. There is an urgent need to classify which practices and at what level such projects implement them. It is especially true if we consider CI as a set of practices, where the benefits and challenges related to CI are directly related to the usage of such a set of practices.

\subsection{Research Opportunities \label{disc: clarify}}

Beyond the research opportunities already discussed in the themes of Section~\ref{rq_themes}, this subsection discusses existing gaps in the research on continuous integration and apparent contradictions among the findings, especially focusing on the themes ``integration patterns'' and ``development activities''.

\subsubsection{Integration Patterns \label{disc: clarify_integration}}

\begin{figure}[h!]
  \centering
  \includegraphics[scale=.54]{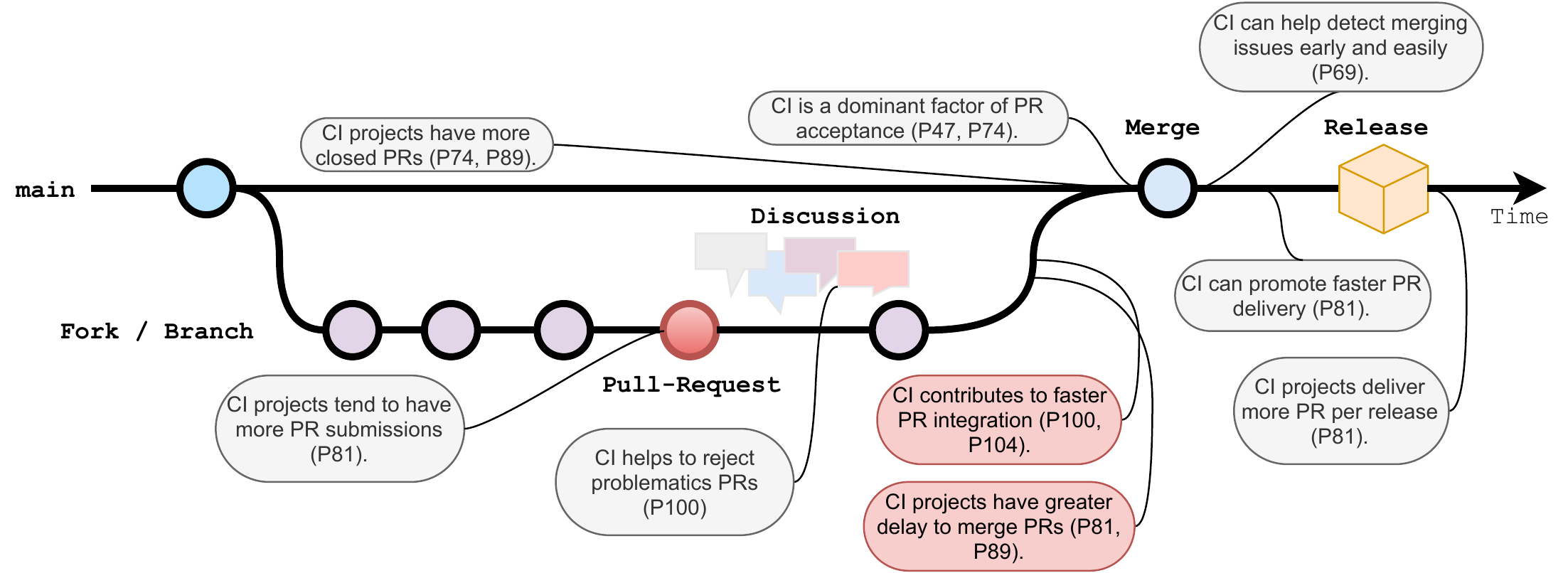}
  \caption{Claims related to the effects of CI on pull requests life cycle.}
  \label{fig:ci-prs}
\end{figure}

Regarding the integration patterns theme, we observe that CI may influence the processing of pull requests in different stages. Figure~\ref{fig:ci-prs} shows the claims related to how CI influences the processing of pull-requests. Figure~\ref{fig:ci-prs} also shows to which stage of the pull-request life-cycle a claim refers. P81 reveals evidence that projects tend to have more pull request (PR) submissions after they adopt CI. P47 and P74 state that CI influences PR acceptance, and CI projects tend to have more closed PRs. After merging, CI is a helpful tool to detect merging issues earlier (P69). Moreover, P81 found that CI projects deliver more PRs and more rapidly.

All of these claims represent the following codes: (i) ``CI is related to a positive impact on pull request life-cycle'' having seven studies supporting studies (P47, P69, P74, P81, P89, P100, P104); and (ii) ``CI is related to negative impact on pull requests life-cycle'' with two studies supporting (P81, P89). There is an apparent contradiction regarding the time to integrate a pull request. Four studies investigate the time to merge pull requests and its relation with continuous integration --- P104 from 2015, P100 from 2016, P81 from 2018, and P89 from 2017. 

In 2015, P104 investigated 103,284 pull requests from 40 different projects using multiple linear regression models to evaluate the time to merge pull requests. P104 observed that CI shortens the time to merge pull requests. Later, in 2016, another MSR study (P100), including 1,529,291 builds and 653,404 pull requests, P100 observed that CI build statuses can influence the development team to merge pull requests more quickly.

On the other hand, P89, which studied the time to merge pull requests using an Regression Discontinuity Design (RDD) model on 77 projects, found that, on average, pull requests have a trend to take a longer to be merged as the project matures, with CI having no apparent impact in this trend, i.e., CI projects keep increasing the time to merge PRs regardless of their adoption of CI. Finally, in 2018, P81 analyzed 87 projects and concluded that projects may take longer to merge pull requests after adopting CI. The difference is small but statistically significant.

This apparent contradiction between P81 and P89 against P100 and P104 might be related to several factors, including (as observed earlier) how these studies determine whether projects are using CI or not. While P81 and P104 consider the start of CI adoption as the moment when the first automated build is created in a CI Service, P100 does not identify a certain moment to identify when CI has been adopted. Instead, P100 grouped pull requests into two categories: with and without build information from the CI server. We conjecture that the main factors contributing to such apparent contradiction might be: (i) the age of projects influencing the longevity of pull requests according to the above-mentioned P89 findings (i.e., the older a project is, the longer the merge delay); (ii) The P81 finding regarding an increasing trend in PR submissions after CI adoption may explain why there is an extra time to process pull requests. Therefore, we argue that more studies are necessary to investigate the claims related to CI and pull-request lifetime considering these possible confounding factors, such as project age, the number of opened PRs, among others.

\subsubsection{Development Activities \label{discussion_development_act}}

From the ``Development Activities'' theme, we find opportunities to construct a deep understanding of some reported phenomena, such as what we highlight as ``confidence contradicting claims'' and ``productivity contradicting claims''. The ``confidence contradicting claims'' is marked by studies that make claims related to the code ``CI may generate a false sense of confidence'', while some studies raise claims under the code ``CI is associated with confidence improvement''.  The ``productivity contradicting claims'' refers to studies claiming that ``CI is related to an increase in productivity or efficiency'', while some studies claim that ``CI is associated with a decreased perception of productivity''.

\textbf{Confidence contradicting claims\label{confidence_paradox}}. Table~\ref{tab:confidence_claims} shows the claims related to developer confidence. Six studies provide support to conclude that CI improves developer confidence (P39, P97, P100, P106, P58, P93). On the other hand, two studies claim that CI can promote a false sense of confidence (P58, P106).

\begin{table}[H]
  \caption{Claims related to the effects of CI developer confidence.}
  \label{tab:confidence_claims} 
  \begin{tabular}{ll}
  \hline
  \multicolumn{2}{c}{\textbf{CODE: CI is associated with confidence improvement}}                                           \\ \hline
  \multicolumn{1}{c}{\textbf{Claim}}                                           & \multicolumn{1}{c}{\textbf{Studies}} \\ \hline
  CI increases the confidence about the quality.                               & P39                                  \\
  CI makes the team less worried about breaking build.                         & P97, P100                            \\
  \begin{tabular}[c]{@{}l@{}}CI improves the developers confidence to perform the \\required code changes.\end{tabular} & P106, P58, P93                       \\ \hline
  \multicolumn{2}{c}{\textbf{CODE: CI may generate a false sense of confidence}}                                            \\ \hline
  \multicolumn{1}{c}{\textbf{Claim}}                                           & \multicolumn{1}{c}{\textbf{Studies}} \\ \hline
  The false sense of confidence is a recurring problem in CI.                  & P58, P106                            \\
  Flaky tests may challenge CI projects.                                       & P106                                 \\ \hline
  \end{tabular}
\end{table}

P93 theorizes that CI allows programmers to assume themselves as single-programmers in a project, supporting an improvement in confidence. For instance, by relying on the lower number of new inconsistencies expected in each integration cycle, the developer can behave as if they were the only person modifying the code, reducing the cognitive tractability of programming. In the same line, P58 and P106 surveyed 158 CI users and reported the perception of respondents that CI provides more confidence to perform the required code changes.

Other studies may help to understand this boost in confidence better. Developers seem to delegate quality assurance to CI service and rely on its feedback. P39, an experience report, sheds light on improved confidence in product quality after CI adoption due to test automatization. P100 reports a survey with 407 respondents and reveals that the most common reason to use CI is the expectation that it makes developers less worried about breaking builds. After a triangulation between an interview and two surveys, P97 reports the same finding.

On the other hand, P58 and P106 also shed light on a reported problem of a false sense of confidence. This situation occurs when developers rely on an environment that may suffer from low quality or insufficient tests. The lack of balance between developer trust and the environment's trustworthiness determines the occurrence or not of overconfidence. The environment may provide a baseless trust and suggest an opportunity for practitioners and researchers to investigate and supply developers with objective criteria and guidance to define a reliable CI environment to avoid the mentioned false sense of confidence. For example, what minimum set of practices or metric values should we achieve before having a reliable CI environment and feedback that can be trusted? 

Additionally, these studies raise substantially different aspects of confidence: (i) confidence in the product quality (P39); (ii) personal confidence to perform tasks (P106, P58, P93); and (iii) confidence in the process reliability (P97, P100, P58, P106). Nonetheless, none of the mentioned studies addressed the confidence question directly, and therefore did not provide a theoretical base to analyze confidence. The studies, in general, registered developers' perceptions, leaving room to further investigation and theory formulation about developer's confidence and the role of CI.

\textbf{Productivity contradicting claims\label{productivity_paradox}}. Table~\ref{tab:productivity_claims} shows the claims related to development productivity. There are 12 claims in 11 studies supporting the code ``CI is related to productivity and efficiency increase'', and one study claiming that ``CI is associated with a decreased perception of productivity''.

\begin{table}[H]
  \caption{Claims related to the effects of CI on development productivity.}
  \label{tab:productivity_claims} 
  \begin{tabular}{ll}
  \hline
  \multicolumn{2}{c}{\textbf{CODE: CI is related to productivity and efficiency increasing}}                                                                                      \\ \hline
  \multicolumn{1}{c}{\textbf{Claim}}                                                                                                 & \multicolumn{1}{c}{\textbf{Studies}} \\ \hline
  \begin{tabular}[c]{@{}l@{}}CI increases development efficiency (due to the \\automation of tasks and fast feed back)\end{tabular} & P39, P52, P73, P59, P31, P97         \\
  \begin{tabular}[c]{@{}l@{}}CI is associated with external contributors having \\ fewer pull requests rejected.\end{tabular}        & P74                                  \\
  CI decreases the debug time.                                                                                                       & P100, P79                            \\
  CI allows quickly grow of source code.                                                                                             & P102                                 \\
  CI speed up development practice.                                                                                                  & P106                                 \\
  \begin{tabular}[c]{@{}l@{}}CI Reduced integration problems allowing \\the team to deliver software more rapidly\end{tabular}       & P79                                  \\ \hline
  \multicolumn{2}{c}{\textbf{CODE: CI is associated with a decreased perception of productivity}}                                                                                 \\ \hline
  \multicolumn{1}{c}{\textbf{Claim}}                                                                                                 & \multicolumn{1}{c}{\textbf{Studies}} \\ \hline
  \begin{tabular}[c]{@{}l@{}}The CI adoption leads to a worsening in the \\perceived team productivity.\end{tabular}                  & P91                                  \\ \hline
  \end{tabular}
\end{table}

P52 performs a case study with four projects and validates the hypothesis that CI contributes to an increase in the developer productivity due to parallel development and reducing tasks before checking in (i.e., committing). Through another case study, P59 confirms this claim, while P39 and P31 share different experience reports that record an increase in development efficiency and throughput per developer, respectively.

P73 reports interviews, and P97 presents a triangulation between an interview and two surveys. They both confirm the perceptions that CI increases productivity. By mining software repositories from 246 projects, P74 finds that external contributors tend to have fewer pull requests rejected if CI is adopted. Other studies such as P100, P102, and P106 also bring results corroborating this code.

In opposition to these studies and findings, P91 investigates the links between agile practices, interpersonal conflict, and perceived productivity. P91 presents a survey with 68 software developers. P91 grounds its research method in the Integrated Model of Group Development (IMGD) --- a theory on group development \cite{wheelan1996} and a tool (questionnaire) to employ psychological measurement of the stage where a group is in a developmental perspective. Moreover, P91 applies two other surveys to measure agile practices and the perceived productivity. It concludes:
\begin{quotation}
  \textit{``I have also shown that with higher scores on Continuous Integration and Testing came lower scores on this perceived productivity measurement. That means that the more continuous integration and testing the team conducts, the worse is the perceived team productivity. However, I do not have any external measurement of the productivity of the teams and can not draw conclusions on the actual productivity [...]''}. (p. 4)  
\end{quotation}

While P91 get productivity as developer perceives (an internal measurement), other studies quantify productivity by the time spent, e.g., adding features vs. debugging (P73, P79, P100), time saved (P52, P59, P106), integrator productivity to merge pull requests (P74), others by developer throughput (P31, P102). Thus, we can suppose that the measurement strategy of productivity could explain the difference in the findings from P91. However, P97 also registers developers' perceptions and finds a positive influence of CI.

That way, considering that 11 studies are going in a direction claiming that ``CI is related to productivity and efficiency increasing'' and only one study declaring a worsening in the perceived productivity, we are led to consider the participants of these studies. While P91 surveys 68 software developers from three big companies (a telecommunications equipment and services company, an aerospace and defense company, and an automotive parts manufacturing company), P97 surveyed 574 developers (51 from one software engineering solutions company and 523 from a broad group on the internet). 

Such a difference in findings may be due to the smaller number of participants in P91, or the difference in the domains of their companies as well as their CI practices~\cite{stahl2014,zhao2017,viggiato2019}. Nevertheless, as mentioned in P91, it is essential to point out that the perceived productivity may be affected by subjective factors, such as the kind of work performed by the developer. For example, code review may be necessary from an organizational perspective but can be seen as not as productive by a particular developer, decreasing their perceived productivity.

\section{Threats to validity \label{threats}}

The goal of our SLR is to provide a summary of the effects of CI on the software development phenomena. We follow the guidelines provided by Kichenham \& Charters \cite{kitchenham2007} to develop our review protocol while defining strategies to mitigate possible bias. However, as it happens to every study, our SLR is not without flaws and, in this section, we discuss the limitations of our study. 

\subsection{Search Strategy}

The search strategy may have bias or limitations on its search string and expression power, the limitations on search engines, and publication bias, i.e., positive results are more 
likely to be published than negative \cite{kitchenham2007}. To mitigate the search string threats, we apply several identified synonyms to reach the effects of continuous integration, 
the intervention studied. In addition, aiming to reduce the limitations of search engines, we use six different search engines including five formal databases and one index engine, following the recommendations from Chen et al. \cite{chen2010}, thereby including 
journals and conferences publications---which contributes to publication bias mitigation.

\subsection{Screening Papers}

The screening and selection phase (see Section \ref{screening}) follows the inclusion and exclusion criteria defined during the protocol definition, as recommended by Kitchenham \& Charters \cite{kitchenham2007} 
to mitigate the selection bias. In addition, the decision relied on the evaluation of two researchers and the agreement was measured using the Cohen Kappa statistic \cite{cohen}. A substantial 
agreement was achieved both in the first screening (0.72) and in the snowballing phase (0.76). The disagreements were read and arbitrated by a third researcher.

\subsection{Data Extraction \label{threats_extraction}}

To reduce the possibility of bias in the data extraction phase, we proceed the following steps (see Section \ref{extraction}) to mitigate it. First, the meta-data was retrieved in 
an automated process using the data obtained in Mendeley - the reference management tool adopted. To avoid mistakes or missing information, the processed meta-data was manually inspected by one researcher. Second, the 
definition of extraction form (see Table \ref{tab:extraction}) was available in the review protocol and in a web host to all three readers. Third, to decrease the chances of inattention, lack of understanding, or any other reason for mistaken data collection, the reading of each paper was performed by two 
researchers that filled the extraction form independently. Fourth, to treat the disagreements in the extraction and also avoid bias, each pair discussed the extracted data to achieve a consensus.

\subsection{Quality Assessment \label{threats_quality}}

The quality assessment stage was performed based on a quality checklist composed of eleven questions. The threats in this phase have the potential to reflect on data extraction and data synthesis in such a way that: (i) the researchers may do not comprehend well the questions; or (ii) the questions may do not express sufficiently the quality of the papers. To mitigate this, we: (i) developed a questionnaire inspired by the previous experiences reported by Dybå and Dingsøyr \cite{dyba2008} and Kitchenham and Charters \cite{kitchenham2007}; (ii) the checklist covers four distinct quality aspects (quality of reporting, rigour, credibility, and relevance);  (iii) we ran two rounds with pilot papers with three researchers together to assess the understanding of the quality checklist; then we (iv) provide the most of questions with instructions, i.e., minor questions to support their assessment and answer.

\subsection{Data Synthesis}

As described in Section~\ref{synthesis}, our study explores quantitative (RQ1 and RQ3) and qualitative synthesis (RQ2). In the quantitative synthesis of RQ1 and RQ3, we present a summarization to create a landscape of the studies and point out some directions to researchers. The main threat in these syntheses is related to the chosen criteria. 

First, in RQ1, while investigating how primary studies identify or classify their subjects as a CI project, we found no clear definition of which practices determine whether a given project uses CI or not. Studies revealed that there are many variants of CI implementation~\cite{stahl2014,zhao2017,viggiato2019}. Therefore, our chosen criteria to identify whether a project uses CI or not may not perfectly match CI usage for every context. Nevertheless, we decided to use the prescriptive list of practices from Duvall et al.~\cite{duvall2013}, and Fowler~\cite{fowler2006} because they have been the most used definition of CI in existing research so far and present a prescriptive list of practices. In this way, we adopted a set of seven generic practices inspired in their lists. Second, in the RQ3, we analyze the studies quality and methodologies relying on our data extraction and quality assessment, then subject to risks presented in Sections \ref{threats_extraction} and 
\ref{threats_quality}.

In the qualitative analysis of RQ2, we follow the guidelines of Cruzes \& Dyba \cite{cruzes2011} to perform a thematic synthesis. In the coding phase, to mitigate the threats of confirmation bias, we first use an inductive approach performed by two researchers to define the set of codes. Second, to avoid a wrong grouping, two researchers coded all the extracted segments independently (this step also achieved a substantial Kappa agreement rate - 0.73), and a third researcher resolves the disagreements. Finally, all the authors discuss and agree with the translation of the codes into the presented themes.

In Sections \ref{rq_themes} and \ref{disc: clarify}, we assess the trustworthiness of the synthesis in terms of type of studies, number of occurrences, and the relationship between the findings of different primary studies. 

\section{Conclusion \label{conclusion}}

We perform a systematic literature review (SLR) on the effects of continuous integration on the software development phenomena. Our main goal is to summarise the existing empirical evidence and body of knowledge regarding CI to support a better decision process, avoiding overestimating or underestimating the results and costs of CI adoption. We collect and analyze empirical evidence from 101 primary studies ranging from 2003 to 2019, conducting quantitative and qualitative analyses. We hope our study can support an evidence-based practice by development teams and organizations to build work policies. Our study can also serve as a map regarding which claims related to CI should be more thoroughly studied in the future (i.e., given the rigour of the state-of-art studies).


\subsection{Results and Implications \label{conclusion_implications}}

The collation of findings related to the effects of CI and their accompanying evidence (see Sections \ref{sub: rq_criteria}, \ref{rq_themes}, \ref{rq_method}, and \ref{discussion}) can be useful for researchers and practitioners. In Sections \ref{sub: rq_criteria} we show that 42.5\% of the primary studies did not present explicit criteria to identify projects that use CI. We also found that 15.8\%  used only one criterion (e.g., more than half of studies used  ``automated builds'' as a criterion). This finding reveals a weakness in our current empirical literature since identifying whether a project uses CI or not is at the core of how we analyze the effects (positive or negative) of using CI. As an implication, we believe that there are plenty of research opportunities to re-evaluate existing analyses by using more robust criteria to identify CI projects. For example, checking whether they use automated builds and also how frequently they perform commits.

Regarding the criteria applied to check whether participants use CI or not (Sections \ref{sub: rq_criteria} and \ref{discussion_ci_criteria}), our findings reveal the need for performing other checks during interviews or surveys related to the adherence of CI beyond the self-declaration, e.g., {\em ``on a scale of 1 to 7, how would you classify that your project adheres to CI?''}. Studies may consider, for example, checking which practices the subjects use in their CI environment before classifying them as CI projects. 

Sections \ref{sec:developer_work} and \ref{discussion_development_act}
discuss our findings related to the effects of CI on {\em development activities}. We find evidence for the association between CI and improved productivity, efficiency, and developer confidence. On the other hand, other findings suggest that CI may introduce complexity to the project, requiring more effort and discipline from developers, negatively impacting developers’ perceived productivity. Some studies also discuss the false sense of confidence, i.e., when developers blindly rely on flaky tests.

Continuous integration benefits the {\em software process} (see Section \ref{sec:software_process}) by promoting faster iterations, more stability, predictability, and transparency in the development process. Although CI may incur technical challenges to the team (e.g., creating a reliable automated build environment), CI is mentioned as a success factor in software projects. 
Regarding {\em quality assurance}, Section \ref{sec:quality_assurance} reveals evidence on the association between CI and better testing. The studies demonstrate a perceived provision of transparency and continuous quality inspections when CI is adopted. 

With respect to {\em integration patterns} (Section \ref{sec:integration_patterns}) our study indicates that CI  positively influences the way developers perform commits (e.g., increasing the frequency and decreasing the size of commits). CI can also benefit the pull-based development by improving and accelerating the integration process. However, there are also studies reporting that CI may prolong the pull request lifetime. Section \ref{disc: clarify_integration} discusses in detail the way CI impacts differently in each stage of the pull-request life-cycle.

Regarding {\em issues \& defects} (Section \ref{sec:issues_defects}), we find that studies credit CI to an improvement in the time to find and fix issues. They also report a decrease in defects reported. About {\em build patterns} the studies reveal that CI impacts the build process (Section \ref{sec:build_patterns}), promoting good practices related to build health and contributing to an increase in successful builds.

Lastly, regarding RQ3, Section ~\ref{sub: projects} shows that there is a wide variety in the primary studies (in terms of domains and subjects). The number of studies making their datasets available is growing over the past few years (Section \ref{sub: reproducibility}). The studies from which we extract claims (38  out of 101) have a notable overall quality (median score of 9 out of 11 --- \(\displaystyle \tilde{Qscore} = 9\)), mainly those that use MSR and mixed-methods as their methodologies, both with \(\displaystyle \tilde{Qscore} = 10\), while survey researches obtain a \(\displaystyle \tilde{Qscore} = 9\). Most of the claims (70.4\%) emerge from these three study types.

\subsection{Open questions for Practitioners and Researchers \label{conclusion_open}}

Given our observed results, we believe that continuous integration has plenty of room for future empirical studies and new tools to address open questions or strengthen the current body of knowledge. Section ~\ref{sub: rq_criteria}, for example, shows that a community effort to build a solid foundation about how to classify projects using CI could be useful to further empirical studies, e.g., a ``CI maturity score'' could be conceived, or a consensual set of minimal practices could be established by researchers and practitioners. 

Sections \ref{sec:developer_work} and \ref{discussion_development_act} highlight that researchers should stay attentive to how factors such as productivity and confidence are measured since there is significant diversity among primary studies. For example, some studies assess the developer perception of productivity, while others consider the time to merge a pull request. Other examples are studies assessing developers’ confidence, in which some investigate the developers’ confidence in performing certain tasks, while others analyze confidence in terms of trusting CI. The {\em development activities} theme reveals that there is a need for guidelines and metrics to inform practitioners about the reliability of their CI environment, avoiding the false sense of confidence phenomenon. This phenomenon has a link with the quality of the tests and their consequent reliability. In Section \ref{sec:quality_assurance}, regarding {\em quality assurance}, we find evidence for the association between CI and an increase in volume and coverage of tests. However, more studies are necessary to understand the relationship between test effort and test quality in CI.

Section \ref{sec:software_process} discusses open challenges to practitioners, researchers, and tool builders. Multiple studies report the difficulty faced by developers with the technical activities, such as configuring the build environment. Practitioners and tool builders may consider such challenges and elaborate strategies and tools to mitigate them. We propose that further studies are necessary to better understand the trade-offs between adopting CI and overcoming its inherent challenges (e.g., trade-offs between automation, technical challenges, and perceived productivity as discussed in Sections \ref{sec:developer_work}, \ref{sec:software_process}, and \ref{discussion_development_act}).

The results of RQ3 show that 29.7\% of the included studies investigate domain-specific projects (section~\ref{sub: projects}), which highlights the need for studying whether CI is better adopted in certain domains (e.g., web application, embedded systems, finance, among others) \cite{stahl2014,viggiato2019}. Section~\ref{sub: results_methods} reveals that researchers should be aware of the low amount of studies applying comparison or control groups to assess their findings and suggests that more diverse and complementary studies may be necessary for {\em quality assurance}. For example, MSR studies assessing the evolution of the test code in project repositories.

\begin{acknowledgements}
  This work is partially supported by INES (\url{www.ines.org.br}), CNPq grants 465614/2014-0 and 425211/2018-5, CAPES grant 88887.136410/2017- 00, and FACEPE grants APQ-0399-1.03/17 and PRONEX APQ/0388-1.03/14.
\end{acknowledgements}

\textbf{Data Availability} 
All data is available via an online appendix: \url{https://doi.org/10.5281/zenodo.4545623}.

%
\section*{Declarations}
\textbf{Conflict of interests} 
The authors declare that they have no conflict of interest.


\begin{appendices}

\section{Demographic attributes \label{appendix_demographic}}
This appendix shows the demographic data of our primary studies. We discuss the evolution of studies over the years and describe the authors' information next.

\begin{figure}[H]
  \centering
  \includegraphics[scale=.4]{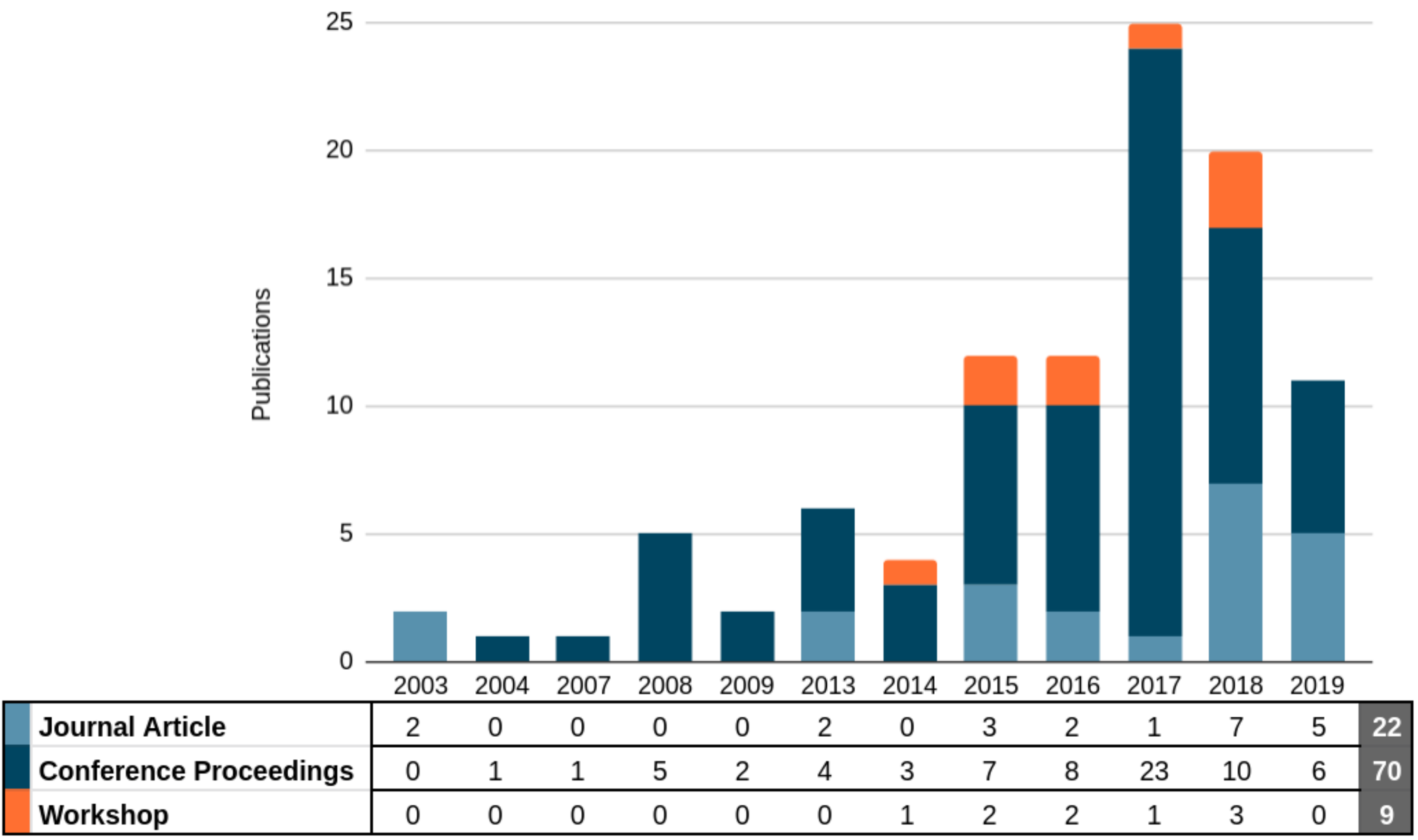}
  \caption{Publications by year and type of venue.}
  \label{fig:pub_year_type}
\end{figure}

\textbf{\textit{Evolution of studies.}} CI emerged in the context of eXtreme Programming, a software development methodology that increased and became popular in the late 90s and early 00s \cite{fowler2013}.  Indeed, we identify the first research efforts on CI in 2003. Figure~\ref{fig:pub_year_type} shows an increasing number of publications over the years, especially in the last five years. The majority of papers have been published in conference proceedings (69 papers, i.e., 67.2\%), followed by 23 papers published in journals (i.e. 22.5\%). 10 other studies have been published in workshops in the last years (i.e. 9.8\%).

\begin{figure}[h!]
  \centering
  \includegraphics[scale=.53]{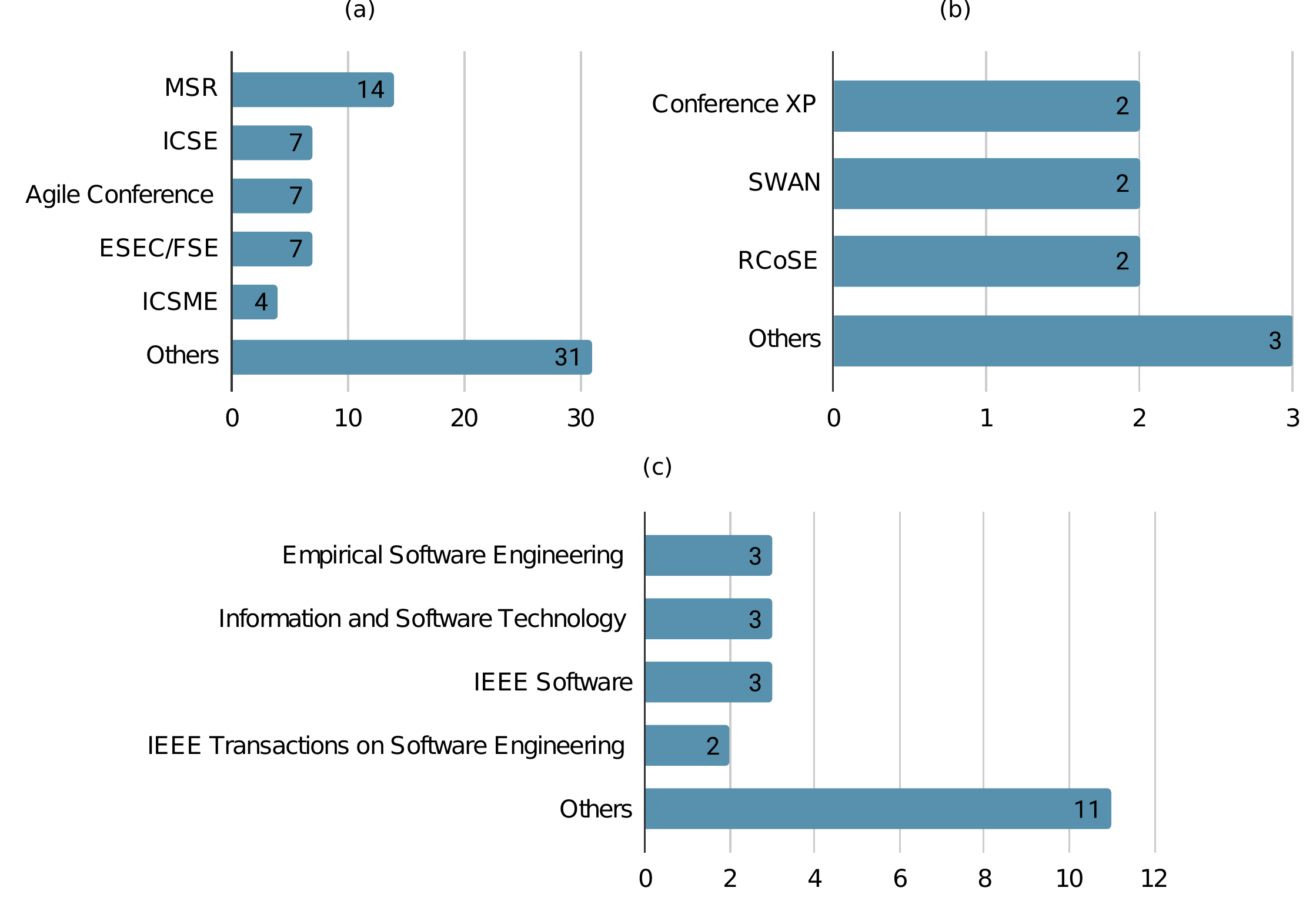}
  \caption{Publications in main venues on (a) conferences, (b) Workshops, and (c) Journals.}
  \label{fig:pub_venues}
\end{figure}

Our primary studies have been published in 29 distinct conferences, 15 journals, and 7 workshops. Figure~\ref{fig:pub_venues} (a) shows that MSR (IEEE International Working Conference on Mining Software Repositories), ICSE (International Conference on Software Engineering), Agile Conference, and ESEC/FSE (European Software Engineering Conference and ACM SIGSOFT Symposium on the Foundations of Software Engineering) are the conferences with the highest number of primary studies. Since we aim to collate the most claims possible related to CI, we do not necessarily focus on the goals of a venue (e.g., magazine-based publication). In a later stage, we analyse the rigour of the studies from which we find claims.

As for workshops, Figure~\ref{fig:pub_venues} (b) shows Conference XP (Scientific Workshops Proceedings), SWAN (International Workshop on Software Analytics), and RCoSE (International Workshop on Rapid Continuous Software Engineering) as the most frequent venues. Figure~\ref{fig:pub_venues} (c) shows that the journals with the highest frequency are Empirical Software Engineering, Information and Software Technology, and IEEE Software. 

\textbf{\textit{Paper Authors.}} The primary studies have 259 different authors involved altogether. Table \ref{tab:author_ranking} shows a ranking with those having the highest number of publications included as a primary study in our SLR. Jan Bosch is the most frequent author and all of the top 6 researchers remain active over the last years. Having described the demographic data of our primary studies, we now describe our obtained results.

\begin{table}[H]
\caption{Ranking of authors per publication number and his publications.}
\label{tab:author_ranking} 
\begin{tabular}{llcl}
\hline
\multicolumn{2}{c}{\textbf{Researcher}} & \textbf{\#} & \multicolumn{1}{c}{\textbf{Publications}}                                                                            \\ \hline
1        & Jan Bosch                     & 8           & \begin{tabular}[c]{@{}l@{}}2013 (P52), 2014 (P19), 2016 (P60, P38), \\ 2017 (P86, P43, P40), 2019 (P85)\end{tabular} \\ \hline
2        & Daniel Ståhl                 & 7           & \begin{tabular}[c]{@{}l@{}}2013 (P52), 2014 (P19), 2016 (P38, P60), \\ 2017 (P86, P40), 2019 (P85)\end{tabular}      \\ \hline
3        & Bogdan Vasilescu             & 6           & \begin{tabular}[c]{@{}l@{}}2014 (P41), 2015 (P104, P74), 2017 (P89), \\ 2018 (P55), 2019 (P4)\end{tabular}           \\ \hline
4        & Massimiliano Di Penta        & 5           & 2017 (P54, P8, P29), 2019 (P18, P7)                                                                                  \\ \hline
         & Carmine Vassallo             & 5           & 2017 (P29, P8), 2018 (P44, P28), 2019 (P18)                                                                          \\ \hline
         & Torvald Mårtensson           & 5           & 2016 (P38), 2017 (P43, P40, P86), 2019 (P85)                                                                         \\ \hline
\end{tabular}
\end{table}


\section{Selected Studies}
\label{appendix_studies}

\begin{longtable}{| p{.05\textwidth} | p{.30\textwidth} | p{.30\textwidth} | p{.1\textwidth} | p{.25\textwidth} |} 
  \hline
  \textbf{ID} & \textbf{Title} & \textbf{Author(s)} & \textbf{Year} & \textbf{Venue}  \\
  \hline
  P2 & (No) influence of continuous integration on the commit activity in GitHub projects & Stephan Diehl, Daniel Anastasiou, Jascha Knack, Sebastian Baltes, Ralf Tymann & 2018 & SWAN - International Workshop on Software Analytics \\ \hline
  P3 & A brief study on build failures in continuous integration: Causation and effect & Bharavi Mishra, Saket Kumar Singh, Romit Jain & 2018 & ICACIE - Progress in Advanced Computing and Intelligent Engineering \\ \hline
  P4 & A conceptual replication of continuous integration pain points in the context of Travis CI & Bogdan Vasilescu, David Gray Widder, Michael Hilton, Christian Kästner & 2019 & ESEC/FSE Joint Meeting European Software Engineering Conference and Symposium on the Foundations of Software Engineering \\ \hline
  P5 & A Current Study on the Limitations of Agile Methods in Industry Using Secure Google Forms & Ashish Agrawal, L. S. Maurya, Mohd Aurangzeb Atiq & 2016 & International Conference on Information Security and Privacy \\ \hline
  P6 & A Hundred Days of Continuous Integration & Ade MillerAde Miller & 2008 & Agile Conference \\ \hline
  P7 & A Study on the Interplay between Pull Request Review and Continuous Integration Builds & Massimiliano Di Penta, Canfora Gerardo, Gabriele Bavota, Fiorella Zampetti & 2019 & SANER \\ \hline
  P8 & A Tale of CI Build Failures: An Open Source and a Financial Organization Perspective & Carmine Vassallo, Gerald Schermann, Fiorella Zampetti, Daniele Romano, Philipp Leitner, Andy Zaidman, Massimiliano Di Penta, Sebastiano Panichella & 2017 & ICSME - International Conference on Software Maintenance and Evolution \\ \hline
  P9 & Agile systems development and stakeholder satisfaction: a South African empirical study & Jason Cohen, Carlos Ferreira & 2008 & SAICSIT \\ \hline
  P10 & An empirical analysis of build failures in the continuous integration workflows of Java-based open-source software & Stefan Schulte, Thomas Rausch, Waldemar Hummer, Philipp Leitner & 2017 & MSR - International Conference on Mining Software Repositories \\ \hline
  P11 & An empirical study of activity, popularity, size, testing, and stability in continuous integration & Saket Vishwasrao, Francisco Servant, Aakash Gautam & 2017 & MSR - International Conference on Mining Software Repositories \\ \hline
  P12 & An empirical study of the long duration of continuous integration builds & Ying Zou, Daniel Alencar da Costa, Taher Ahmed Ghaleb & 2019 & Empirical Software Engineering \\ \hline
  P13 & An empirical study of the personnel overhead of continuous integration & Shane McIntosh, Eduardo Coronado-Montoya, Marco Manglaviti, Keheliya Gallaba & 2017 & MSR - International Conference on Mining Software Repositories \\ \hline
  P14 & Analyzing the effects of test driven development in GitHub & Abram Hindle, Neil Borle, Meysam Feghhi, Eleni Stroulia, Russ Greiner & 2018 & Empirical Software Engineering \\ \hline
  P15 & Analyzing the impact of social attributes on commit integration success & Mauricio Soto, Zack Coker, Claire Le Goues & 2017 & MSR - International Conference on Mining Software Repositories \\ \hline
  P16 & Angry-builds: an empirical study of affect metrics and builds success on github ecosystem & Michele Marchesi, David Bowes, Giuseppe Destefanis, Marco Ortu, Andrea Pinna, Roberto Tonelli & 2018 & Conference XP \\ \hline
  P17 & Applying Continuous Integration for Reducing Web Applications Development Risks & Fang Yie Leu, Sen Tarng Lai & 2015 & BWCCA - International Conference on Broadband and Wireless Computing, Communication and Applications \\ \hline
  P18 & Automated reporting of anti-patterns and decay in continuous integration & Sebastian Proksch, Harald C. Gall, Massimiliano Di Penta, Carmine Vassallo & 2019 & ICSE - International Conference on Software Engineering \\ \hline
  P19 & Automated software integration flows in industry: a multiple-case study & Daniel Ståhl, Jan Bosch & 2014 & ICSE - International Conference on Software Engineering \\ \hline
  P20 & Build waiting time in continuous integration: an initial interdisciplinary literature review & Mika Mantyla, Eero Laukkanen & 2015 & RCoSE - International Workshop on Rapid Continuous Software Engineering \\ \hline
  P21 & Building lean continuous integration and delivery pipelines by applying DevOps principles: a case study at Varidesk & Vidroha Debroy, Senecca Miller, Lance Brimble & 2018 & ESEC/FSE Joint Meeting European Software Engineering Conference and Symposium on the Foundations of Software Engineering \\ \hline
  P22 & Building lean thinking in a telecom software development organization: strengths and challenges & Pasi Kuvaja, Pilar Rodríguez, Kirsi Mikkonen, Markku Oivo, Juan Garbajosa & 2013 & ICSSP - International Conference on Software and System Process \\ \hline
  P23 & Built to last or built too fast? evaluating prediction models for build times & Ekaba Bisong, Eric Tran, Olga Baysal & 2017 & MSR - International Conference on Mining Software Repositories \\ \hline
  P24 & Challenges When Adopting Continuous Integration: A Case Study & Mikael Dienér, Richard Berntsson Svensson, Adam Debbiche & 2014 & International Conference on Product-Focused Software Process Improvement \\ \hline
  P25 & Characterizing the influence of continuous integration: empirical results from 250+ open source and proprietary projects & Akond Rahman, Amritanshu Agrawal, Rahul Krishna, Alexander Sobran & 2018 & SWAN - International Workshop on Software Analytics \\ \hline
  P27 & Comparison of release engineering practices in a large mature company and a startup & Eero Laukkanen, Casper Lassenius, Juha Itkonen, Maria Paasivaara & 2018 & Empirical Software Engineering \\ \hline
  P28 & Continuous code quality: are we (really) doing that? & Alberto Bacchelli, Harald C. Gall, Fabio Palomba, Carmine Vassallo & 2018 & ASE - International Conference on Automated Software Engineering \\ \hline
  P29 & Continuous Delivery Practices in a Large Financial Organization & Andy Zaidman, Carmine Vassallo, Fiorella Zampetti, Daniele Romano, Moritz Beller, Annibale Panichella, Massimiliano Di Penta & 2017 & ICSME - International Conference on Software Maintenance and Evolution \\ \hline
  P30 & Continuous Delivery: Huge Benefits, but Challenges Too & Lianping Chen & 2015 & IEEE Software \\ \hline
  P31 & Continuous Delivery? Easy! Just Change Everything (Well, Maybe It Is Not That Easy) & Steve Neely, Steve Stolt & 2013 & Agile Conference \\ \hline
  P32 & Continuous deployment and schema evolution in SQL databases & Michael De Jong, Arie Van Deursen & 2015 & RELENG - International Workshop on Release Engineering \\ \hline
  P33 & Continuous deployment at Facebook and OANDA & Michael Gentili, Kent Beck, Laurie Williams, Michael Stumm, Tony Savor, Mitchell Douglas & 2016 & ICSE - International Conference on Software Engineering \\ \hline
  P34 & Continuous deployment of mobile software at facebook (showcase) & Elisa Shibley, Chuck Rossi, Kent Beck, Shi Su, Michael Stumm, Tony Savor & 2016 & ESEC/FSE Joint Meeting European Software Engineering Conference and Symposium on the Foundations of Software Engineering \\ \hline
  P36 & Continuous Integration and Delivery for HPC: Using Singularity and Jenkins & Zebula Sampedro, Aaron Holt, Thomas Hauser & 2018 & PEARC - Practice and Experience on Advanced Research Computing \\ \hline
  P37 & Continuous Integration and Quality Assurance: a case study of two open source projects & Jesper Holck, Niels Jørgensen & 2003 & Australasian Journal of Information Systems \\ \hline
  P38 & Continuous Integration Applied to Software-Intensive Embedded Systems – Problems and Experiences & Torvald Mårtensson, Daniel Ståhl, Jan Bosch & 2016 & International Conference on Product-Focused Software Process Improvement \\ \hline
  P39 & Continuous Integration for Web-Based Software Infrastructures: Lessons Learned on the webinos Project & John Lyle, Tao Su, Andrea Atzeni, Shamal Faily, Habib Virji, Christos Ntanos, Christos Botsikas & 2013 & Haifa Verification Conference \\ \hline
  P40 & Continuous Integration Impediments in Large-Scale Industry Projects & Torvald Mårtensson, Jan Bosch, Daniel Ståhl & 2017 & ICSA - IEEE International Conference on Software Architecture \\ \hline
  P41 & Continuous Integration in a Social-Coding World: Empirical Evidence from GitHub & Bogdan Vasilescu, Mark G J Van Den Brand, Jules Wulms, Stef Van Schuylenburg, Alexander Serebrenik & 2014 & ICSME - International Conference on Software Maintenance and Evolution \\ \hline
  P42 & Continuous Integration in Open Source Software Development & Amit Deshpande, Dirk Riehle & 2008 & IFIP International Federation for Information Processing \\ \hline
  P43 & Continuous Integration is Not About Build Systems & Torvald Mårtensson, Par Hammarstrom, Jan Bosch & 2017 & SEAA - Euromicro Conference on Software Engineering and Advanced Applications \\ \hline
  P44 & Continuous Refactoring in CI: A Preliminary Study on the Perceived Advantages and Barriers & Carmine Vassallo, Fabio Palomba, Harald C. Gall & 2018 & ICSME - International Conference on Software Maintenance and Evolution \\ \hline
  P45 & Continuous software engineering and beyond: trends and challenges & Brian Fitzgerald, Klaas Jan Stol & 2014 & RCoSE - International Workshop on Rapid Continuous Software Engineering \\ \hline
  P46 & Contrasting Big Bang with Continuous Integration Through Defect Reports & Daniel Levin, Ana Magazinius, Niklas Mellegard, Hakan Burden, Kenneth Lind & 2018 & IEEE Software \\ \hline
  P47 & Determinants of pull-based development in the context of continuous integration & Cheng Yang, Huaimin Wang, Tao Wang, Gang Yin, Yue Yu & 2016 & Science China Information Sciences \\ \hline
  P48 & DevOps: A Definition and Perceived Adoption Impediments & Kristian Nybom, Jens Smeds, Ivan Porres & 2015 & International Conference on Agile Software Development \\ \hline
  P49 & Effect of Continuous Integration on Build Health in Undergraduate Team Projects & Suzanne M. Embury, Christopher Page & 2017 & Conference on Software Engineering Education and Training \\ \hline
  P50 & Effectiveness of Test-Driven Development and Continuous Integration: A Case Study & Yoni Meijberg, Chintan Amrit & 2018 &  IT Professional \\ \hline
  P51 & Enabling Agile Testing through Continuous Integration & Sean Stolberg & 2009 & Agile Conference \\ \hline
  P52 & Experienced benefits of continuous integration in industry software product development: A case study & Jan Bosch, Daniel Ståhl & 2013 & IASTED International Conference on Software Engineering \\ \hline
  P53 & How does contributors' involvement influence the build status of an open-source software project? & Renato O. Santos, Fernando Castor, Gustavo Pinto, Marcel Reboucas & 2017 & MSR - International Conference on Mining Software Repositories \\ \hline
  P54 & How open source projects use static code analysis tools in continuous integration pipelines & Fiorella Zampetti, Rocco Oliveto, Gerardo Canfora, Massimiliano Di Penta, Simone Scalabrino & 2017 & MSR - International Conference on Mining Software Repositories \\ \hline
  P55 & I'm leaving you, Travis: a continuous integration breakup story & Bogdan Vasilescu, Christian Kästner, Michael Hilton, David Gray Widder & 2018 & ICSE - International Conference on Software Engineering \\ \hline
  P56 & Impact of continuous integration on code reviews & Mohammad Masudur Rahman, Chanchal K. Roy & 2017 & MSR - International Conference on Mining Software Repositories \\ \hline
  P57 & Implementation of a DevOps Pipeline for Serverless Applications & Vitalii Ivanov, Kari Smolander & 2018 & International Conference on Product-Focused Software Process Improvement \\ \hline
  P58 & Inadequate testing, time pressure, and (over) confidence: a tale of continuous integration users & Marcel Reboucas, Gustavo Pinto, Fernando Castor & 2017 & CHASE - International Workshop on Cooperative and Human Aspects of Software Engineering \\ \hline
  P59 & Increasing quality and managing complexity in neuroinformatics software development with continuous integration & Yury V. Zaytsev, Abigail Morrison & 2013 & Frontiers in Neuroinformatics \\ \hline
  P60 & Industry application of continuous integration modeling: a multiple-case study & Daniel Ståhl, Jan Bosch & 2016 & ICSE - International Conference on Software Engineering \\ \hline
  P62 & Insights into continuous integration build failures & Md Rakibul Islam, Minhaz F. Zibran & 2017 & MSR - International Conference on Mining Software Repositories \\ \hline
  P63 & ISM based identification of quality attributes for agile development & Parita Jain, Laxmi Ahuja, Arun Sharma & 2016 & International Conference on Reliability \\ \hline
  P64 & It's Not the Pants, it's the People in the Pants Learnings from the Gap Agile Transformation -- What Worked, How We Did it, and What Still Puzzles Us & David Goodman, Michael Elbaz & 2008 & Agile Conference \\ \hline
  P65 & Lessons Learned: Using a Static Analysis Tool within a Continuous Integration System &  & 2016 & ISSREW - International Symposium on Software Reliability Engineering Workshops \\ \hline
  P66 & Managing to release early, often and on time in the OpenStack software ecosystem & José Apolinário Teixeira, Helena Karsten & 2019 & Journal of Internet Services and Applications \\ \hline
  P67 & Measurement and Impact Factors of Speed of Reviews and Integration in Continuous Software Engineering & Wilhelm Meding, Ola Söder, Magnus Bäck, Miroslaw Staron & 2018 & Foundations of Computing and Decision Sciences \\ \hline
  P69 & Moving from Closed to Open Source: Observations from Six Transitioned Projects to GitHub & Pavneet Singh Kochhar, Nachiappan Nagappan, Eirini Kalliamvakou, Christian Bird, Thomas Zimmermann & 2019 & IEEE Transactions on Software Engineering \\ \hline
  P70 & On the interplay between non-functional requirements and builds on continuous integration & Marcelo De A. Maia, Cricia Z. Felicio, Klerisson V.R. Paixao, Fernanda M. Delfim & 2017 & MSR - International Conference on Mining Software Repositories \\ \hline
  P71 & On the journey to continuous deployment: Technical and social challenges along the way & Gerry Gerard Claps, Richard Berntsson Svensson, Aybüke Aurum & 2015 & Information and Software Technology \\ \hline
  P72 & Oops, my tests broke the build: an explorative analysis of Travis CI with GitHub & Moritz Beller, Andy Zaidman, Georgios Gousios & 2017 & MSR - International Conference on Mining Software Repositories \\ \hline
  P73 & Practitioners' eye on continuous software engineering: An interview study & Jan Ole Johanssen, Anja Kleebaum, Bernd Bruegge, Barbara Paech & 2018 & ICSSP - International Conference on Software and System Process \\ \hline
  P74 & Quality and productivity outcomes relating to continuous integration in GitHub & Vladimir Filkov, Bogdan Vasilescu, Yue Yu, Huaimin Wang, Premkumar Devanbu & 2015 & ESEC/FSE Joint Meeting European Software Engineering Conference and Symposium on the Foundations of Software Engineering \\ \hline
  P75 & Scaling Continuous Integration & R. Owen Rogers & 2004 & International Conference on Extreme Programming and Agile Processes in Software Engineering \\ \hline
  P76 & Screening heuristics for project gating systems & Edi Shmueli, Zahy Volf & 2017 & ESEC/FSE Joint Meeting European Software Engineering Conference and Symposium on the Foundations of Software Engineering \\ \hline
  P77 & Sentiment analysis of Travis CI builds & Bruno Silva, Rodrigo Souza & 2017 & MSR - International Conference on Mining Software Repositories \\ \hline
  P78 & Software artefacts consistency management towards continuous integration: A roadmap & I. Perera, D. A. Meedeniya, I. D. Rubasinghe & 2019 & International Journal of Advanced Computer Science and Applications \\ \hline
  P79 & Software Quality Improvement Practices in Continuous Integration & Selin Aydin, İlgi Keskin Kaynak, Evren Çilden & 2019 & European Conference on Software Process Improvement \\ \hline
  P80 & Stakeholder Perceptions of the Adoption of Continuous Integration -- A Case Study & Maria Paasivaara, Teemu Arvonen, Eero Laukkanen & 2015 & Agile Conference \\ \hline
  P81 & Studying the impact of adopting continuous integration on the delivery time of pull requests & Joao Helis Bernardo, Uirá Kulesza, Daniel Alencar da Costa & 2018 & ICSE - International Conference on Software Engineering \\ \hline
  P82 & Successful extreme programming: Fidelity to the methodology or good teamworking? & Stephen Wood, George Michaelides, Chris Thomson & 2013 & Information and Software Technology \\ \hline
  P83 & Synthesizing Continuous Deployment Practices Used in Software Development & Chris Parnin, Akond Rahman, Eric Helms, Laurie Williams & 2015 & Agile Conference \\ \hline
  P84 & Team Pace Keeping Build Times Down & Graham Brooks & 2008 & Agile Conference \\ \hline
  P85 & Test activities in the continuous integration and delivery pipeline & Daniel Ståhl, Torvald Mårtensson, Jan Bosch & 2019 & Journal of Software: Evolution and Process \\ \hline
  P86 & The continuity of continuous integration: Correlations and consequences & Jan Bosch, Torvald Mårtensson, Daniel Ståhl & 2017 & Journal of Systems and Software \\ \hline
  P87 & The effects of individual XP practices on software development effort & Paul Rodrigues, Prakash Ramaswamy, S Kuppuswami, K Vivekanandan & 2003 & ACM SIGSOFT Software Engineering Notes \\ \hline
  P88 & The highways and country roads to continuous deployment & Marko Leppänen, Mika V Mäntylä, Juha Itkonen, Veli-Pekka Eloranta, Max Pagels, Simo Mäkinen, Tomi Männistö & 2015 & IEEE Software \\ \hline
  P89 & The impact of continuous integration on other software development practices: a large-scale empirical study & Vladimir Filkov, Yuming Zhou, Alexander Serebrenik, Yangyang Zhao, Bogdan Vasilescu & 2017 & ASE - International Conference on Automated Software Engineering \\ \hline
  P90 & The impact of the adoption of continuous integration on developer attraction and retention & Keheliya Gallaba, Yash Gupta, Yusaira Khan, Shane McIntosh & 2017 & MSR - International Conference on Mining Software Repositories \\ \hline
  P91 & The links between agile practices, interpersonal conflict, and perceived productivity & Lucas Gren & 2017 & EASE - Conference on Evaluation and Assessment in Software Engineering \\ \hline
  P92 & An empirical study examining the usage and perceived importance of XP practices & Jessica Zhang, Ann Fruhling & 2007 & AMCIS - Americas Conference on Information Systems \\ \hline
  P93 & The Tarpit – A general theory of software engineering & Pontus Johnson, Mathias Ekstedt & 2016 & Information and Software Technology \\ \hline
  P94 & Towards Agile Testing for Railway Safety-critical Software & Jin Guo, Yaxin Cao, Chang Rao, Yao Li, Nan Li, Jeff Lei & 2016 & Conference XP \\ \hline
  P95 & Towards Architecting for Continuous Delivery & Lianping Chen & 2015 & ICSA - IEEE International Conference on Software Architecture \\ \hline
  P96 & Towards quality gates in continuous delivery and deployment & Gerald Schermann, Jürgen Cito, Harald C. Gall, Philipp Leitner & 2016 & ICPC \\ \hline
  P97 & Trade-offs in continuous integration: assurance, security, and flexibility & Danny Dig, Michael Hilton, Nicholas Nelson, Timothy Tunnell, Darko Marinov & 2017 & ESEC/FSE Joint Meeting European Software Engineering Conference and Symposium on the Foundations of Software Engineering \\ \hline
  P98 & Transparency and contracts: continuous integration and delivery in the automotive ecosystem & Eric Knauss, Rob Van Der Valk, Patrizio Pelliccione, Rogardt Heldal, Patricia Lago, Jacob Juul & 2018 & ICSE - International Conference on Software Engineering \\ \hline
  P99 & Understanding similarities and differences in software development practices across domains & Pooyan Jamshidi, Christian Kästner, Markos Viggiato, Eduardo Figueiredo, Johnatan Oliveira & 2019 & ICGSE - International Conference on Global Software Engineering \\ \hline
  P100 & Usage, costs, and benefits of continuous integration in open-source projects & Timothy Tunnell, Michael Hilton, Kai Huang, Darko Marinov, Danny Dig & 2016 & ASE - International Conference on Automated Software Engineering \\ \hline
  P101 & Use and Misuse of Continuous Integration Features: An Empirical Study of Projects that (mis)use Travis CI & Keheliya Gallaba, Shane McIntosh & 2018 & IEEE Transactions on Software Engineering \\ \hline
  P102 & Using continuous integration and automated test techniques for a robust C4ISR system & Eray Tüzün, Erdoǧan Gelirli, H. Mehmet Yüksel, Emrah Biyikli, Buyurman Baykal & 2009 & ISCIS - International Symposium on Computer and Information Sciences \\ \hline
  P103 & Vulnerabilities in Continuous Delivery Pipelines? A Case Study & Christina Paule, Thomas F. Dullmann, Andre Van Hoorn & 2019 & ICSA - IEEE International Conference on Software Architecture \\ \hline
  P104 & Wait for it: determinants of pull request evaluation latency on GitHub & Yue Yu, Bogdan Vasilescu, Premkumar Devanbu, Vladimir Filkov, Huaimin Wang & 2015 & MSR - International Conference on Mining Software Repositories \\ \hline
  P105 & Why modern open source projects fail & Jailton Coelho, Marco Tulio Valente & 2017 & ESEC/FSE Joint Meeting European Software Engineering Conference and Symposium on the Foundations of Software Engineering \\ \hline
  P106 & Work practices and challenges in continuous integration: A survey with Travis CI users & Rodrigo Bonifacio, Marcel Reboucas, Gustavo Pinto, Fernando Castor & 2018 & Software - Practice and Experience \\ \hline

  \caption{Primary Studies selected in the review.} 
  \label{tab:selected_studies}
\end{longtable}

\end{appendices}


\newpage



\authorbiography[scale=0.15,width={3cm},imagewidth=3cm,wraplines=12,imagepos=l,overhang=0pt]{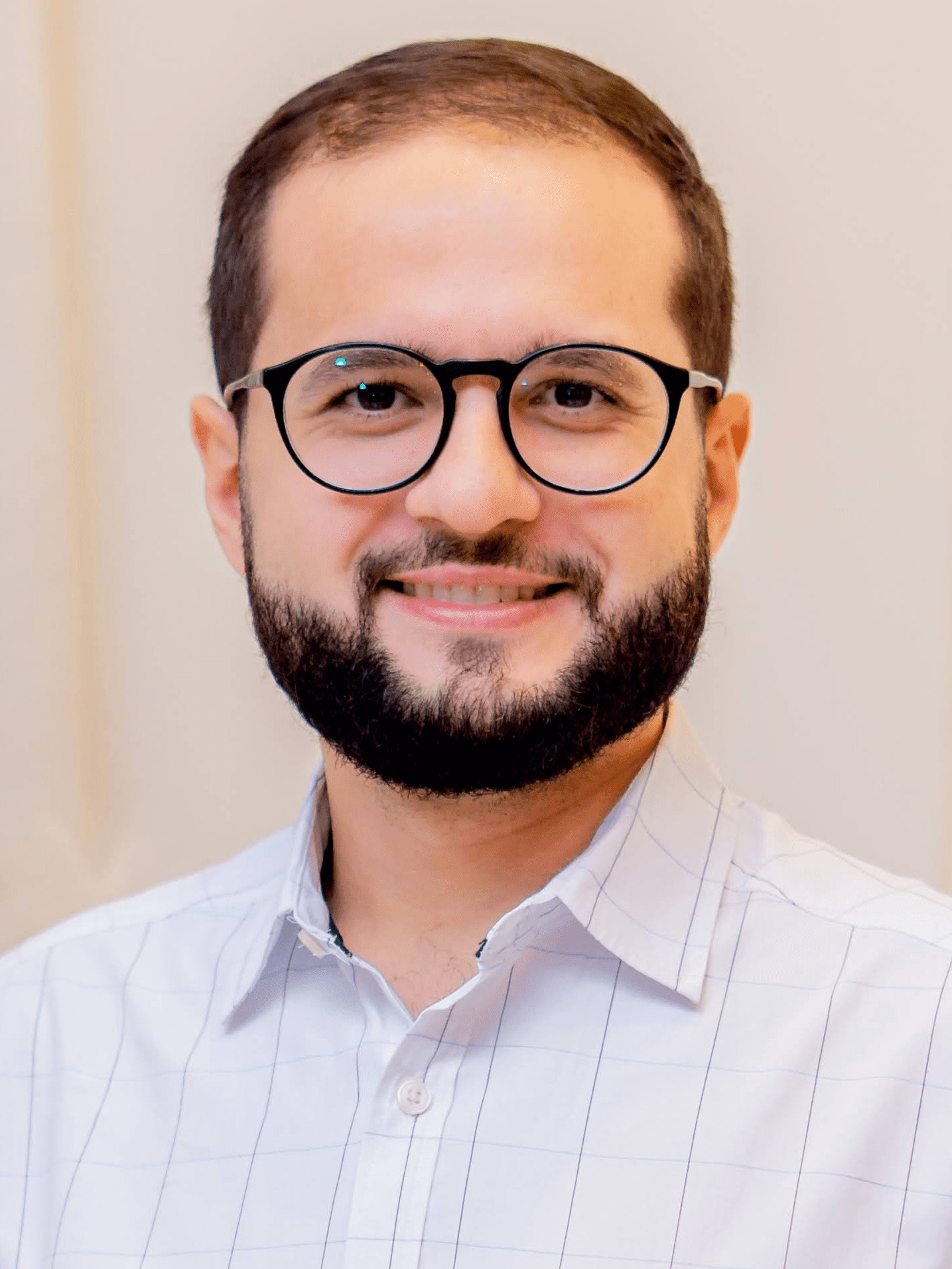}{\\Eliezio Soares}{
    Eliezio Soares is an adjunct professor at the  Federal Institute of Education, Science, and Technology of Rio Grande do Norte (IFRN), with experience in systems development. He is currently a PhD student at the Graduate Program in Systems and Computing of the Federal University of Rio Grande do Norte (UFRN), Brazil. In 2014 received a Master's degree in Systems and Computing from the same program. His research interests include empirical software engineering, mining software repositories, and continuous integration.
    \vspace{7mm}
    }%

\authorbiography[scale=0.2,width={3cm},imagewidth=3cm,wraplines=12,imagepos=l,overhang=0pt]{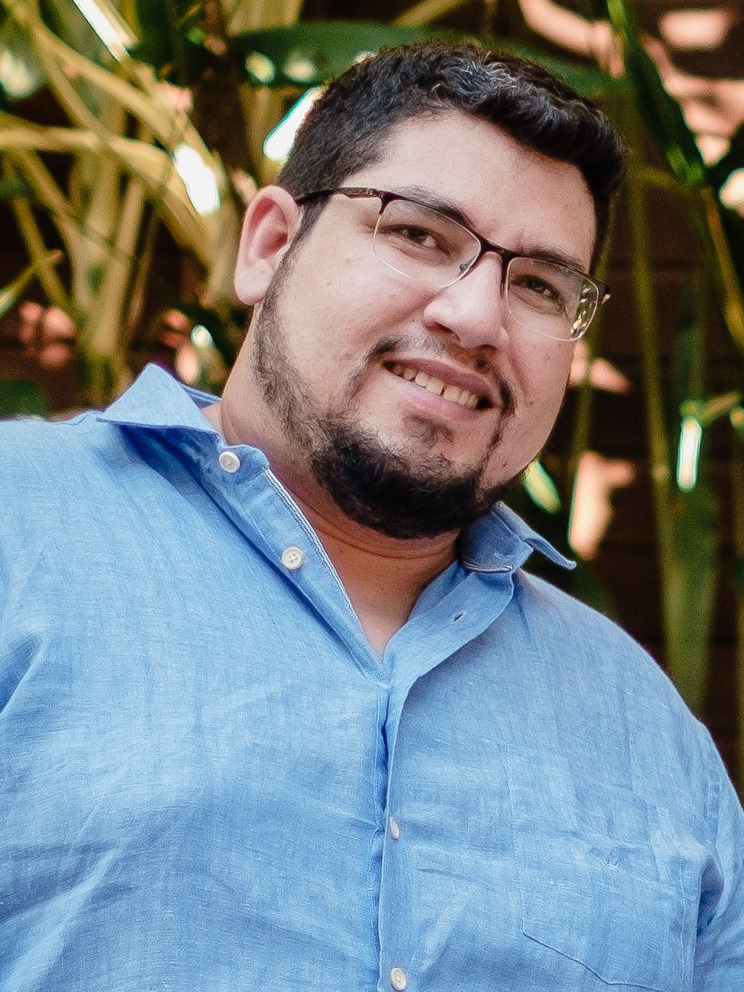}{\\Gustavo Sizílio}{
    Gustavo Sizílio Nery is an adjunct professor at the  Federal Institute of Education, Science, and Technology of Rio Grande do Norte with broader experience in Computer Science with an emphasis in Analysis and Systems Development and Software Engineering. Holds a Ph.D. in Computer Science at the Graduate Program in Systems and Computer of the Federal University of Rio Grande do Norte / UFRN with a Master's degree in the same program. His research interests are centered on empirical software engineering, mining software repositories, business intelligence, and applied data science.
    \vspace{2mm}
    }%

\authorbiography[scale=0.2,width={3cm},imagewidth=3cm,wraplines=12,imagepos=l,overhang=0pt]{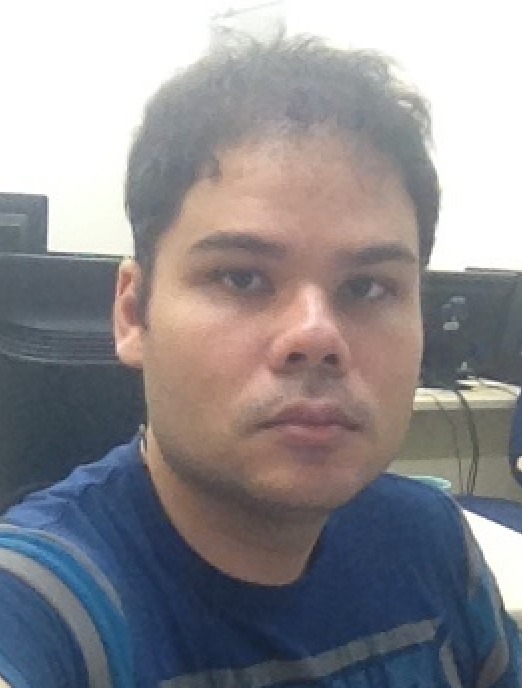}{\\Jadson Santos}{
    Jadson Santos is an Information Technology Analyst at the Federal University of Rio Grande do Norte (UFRN). He is graduated in Computer Engineering in 2005 from the same university. In 2016 received a Master’s degree in Systems and Computing in the Graduate Program of the Federal University of Rio Grande do Norte. He is currently PhD student in the same program. He has sixteen years of experience in developing corporate systems in Java. His main research interests are Continuous Integration, Continuous Delivery, DevOps, and Software Architecture. 
    \vspace{6mm}
    }%

\authorbiography[scale=0.1,width={3cm},imagewidth=3cm,wraplines=12,imagepos=l,overhang=0pt]{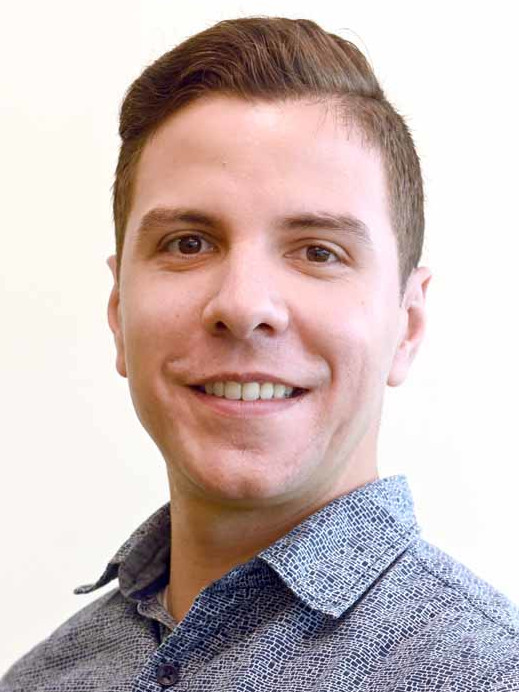}{\\Daniel Alencar}{
Daniel Alencar da Costa received the PhD degree in computer science from the Federal University of Rio Grande do Norte (UFRN) in 2017 followed by a postdoctoral fellowship at Queen's University, Canada, from 2017 to late 2018. He is a Senior Lecturer in the University of Otago, New Zealand. His research goals are to advance the body of knowledge of software engineering methodologies through empirical studies and to develop tools that can facilitate the software engineering practice.
\vspace{7mm}
}%

\authorbiography[scale=0.2,width={3cm},imagewidth=3cm,wraplines=12,imagepos=l,overhang=0pt]{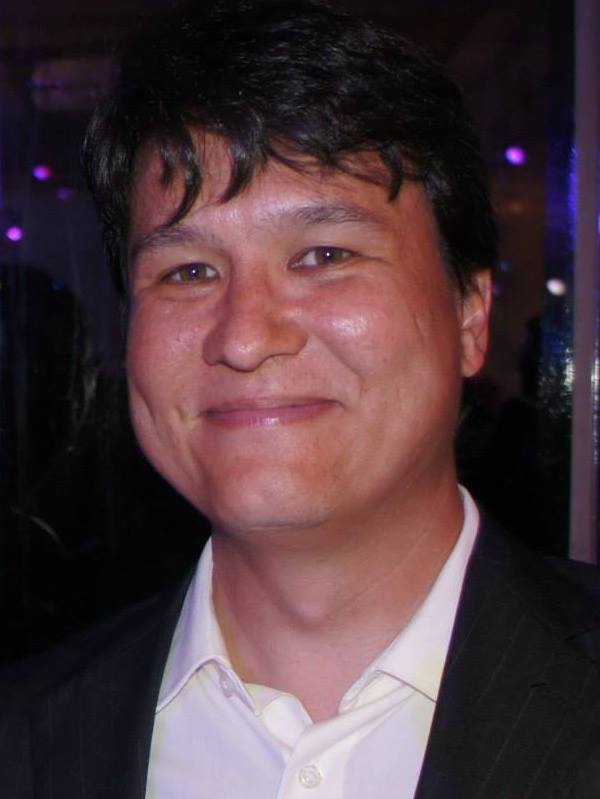}{\\Uirá Kulesza}{
Uirá Kulesza received the bachelor's degree in computer science from Federal University of Campina Grande, the MSc degree in computer science from the University of São Paulo, and the PhD degree in computer science from Pontifical Catholic University of Rio de Janeiro (PUC-Rio) in cooperation with University of Waterloo. He is an associate professor in the Department of Informatics and Applied Mathematics (DIMAp), Federal University of Rio Grande do Norte (UFRN), Brazil. He leads the Collaborative \& Automated Software Engineering (CASE) Research Group and Lab at the Digital Metropolis Institute (IMD) at UFRN (\url{http://caseufrn.github.io/}). His main research interests include software evolution, software architecture, software analytics, empirical software engineering, and DevOps education.  More about him and his work is available online at \url{http://www.dimap.ufrn.br/~uira}.
}%


\Authorbiography[totoc=true,BiographyName={Authors}] 




\begin{thebibliography}{9}


\bibitem{baker2016}BAKER M (2016) Reproducibility crisis. Nature, v. 533, n. 26, p. 353-66.

\bibitem{beck2005} Beck K, Andres C (2005) Extreme programming explained: embrace change. Addison-Wesley, Boston

\bibitem{Bernardo2018} Bernardo JH, Costa DAD, Kulesza U (2018) Studying the impact of adopting continuous integration on the delivery time of pull requests. Proceedings of the 15th International Conference on Mining Software Repositories - MSR 18. doi: 10.1145/3196398.3196421

\bibitem{travistorrent} Beller Moritz, Gousios Georgios, Zaidman Andy (2017) TravisTorrent: Synthesizing Travis CI and GitHub for Full-Stack Research on Continuous Integration. 
Proceedings of the 14th working conference on mining software repositories. 
\url{http://www.st.ewi.tudelft.nl/~mbeller/publications/2017_beller_gousios_zaidman_travistorrent_synthesizing_travis_ci_and_github_for_full-stack_research_on_continuous_integration.pdf}
Accessed 22 July 2020.


\bibitem{citheatre2017}(2017) CI theatre. In: ThoughtWorks. https://www.thoughtworks.com/radar/techniques/ci-theatre. Accessed 25 Aug 2020

\bibitem{cohen} Cohen J (1968) Weighted kappa: Nominal scale agreement provision for scaled disagreement or partial credit. 
Psychological Bulletin 70:213–220. doi = {10.1037/h0026256}

\bibitem{chen2010} Chen L, Babar MA, Zhang H (2010) Towards an Evidence-Based Understanding of Electronic Data Sources. doi = {10.14236/ewic/ease2010.17}

\bibitem{cruzes2011}Cruzes DS, Dyba T (2011) Recommended Steps for Thematic Synthesis in Software Engineering. 2011 International Symposium on Empirical Software Engineering and Measurement. doi: 10.1109/esem.2011.36

\bibitem{graziotin2015} Graziotin D, Wang X,  Abrahamsson P (2015) Do feelings matter? On the correlation of affects and the self‐assessed productivity in software engineering. Journal of Software: Evolution and Process, 27(7), 467-487. doi: 10.1002/smr.1673

\bibitem{Debbiche2014}Debbiche A, Dienér M, Svensson RB (2014) Challenges When Adopting Continuous Integration: A Case Study. Product-Focused Software Process Improvement Lecture Notes in Computer Science 17–32. doi: 10.1007/978-3-319-13835-0

\bibitem{debroy2018} Debroy V, Miller S, Brimble L (2018) Building lean continuous integration and delivery pipelines by applying DevOps principles: a case study at Varidesk. Proceedings of the 2018 26th ACM Joint Meeting on European Software Engineering Conference and Symposium on the Foundations of Software Engineering - ESEC/FSE 2018. doi: 10.1145/3236024.3275528

\bibitem{dikert2016} Dikert K, Paasivaara M, Lassenius C (2016) Challenges and success factors for large-scale agile transformations: A systematic literature review. Journal of Systems and Software 119:87–108. doi: 10.1016/j.jss.2016.06.013

\bibitem{duvall2013} Duvall PM, Matyas S, Glover A (2013) Continuous integration improving software quality and reducing risk. Addison-Wesley, Upper Saddle River, NJ

\bibitem{duvall2018} Duvall PM (2018) Continuous Delivery - Patterns and Anti-Patterns in the Software Lifecycle. In: dzone.com. https://dzone.com/refcardz/continuous-delivery-patterns. Accessed 7 Feb 2021

\bibitem{dyba2005} Dybå T, Kitchenham B, Jorgensen M (2005) Evidence-based software engineering for practitioners. IEEE Software 22:58–65. doi: 10.1109/ms.2005.6

\bibitem{dyba2007} Dyba T, Dingsoyr T, Hanssen GK (2007) Applying Systematic Reviews to Diverse Study Types: An Experience Report. First International Symposium on Empirical Software Engineering and Measurement (ESEM 2007). doi: 10.1109/esem.2007.59

\bibitem{dyba2008} Dybå T, Dingsøyr T (2008) Strength of evidence in systematic reviews in software engineering. Proceedings of the Second ACM-IEEE international symposium on Empirical software engineering and measurement - ESEM 08. doi: 10.1145/1414004.1414034

\bibitem{dyba2008a} Dybå T, Dingsøyr T (2008) Empirical studies of agile software development: A systematic review. Information and Software Technology 50:833–859. doi: 10.1016/j.infsof.2008.01.006

\bibitem{easterbrook2008} Easterbrook S, Singer J, Storey M A, Damian D (2008) Selecting empirical methods for software engineering research. In Guide to advanced empirical software engineering (pp. 285-311). Springer, London.

\bibitem{Embury2019} Embury SM, Page C (2019) Effect of Continuous Integration on Build Health in Undergraduate Team Projects. Software Engineering Aspects of Continuous Development and New Paradigms of Software Production and Deployment Lecture Notes in Computer Science 169–183. doi: 10.1007/978-3-030-06019-0

\bibitem{fauci2020} Fauci AS, Lane HC, Redfield RR (2020) Covid-19 — Navigating the Uncharted. New England Journal of Medicine 382:1268–1269. doi: 10.1056/nejme2002387

\bibitem{felidre2019} Felidre W, Furtado L, Costa DAD, et al (2019) Continuous Integration Theater. 2019 ACM/IEEE International Symposium on Empirical Software Engineering and Measurement (ESEM). doi: 10.1109/esem.2019.8870152

\bibitem{fitzgerald2017}Fitzgerald B, Stol K-J (2017) Continuous software engineering: A roadmap and agenda. Journal of Systems and Software 123:176–189. doi: 10.1016/j.jss.2015.06.063

\bibitem{fowler2006} Fowler M (2006) Continuous Integration. In: martinfowler.com. \url{https://martinfowler.com/articles/continuousIntegration.html}. Accessed 29 Jun 2020

\bibitem{fowler2013}Fowler M (2013) ExtremeProgramming. In: martinfowler.com. \url{https://martinfowler.com/bliki/ExtremeProgramming.html}.

\bibitem{fowler2017} Fowler M (2017) ContinuousIntegrationCertification. In: martinfowler.com. \url{https://martinfowler.com/bliki/ContinuousIntegrationCertification.html}. Accessed 26 Jun 2020

\bibitem{Ghaleb2019}Ghaleb TA, Costa DAD, Zou Y (2019) An empirical study of the long duration of continuous integration builds. Empirical Software Engineering 24:2102–2139. doi: 10.1007/s10664-019-09695-9

\bibitem{gousios2015}Gousios G, Zaidman A, Storey M A, Van Deursen A. (2015, May). Work practices and challenges in pull-based development: The integrator's perspective. In 2015 IEEE/ACM 37th IEEE International Conference on Software Engineering (Vol. 1, pp. 358-368). IEEE.

\bibitem{hilton2016} Hilton M, Tunnell T, Huang K, et al (2016) Usage, costs, and benefits of continuous integration in open-source projects. Proceedings of the 31st IEEE/ACM International Conference on Automated Software Engineering - ASE 2016. doi: 10.1145/2970276.2970358

\bibitem{homstrom2006}Holmstrom H, Conchuir E, Agerfalk P, Fitzgerald B (2006) Global Software Development Challenges: A Case Study on Temporal, Geographical and Socio-Cultural Distance. 2006 IEEE International Conference on Global Software Engineering (ICGSE06). doi: 10.1109/icgse.2006.261210

\bibitem{Johanssen2018}Johanssen JO, Kleebaum A, Paech B, Bruegge B (2018) Practitioners eye on continuous software engineering. Proceedings of the 2018 International Conference on Software and System Process - ICSSP 18. doi: 10.1145/3202710.3203150

\bibitem{kitchenham2007} Kitchenham B, Charters S (2007) Guidelines for performing Systematic Literature Reviews in Software Engineering. https://citeseerx.ist.psu.edu/viewdoc/summary?doi=10.1.1.117.471. Accessed 26 June 2020.

\bibitem{kitchenham2004} Kitchenham Barbara A; Dyba Tore, Jorgensen Magne (2004)
Evidence-based software engineering. https://10.1109/ICSE.2004.1317449

\bibitem{laukkanen2017} Laukkanen E, Itkonen J, Lassenius C (2017) Problems, causes and solutions when adopting continuous delivery—A systematic literature review. Information and Software Technology 82:55–79. doi: 10.1016/j.infsof.2016.10.001

\bibitem{leppanen2015} Leppanen M, Makinen S, Pagels M, et al (2015) The highways and country roads to continuous deployment. IEEE Software 32:64–72. doi: 10.1109/ms.2015.50

\bibitem{Meedeniya2019} Meedeniya DA, D. I, Perera I (2019) Software Artefacts Consistency Management towards Continuous Integration: A Roadmap. International Journal of Advanced Computer Science and Applications. doi: 10.14569/ijacsa.2019.0100411

\bibitem{perez2018} Rodríguez-Pérez G, Robles G, González-Barahona JM (2018) Reproducibility and credibility in empirical software engineering: A case study based on a systematic literature review of the use of the SZZ algorithm. Information and Software Technology 99:164–176. doi: 10.1016/j.infsof.2018.03.009

\bibitem{pinto2018} Pinto G, Castor F, Bonifacio R, Rebouças M (2018) Work practices and challenges in continuous integration: A survey with Travis CI users. Software: Practice and Experience 48:2223–2236. doi: 10.1002/spe.2637

\bibitem{ralph2020} Ralph P, Baltes S, Adisaputri G. et al. Pandemic programming. Empir Software Eng 25, 4927–4961 (2020). https://doi.org/10.1007/s10664-020-09875-y

\bibitem{Raoul2018}Vallon R, da Silva Estacio B J, Prikladnicki R, Grechenig T. (2018). Systematic literature review on agile practices in global software development. Information and Software Technology, 96, 161-180.

\bibitem{Rausch2017}Rausch T, Hummer W, Leitner P, Schulte S (2017) An Empirical Analysis of Build Failures in the Continuous Integration Workflows of Java-Based Open-Source Software. 2017 IEEE/ACM 14th International Conference on Mining Software Repositories (MSR). doi: 10.1109/msr.2017.54

\bibitem{robles2010} Robles G (2010) Replicating MSR: A study of the potential replicability of papers published in the Mining Software Repositories proceedings. 2010 7th IEEE Working Conference on Mining Software Repositories (MSR 2010). doi: 10.1109/msr.2010.5463348

\bibitem{rogers2004} Rogers RO (2004) Scaling Continuous Integration. Extreme Programming and Agile Processes in Software Engineering Lecture Notes in Computer Science 68–76.

\bibitem{russo2021} Russo D, Hanel P H P, Altnickel S et al. Predictors of well-being and productivity among software professionals during the COVID-19 pandemic – a longitudinal study. Empir Software Eng 26, 62 (2021). https://doi.org/10.1007/s10664-021-09945-9

\bibitem{shahin2017} Shahin M, Babar MA, Zhu L (2017) Continuous Integration, Delivery and Deployment: A Systematic Review on Approaches, Tools, Challenges and Practices. IEEE Access 5:3909–3943. doi: 10.1109/access.2017.2685629

\bibitem{soares_eliezio_2020} Soares E, Sizilio G, Santos J, Alencar D, Kulesza U (2021) SLR Artifacts - CONTINUOUS INTEGRATION QUALITY IMPACTS (v.1.0.2) [Data set]. In: Zenodo. https://doi.org/10.5281/zenodo.4545623. Accessed 10 Aug 2021

\bibitem{stahl2013} Ståhl D, Bosch J (2013) Experienced Benefits of Continuous Integration in Industry Software Product Development: A Case Study. Artificial Intelligence and Applications / 794: Modelling, Identification and Control / 795: Parallel and Distributed Computing and Networks / 796: Software Engineering / 792: Web-based Education. doi: 10.2316/p.2013.796-012

\bibitem{stahl2014} Ståhl D, Bosch J (2014) Modeling continuous integration practice differences in industry software development. Journal of Systems and Software 87:48–59. doi: 10.1016/j.jss.2013.08.032

\bibitem{stahl2014b} Ståhl D, Bosch J (2014) Automated software integration flows in industry: a multiple-case study. Companion Proceedings of the 36th International Conference on Software Engineering - ICSE Companion 2014. doi: 10.1145/2591062.2591186

\bibitem{vassalo2018} Vassallo C, Palomba F, Gall HC (2018) Continuous Refactoring in CI: A Preliminary Study on the Perceived Advantages and Barriers. 2018 IEEE International Conference on Software Maintenance and Evolution (ICSME). doi: 10.1109/icsme.2018.00068

\bibitem{vassalo2019} Vassallo C, Proksch S, Gall HC, Penta MD (2019) Automated Reporting of Anti-Patterns and Decay in Continuous Integration. 2019 IEEE/ACM 41st International Conference on Software Engineering (ICSE). doi: 10.1109/icse.2019.00028

\bibitem{vasilescu2015} Vasilescu B, Yu Y, Wang H, et al (2015) Quality and productivity outcomes relating to continuous integration in GitHub. Proceedings of the 2015 10th Joint Meeting on Foundations of Software Engineering - ESEC/FSE 2015. doi: 10.1145/2786805.2786850

\bibitem{vasilescu2014} Vasilescu B, Schuylenburg SV, Wulms J, et al (2014) Continuous Integration in a Social-Coding World: Empirical Evidence from GitHub. 2014 IEEE International Conference on Software Maintenance and Evolution. doi: 10.1109/icsme.2014.62

\bibitem{viggiato2019}Viggiato M, Oliveira J, Figueiredo E, et al (2019) Understanding Similarities and Differences in Software Development Practices Across Domains. 2019 ACM/IEEE 14th International Conference on Global Software Engineering (ICGSE). doi: 10.1109/icgse.2019.00013

\bibitem{Volf2017} Volf Z, Shmueli E (2017) Screening heuristics for project gating systems. Proceedings of the 2017 11th Joint Meeting on Foundations of Software Engineering - ESEC/FSE 2017. doi: 10.1145/3106237.3117766

\bibitem{wheelan1996} Wheelan S A, Hochberger J M (1996). Validation studies of the group development questionnaire. Small group research, 27(1), 143-170.

\bibitem{Yu2016} Yu Y, Yin G, Wang T, et al (2016) Determinants of pull-based development in the context of continuous integration. Science China Information Sciences. doi: 10.1007/s11432-016-5595-8

\bibitem{zampetti2017} Zampetti F, Scalabrino S, Oliveto R, et al (2017) How Open Source Projects Use Static Code Analysis Tools in Continuous Integration Pipelines. 2017 IEEE/ACM 14th International Conference on Mining Software Repositories (MSR). doi: 10.1109/msr.2017.2

\bibitem{zhao2017} Zhao Y, Serebrenik A, Zhou Y, et al (2017) The impact of continuous integration on other software development practices: A large-scale empirical study. 2017 32nd IEEE/ACM International Conference on Automated Software Engineering (ASE). doi: 10.1109/ase.2017.8115619

%
%


\end{thebibliography}
\end{document}